\documentclass[a4paper,fleqn,usenatbib]{mnras}

\usepackage{tabularx}
\usepackage{newtxtext,newtxmath}

\usepackage[T1]{fontenc}
\usepackage{ae,aecompl}
\usepackage{float}
\usepackage{xfrac}
\usepackage[caption = false]{subfig}

\usepackage{graphicx}	
\usepackage{amsmath}	
\usepackage{amssymb}	

\title[Ionised outflow properties of local ULIRGs]{Quantifying the AGN-driven outflows in ULIRGs (QUADROS) III: Measurements of the radii and kinetic powers of 8 near-nuclear outflows}

\author[R.A.W. Spence et al.]{
R. A. W. Spence,\thanks{E-mail: rspence1@sheffield.ac.uk}
C. N. Tadhunter,
M. Rose
and J. Rodr\'{i}guez Zaur\'{i}n
\\
Department of Physics \& Astronomy, University of Sheffield,\\ Sheffield, S3 7RH, UK\\
}

\date{Accepted XXX. Received YYY; in original form ZZZ}

\pubyear{2018}

\begin{document}
\label{firstpage}
\pagerange{\pageref{firstpage}--\pageref{lastpage}}
\maketitle

\begin{abstract}
As part of the QUADROS project to quantify the impact of AGN-driven outflows in rapidly evolving galaxies in the local universe, we present observations of 8 nearby ULIRGs ($0.04 < z < 0.2$) taken with the ISIS spectrograph on the \textit{William Herschel Telescope} (WHT), and also summarise the results of the project as a whole. Consistent with \cite{Rose2018}, we find that the outflow regions are compact (0.08 < $R_{[OIII]}$ < 1.5 kpc), and the electron densities measured using the [SII], [OII] trans-auroral emission-line ratios are relatively high  (2.5 < log n$_{e}$(cm$^{-3}$) < 4.5, median log n$_{e}$(cm$^{-3})\sim 3.1$). Many of the outflow regions are also significantly reddened (median E(B-V) $\sim$ 0.5). Assuming that the de-projected outflow velocities are represented by the 5$\rm^{th}$ percentile velocities ($v_{05}$) of the broad, blueshifted components of [OIII]$\lambda$5007, we calculate relatively modest mass outflow rates ($0.1 < \dot{M} < 20$ M$_{\sun}$ yr$^{-1}$, median $\dot{M}$ $\sim$ 2 M$_{\sun}$ yr$^{-1}$), and find kinetic powers as a fraction of the AGN bolometric luminosity ($\dot{F} = \dot{E}/L_{bol}$) in the range 0.02 < $\dot{F}$ < 3 \%, median $\dot{F}$ $\sim$ 0.3 \%). The latter estimates are in line with the predictions of multi-stage outflow models, or single-stage models in which only a modest fraction of the initial kinetic power of the inner disk winds is transferred to the larger-scale outflows. Considering the QUADROS sample as a whole, we find no clear evidence for correlations between the properties of the outflows and the bolometric luminosities of the AGN, albeit based on a sample that covers a relatively small range in $L_{bol}$. Overall, our results suggest that there is a significant intrinsic scatter in outflow properties of ULIRGs for a given AGN luminosity.
\end{abstract}




\section{Introduction}

Over the past two decades, Active Galactic Nuclei (AGN) induced outflows have been detected in merging galaxies across all gas phases (ionised, e.g. \citealt{Rodriguez2013, Harrison2012, Harrison2014}; neutral e.g. \citealt{Rupke2005c, Morganti2016} and molecular e.g. \citealt[]{spoon2013, Cicone2014, Morganti2015}). In order to reproduce the observed correlations between the central black hole (BH) and host galaxy properties, such outflows are now routinely incorporated into hydrodynamical simulations \citep{DiMatteo2005, Springel2005, Johansson2009}. Results from these simulations suggest that the AGN-induced outflows have a substantial effect on the growth of galaxies. However, from an observational perspective, considerable uncertainties remain about their net impact. \par

In the case of warm ionised outflows, which are observed via emission lines at optical wavelengths, this uncertainty arises from a lack of accurate diagnostics of the properties of the outflowing gas. Although much recent attention has been paid to galaxies at high redshifts, which show signs of AGN activity \citep[e.g][]{nesvadba2008, canodiaz2012, Harrison2012, Harrison2014, Carniani2015, Carniani2016, Perna2015}, there is a limit to what can be learnt in detail about the co-evolution of black holes and galaxy bulges in such objects because of both their faintness, and the fact that many important diagnostic lines are shifted out of the optical/near-IR wavelength range. \par

Fortunately, the hierarchical evolution of galaxies is continuing in the local Universe, albeit at a reduced rate. Ultra Luminous Infra-red Galaxies (ULIRGs: $\rm L_{IR} > 10^{12} L_{\sun}$) \citep{Sanders1996}, which almost ubiquitously show signs of being
involved in major mergers (i.e. tidal tails, multiple nuclei), represent local analogues of the extreme starburst galaxies detected in the distant Universe. However they have the considerable advantage of being close enough to study in detail. Furthermore, the subset of ULIRGs with optical AGN show evidence for warm outflows in the form of broad (full-width at half-maximum, FWHM > 1000 km s$^{-1}$) and blueshifted ($|v_{out}|$ > $500$km s$^{-1}$) emission lines \citep[e.g.][]{Spoon2009, Rodriguez2013, Arribas2014}. However, the high degree of uncertainty in estimates of the radial extents, reddening and densities of warm outflows mean their fundamental properties (e.g. mass outflow rates, kinetic powers) have not yet been determined with any accuracy.

To address these issues, we are undertaking a programme that combines high resolution \textit{Hubble Space Telescope} (\textit{HST}) imaging with wide-spectral-coverage spectroscopic observations to measure outflow properties of local ULIRGs - the Quantifying Ulirg Agn-DRiven OutflowS (QUADROS) project. We take advantage of a technique pioneered by \cite{Holt2011} which uses the trans-auroral [SII]$\lambda\lambda$(4067/6725) and [OII]$\lambda\lambda$(3727/7320) emission lines ratios to simultaneously determine the density and reddening. The full QUADROS sample, consisting of 15 objects, is described in Paper I \citep{Rose2018} where we present deep VLT/XShooter spectroscopic observations of 9 objects. Paper II \citep[][submitted]{Tadhunter2018} presents \textit{HST} Advanced Camera for Surveys (ACS) imaging of 8 objects with optical AGN. \par

In this third instalment of the project, we present spectroscopy of a further 7 local ULIRGs in our QUADROS sample, plus one additional ULIRG, observed with the Intermediate dispersion Spectrograph and Imaging System (ISIS) on the William Herschel Telescope (WHT) on La Palma. \S\ref{sec:observations} describes the sample selection, observations and data reduction. In \S\ref{sec:results} we provide measurements of the radii and kinematics of the outflows. We then use the trans-auroral line ratios to  estimate the electron densities and reddening, and hence derive the mass outflow rates and kinetic powers of the warm AGN-driven outflows, which we compare with those for neutral and molecular outflows in \S\ref{sec:powers}. We also investigate possible links between the warm outflows and the properties of the AGN (\S\ref{sec:links}), and  consider whether the densities determined using the high-critical-density trans-auroral emission lines are representative of the warm outflows as a whole (\S\ref{sec:origin}). \par

Throughout the paper we assume a cosmology with $\rm H_{0} = 73 km s^{-1} Mpc^{-1},  \Omega_{0} = 0.27, \Omega_{\Lambda} = 0.73.$

\section{Sample selection, observations and data reduction}
\label{sec:observations}

\subsection{The sample}
\label{sec:sample} 

The full QUADROS sample of 15 objects is a 90\% complete sample of local ULIRGs selected from the 1Jy sample of \citet{Kim1998} that show evidence for warm outflows based on their [OIII] emission line profiles \citep[see][for details]{Rose2018}. Most objects are classified as having type II Seyfert nuclei based on the emission line diagnostic criteria of \cite{Yuan2010}, have right ascensions (RA) in the range 12hr < RA < 02hr, declinations $\rm \delta > -25$ degrees and redshifts $z < 0.175$. However, F15462--0450 from \cite{Rose2018} is a type I AGN. In addition, based on the rise in the continuum at the red end of its optical spectrum (see Figure \ref{fig:spectra}), and detection of broad components to both the H$\alpha$ and Pa$\alpha$ emission lines \citep{Veilleux1999, Rodriguez2013}, F23060+0505 also has a reddened type I AGN component. \par

In paper I, \cite{Rose2018} present an analysis of 7 (mostly southern) objects in the QUADROS sample, plus two additional 
ULIRGs, observed with VLT/XShooter. \par 

Here we present WHT/ISIS observations of a further 7 objects from the sample, plus one additional Seyfert ULIRG -- IRAS F05189--2524 -- which meets our spectral and redshift criteria, but falls outside of the RA and declination ranges for the QUADROS sample. This object was observed to fill a gap in the observing schedule. Note that, for the remaining ULIRG in the sample (IRAS F13428+5608), a detailed study of the off-nuclear ionised emission has been published separately (see \citealt{Spence2016}). The current paper will focus on the near-nuclear outflows, and F13428+5608, which was observed with only a single off-nuclear slit due to its extensive ionised nebula, will be excluded from the discussion. Two of the targets in this sample (F01004--2237 \& F05189--2524) were also observed with HST/STIS (long-slit spectroscopy) and 4 objects (F13428+5608, F14394+5332E, F17044+6720 \& F17179+5444) using HST/ACS narrow-band imaging \citep[see][submitted]{Tadhunter2018}. \par

Table \ref{tab:properties} shows the basic properties of the objects in the sample, and Table \ref{tab:log} gives details of the observations.  \\

\subsection{WHT/ISIS Observations}
\label{ref:obs}

Long-slit spectra for the objects considered in this paper were taken in June 2014 and September 2015 with the ISIS dual-beam spectrograph\footnote{http://www.ing.iac.es/astronomy/instruments/isis} on the 4.2m WHT on La Palma, Spain. These observations were optimised to make accurate measurements of the trans-auroral [SII]$\lambda$4073 and [OII]$\lambda$7320 emission features, in order to facilitate estimation of the electron densities and reddening for the
warm gas \citep{Holt2011,Rose2018}. We used the R300B and R316R gratings on the blue and red arms respectively, along with a dichroic cutting at 6100\AA{}, to achieve spectral coverage of 3600 - 8800\AA. This ensured that the key emission lines were contained in the spectral range that is relatively unvignetted. \par

Use of a 1.5" slit resulted in spectral resolutions of 5.4(5.6)$\pm$0.1\AA{} on the blue arm, and 5.1(5.4)$\pm$0.2(0.1) \AA{} on the red arm for the 2014(2015) observations, as measured using the mean spectral FWHM of several prominent night-sky lines. This corresponds to 272(283)$\pm$5 km s$^{-1}$ at 5938\AA{}, and 222(236)$\pm$9(4) km s$^{-1}$ at 6876\AA{}, for the blue and red arms, respectively. In order to mimimise the effects of differential atmospheric refraction, the objects were observed with the slit aligned along the parallactic angle for the centre of the observations. A 2x2 binning mode was used, resulting in a spatial scale of 0.4 arcsec pix$^{-1}$, and dispersions of 1.73\AA{} and 1.84\AA{} on the blue and red arms respectively.  The integration time per object ranged between 2700 and 6000 seconds per arm. We also took observations of A-type stars at the midpoint of each set of observations in order to facilitate removal of telluric absorption features from the target spectra, as well as three spectro-photometric standard stars per night for accurate flux calibration. \par 

Unfortunately WHT/ISIS does not provide acquisition images that could be used for the calculation of the seeing. The RoboDIMM seeing monitor on La Palma does monitor the seeing each night, however it points continuously towards stars close to the zenith and so is not accurate for the positions of our individual objects. Instead, we have used the spatial FWHM of the telluric standard stars, as measured from our long-slit observations, to estimate the seeing for each object. \par 

To do this we extracted spatial slices across a $\sim$10\AA\, wavelength range close to the wavelengths of the redshifted [OIII]$\lambda$5007 in the 2D spectra of the telluric stars, and fit a single Gaussian to the resultant 1D profiles using DIPSO. This technique has the advantage over 2D DIMM seeing estimates
in that it accounts for the integration of the seeing disk across the spectroscopic slit (see the discussion in \citealt{Rose2018}). The 1D seeing (FWHM) estimates obtained in this way ranged from 1.07 to 1.63 arcsec across the two nights of observations. The estimates for the individual objects are displayed in Table \ref{tab:log}. 

Note that the exposure times for the telluric standard stars were short compared with those of the science targets, so the errors on the 1D Gaussian fits to the telluric star profiles underestimate the true uncertainty on the seeing, which is likely to be dominated by seeing variations over the whole observation period for each target. Therefore we used the RoboDIMM monitor measurements -- taken every two minutes -- to provide an estimate of the likely variation in the seeing over the observation period, and use the standard error on the mean of the RoboDIMM measurements as a more realistic estimate of the uncertainty in the seeing.
\begin{table*}
	\centering
	\caption{Properties of the ULIRGs discussed in this paper. Column (1): object designation in the IRAS Bright Galaxy Survey (\citealt{Soifer1987}); Column (2): optical redshifts measured from the WHT/ISIS spectra in this paper; Columns (3) and (4): right ascensions and declination of the IRAS source position as listed in the SIMBAD astronomical database; Column (5): IR luminosity; Columns (6) and (7): nuclear structure and separation from \citealt{Veilleux2002}; Column (8): Interaction class from \citealt{Veilleux2002}: Class III: pre-merger; class IV: merger; tpl: Triple; Iso: Isolated.
			}
	\label{tab:properties}
	\begin{tabular}{lcccccccr} 
		\hline
		Object name & z & RA & Dec. & log $L_{IR}$ & Nuclear structure & Nuclear separation & IC \\
		IRAS & & (J2000.0) & (J2000.0) & ($L_{\sun}$) & & (kpc) & \\
		\\
		(1) & (2) & (3) & (4) & (5) & (6) & (7) & (8)\\
		\hline\\
		F01004 -- 2237 & 0.11783 $\pm$ 0.00009 & 01 02 49.9 & -- 22 21 57 & 12.28  & Single & -- & IIIb\\
		F05189 -- 2524 & 0.04275 $\pm$ 0.00007 & 05 21 01.4 & -- 25 21 45 & 12.07 & Single & -- & IVb\\
		F13428 + 5608 & 0.03842 $\pm$ 0.00004  & 13 44 42.1 & + 55 53 13 & 12.14 & Double/Multiple & 0.7 & IVb\\
		F14394 + 5332E & 0.10517 $\pm$ 0.00017 & 14 41 04.4 & + 53 20 09 & 12.08 & Multiple & 54.0 & Tpl\\
		F17044 + 6720 & 0.13600 $\pm$ 0.00014 & 17 04 28.5 & + 67 16 28 & 12.17 & Single & -- & IVb\\
		F17179 + 5444 & 0.14768 $\pm$ 0.00015 & 17 18 54.4 & + 54 41 48 & 12.24 & Single & -- & IVb\\
		F23060 + 0505 & 0.17301 $\pm$ 0.00007 & 23 08 34.0 & + 05 21 29 & 12.48 & Single & -- & IVb\\
		F23233 + 2817 & 0.11446 $\pm$ 0.00010 & 23 25 49.4 & + 28 34 21 & 12.04 & Single & -- & Iso.\\
		F23389 + 0303N & 0.14515 $\pm$ 0.00013 & 23 41 30.3 & + 03 17 27 & 12.13 & Double & 5.2 & IIIb\\
		\hline
	\end{tabular}
\end{table*} 

\begin{table*}
	\centering
	\caption{Log of the spectroscopic observations. Column (1): Object designation in IRAS BGS; Column (2): Date of observation; Column (3): Position angle of the spectroscopic slit; Column (4): Total exposure time per arm; Column (5): Range of airmass during observations; Column (6): Estimated 1D seeing measured from the FWHM of a telluric standard star, which was observed at the same time and with approximately the same conditions as the target. The error is deduced from the standard error on the mean of the seeing variation over the observation period, as measured by the RoboDIMM seeing monitor. Column (7): Galactic extinction in A$_{V}$ from \citealt{Schlafly2011}. Column (8): The size of the nuclear extraction aperture in kpc. Column (9): The pixel scale for our adopted cosmology.  
			}
	\label{tab:log}
	\begin{tabular}{lcccccccr} 
		\\
		\hline
		Object name & Date & PA & Integration  & Airmass & Seeing$_{1D}$ FWHM & GA$_{V}$ & Aperture & Scale \\
		IRAS & & (degrees) & (s) &  & (arcsec) & (mag.) & (kpc) & (kpc/")\\
		\\
		(1) & (2) & (3) & (4) & (5) & (6) & (7) & (8) & (9)\\
		\hline\\
		F01004 -- 2237 & 09/2015 & 5 & 6000 & 1.59 -- 1.68 & 1.43 $\pm$ 0.12 & 0.048 & 4.9 & 2.033\\ 
		F05189 -- 2524 & 09/2015 & 340 & 2700  & 1.83 -- 1.98 & 1.63 $\pm$ 0.05 & 0.080 & 4.8 & 0.808\\
		F13428 + 5608 & 06/2014 & 23 & 3000 & 1.15 -- 1.19 & 1.20 $\pm$ 0.35 & 0.022 & -- & --  \\
		F14394 + 5332E & 06/2014 & 130 & 6000 &1.13 -- 1.30 & 1.25 $\pm$ 0.07 & 0.031 & 5.2 & 1.849 \\
		F17044 + 6720 & 06/2014 & 145 & 6000 & 1.29 -- 1.40 & 1.61 $\pm$ 0.07 & 0.081 & 4.6 & 2.300\\
		F17179 + 5444 & 06/2014 & 105 & 6000 & 1.25 -- 1.57 & 1.60 $\pm$ 0.06 & 0.082 & 4.9 & 2.471 \\
		F23060 + 0505 & 09/2015 & 325 & 5400 & 1.11 -- 1.28  & 1.07 $\pm$ 0.11 & 0.173 & 5.6 & 2.813\\
		F23233 + 2817 & 09/2015 & 282 & 6000 & 1.28 -- 1.83 & 1.17 $\pm$ 0.15 & 0.331 & 4.7 & 1.972\\
		F23389 + 0303N & 09/2015 & 0 & 5400 & 1.11 -- 1.14 & 1.44 $\pm$ 0.10 & 0.145 & 4.9 & 2.427\\
		\hline
	\end{tabular}
\end{table*}

\subsubsection{Data Reduction}
\label{sec:reduction}

The data were reduced (bias-subtracted, flat-field corrected, cleaned of cosmic rays, wavelength calibrated and flux calibrated) and straightened before extraction of the individual spectra using standard packages in IRAF, and the STARLINK packages FIGARO and DIPSO. The wavelength calibration accuracy, determined using the mean shift between the published\footnote{http://www.eso.org/observing/dfo/quality/UVES/txt/sky/} and measured wavelengths of night-sky emission lines, was $\sim$0.5(0.2)\AA{} and $\sim$0.5(0.3)\AA{} for the 2014(2015) blue and red arms, respectively. The estimated uncertainty for the relative flux calibration across the full spectral range of the observations was $\pm$5\%, based on the comparison of the response curves of the three spectrophotometric standard stars observed in each run. An important detail to note is that the pixel scales of the blue and red arms of ISIS are different (0.4" and 0.44" respectively). This was corrected using the ISTRETCH command within FIGARO, resulting in a common pixel scale of 0.4" across both arms. The data from the red arm were also corrected for telluric absorption features using the A-type stars observed at similar time and airmass as each science target. 

\subsubsection{Aperture Extraction}
\begin{figure*}
	\vspace{-5mm}
	\subfloat{\includegraphics[width = 3in]{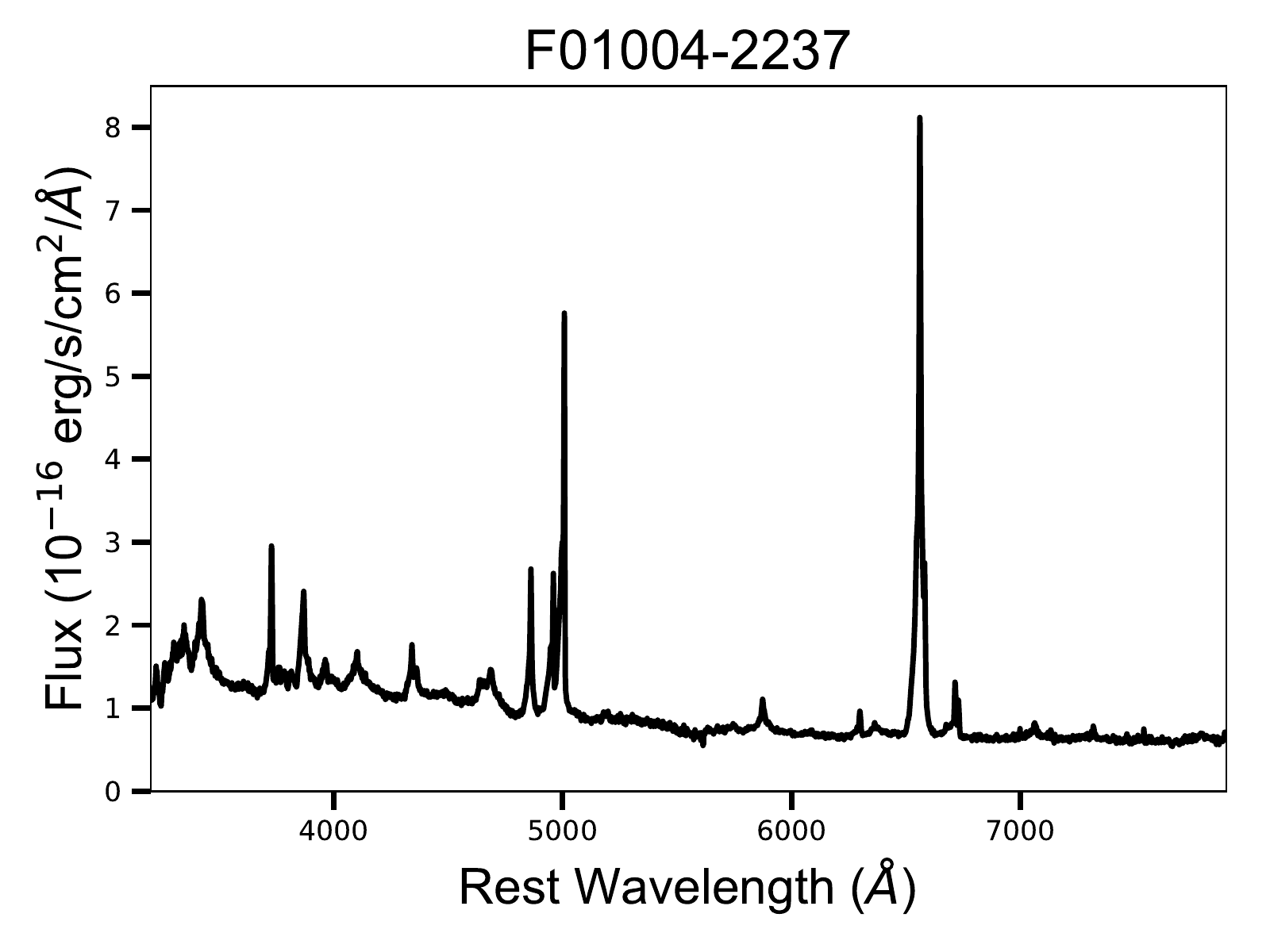}} 
	\subfloat{\includegraphics[width = 3in]{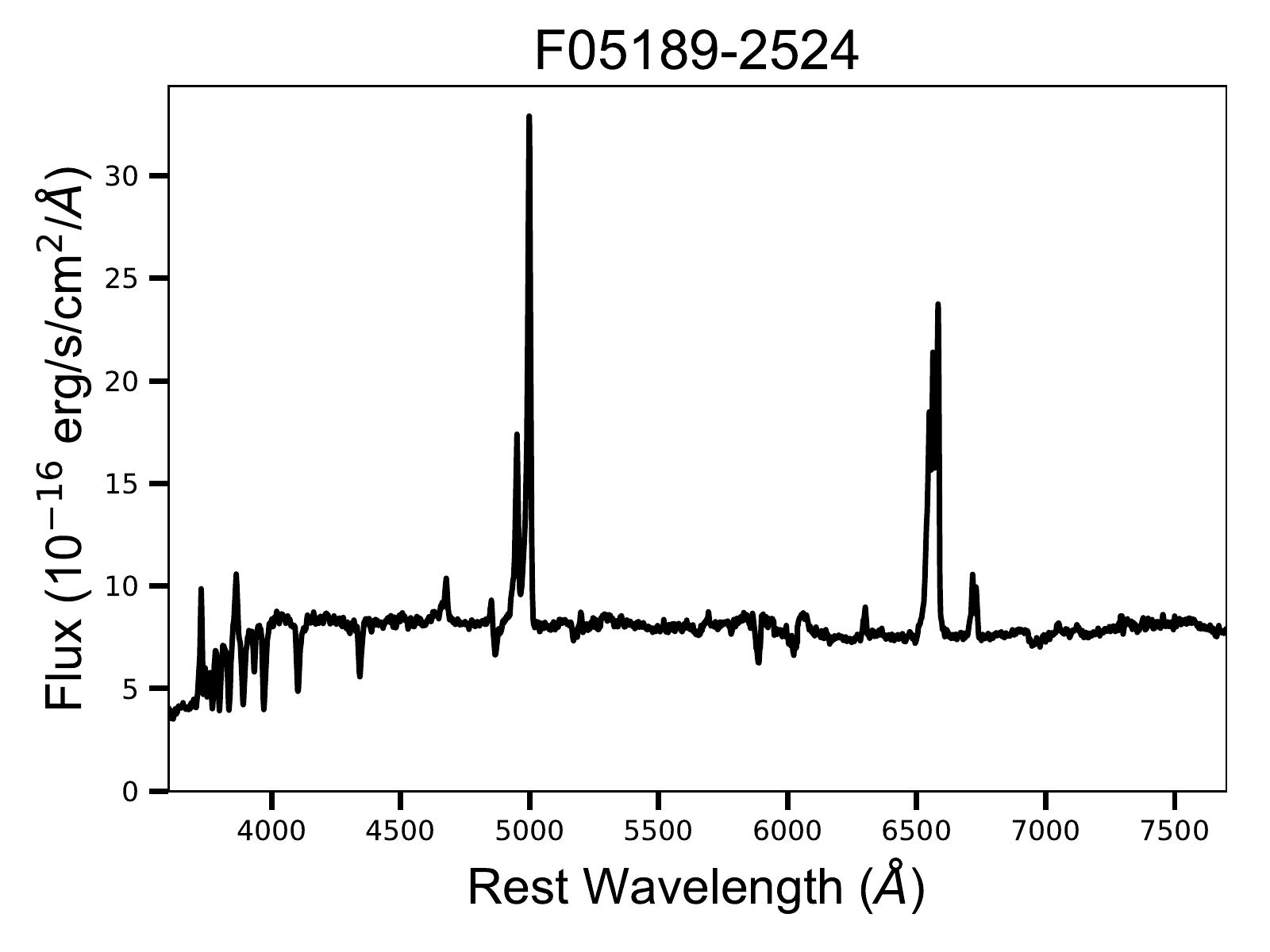}}\\  
	\vspace{-3.0mm}
	\subfloat{\includegraphics[width = 3in]{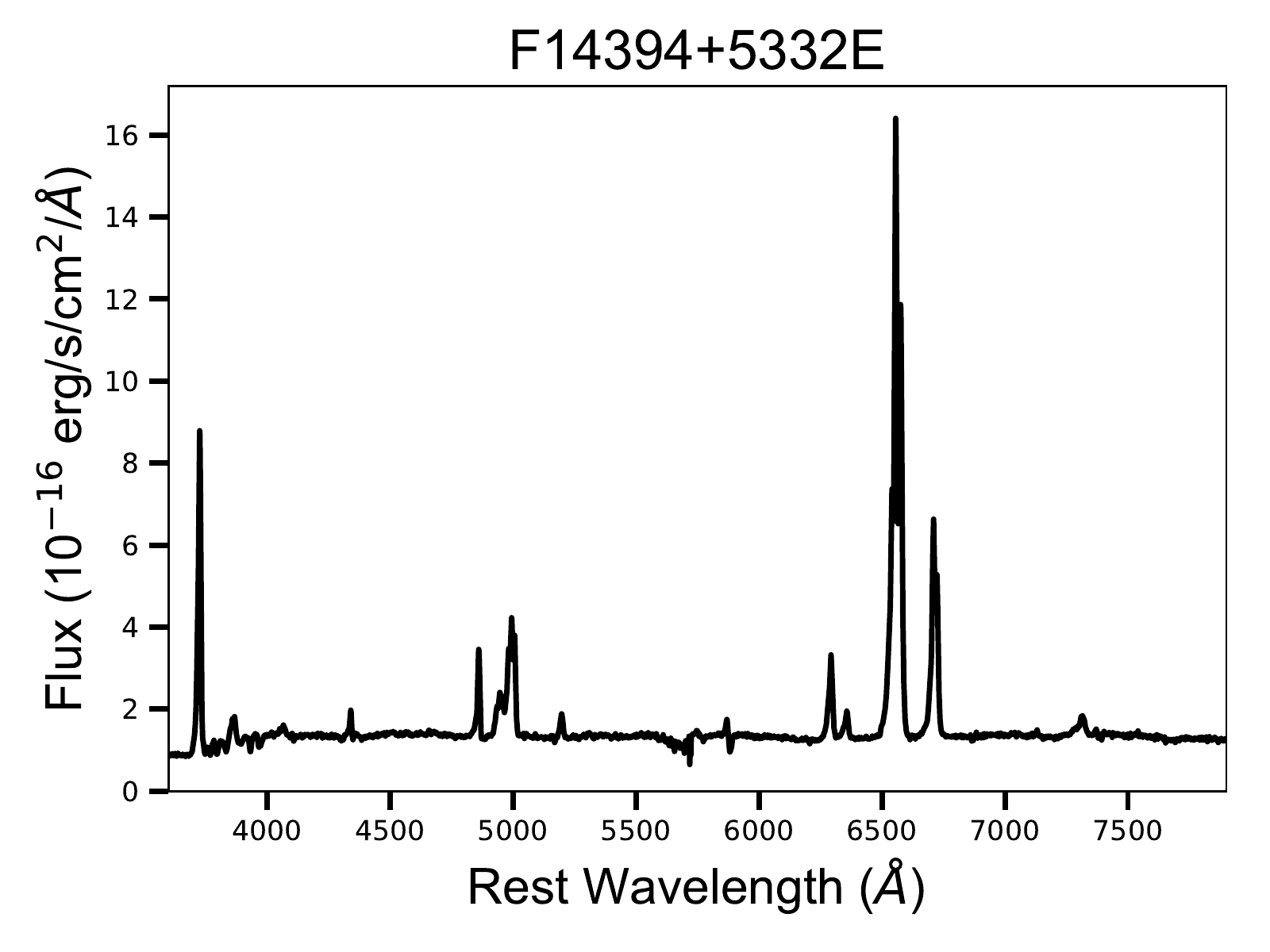}}
	\subfloat{\includegraphics[width = 3in]{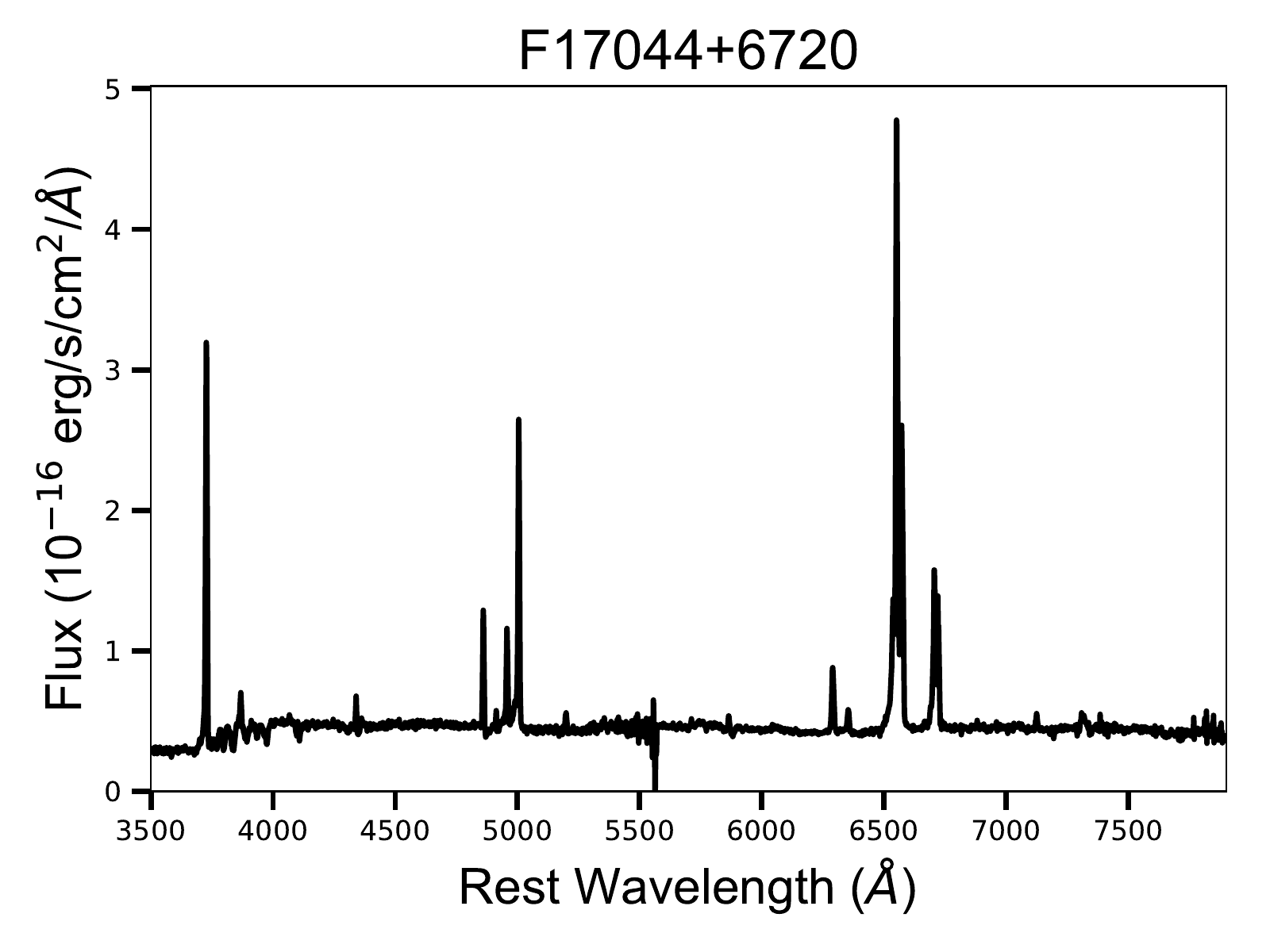}}\\
	\vspace{-3.0mm}
	\subfloat{\includegraphics[width = 3in]{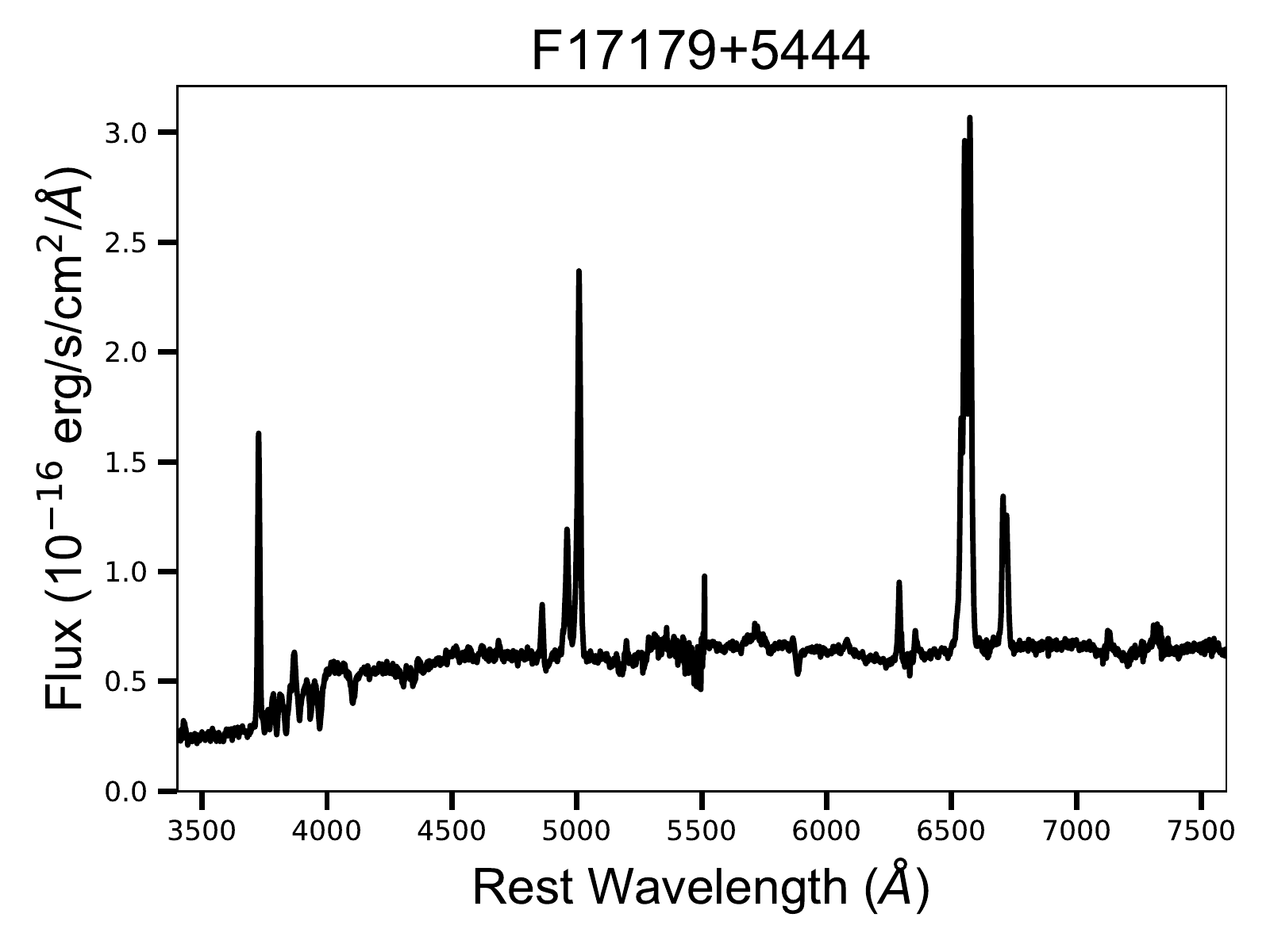}}
	\subfloat{\includegraphics[width = 3in]{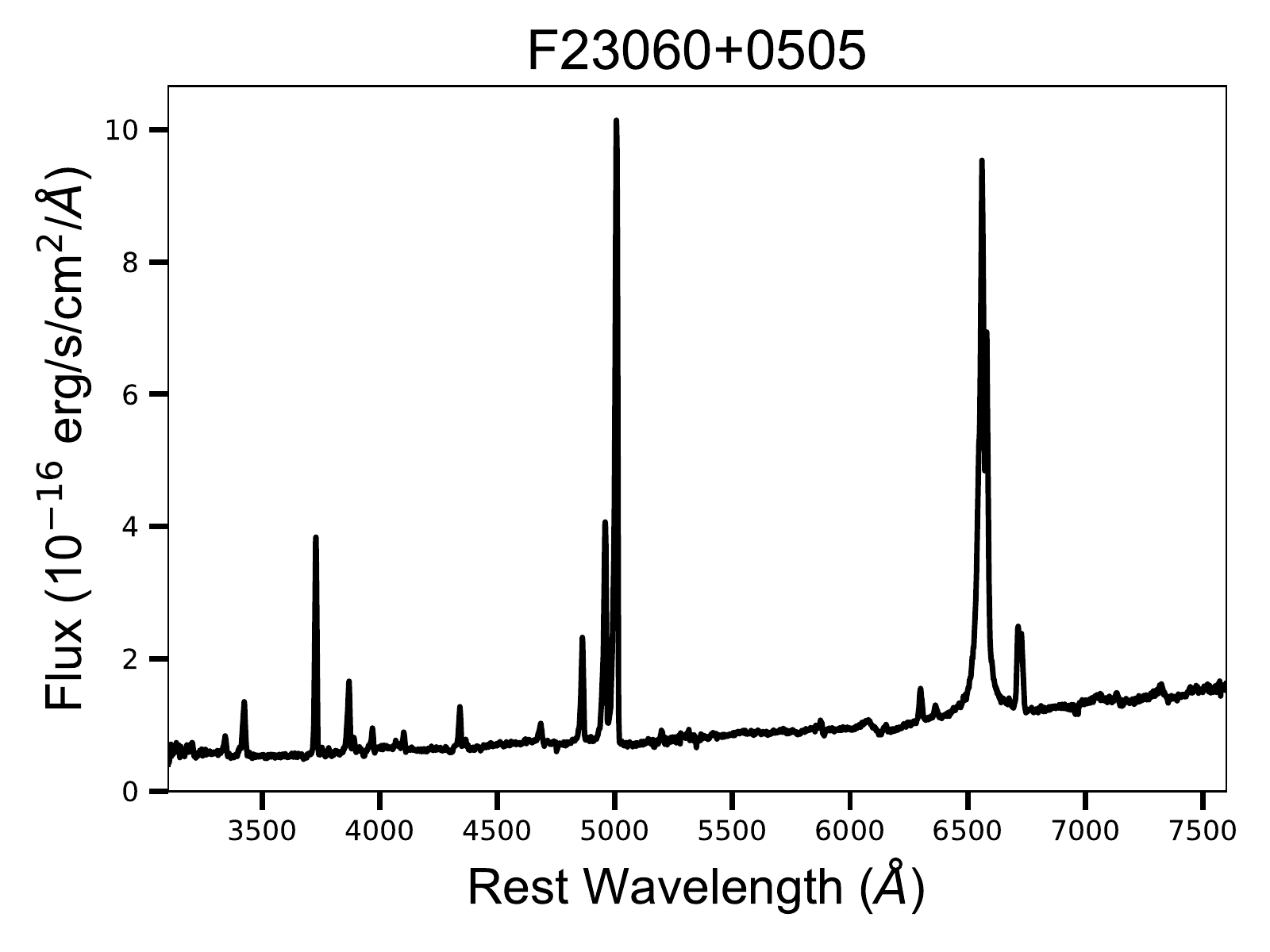}}\\ 
	\vspace{-3.0mm}
	\subfloat{\includegraphics[width = 3in]{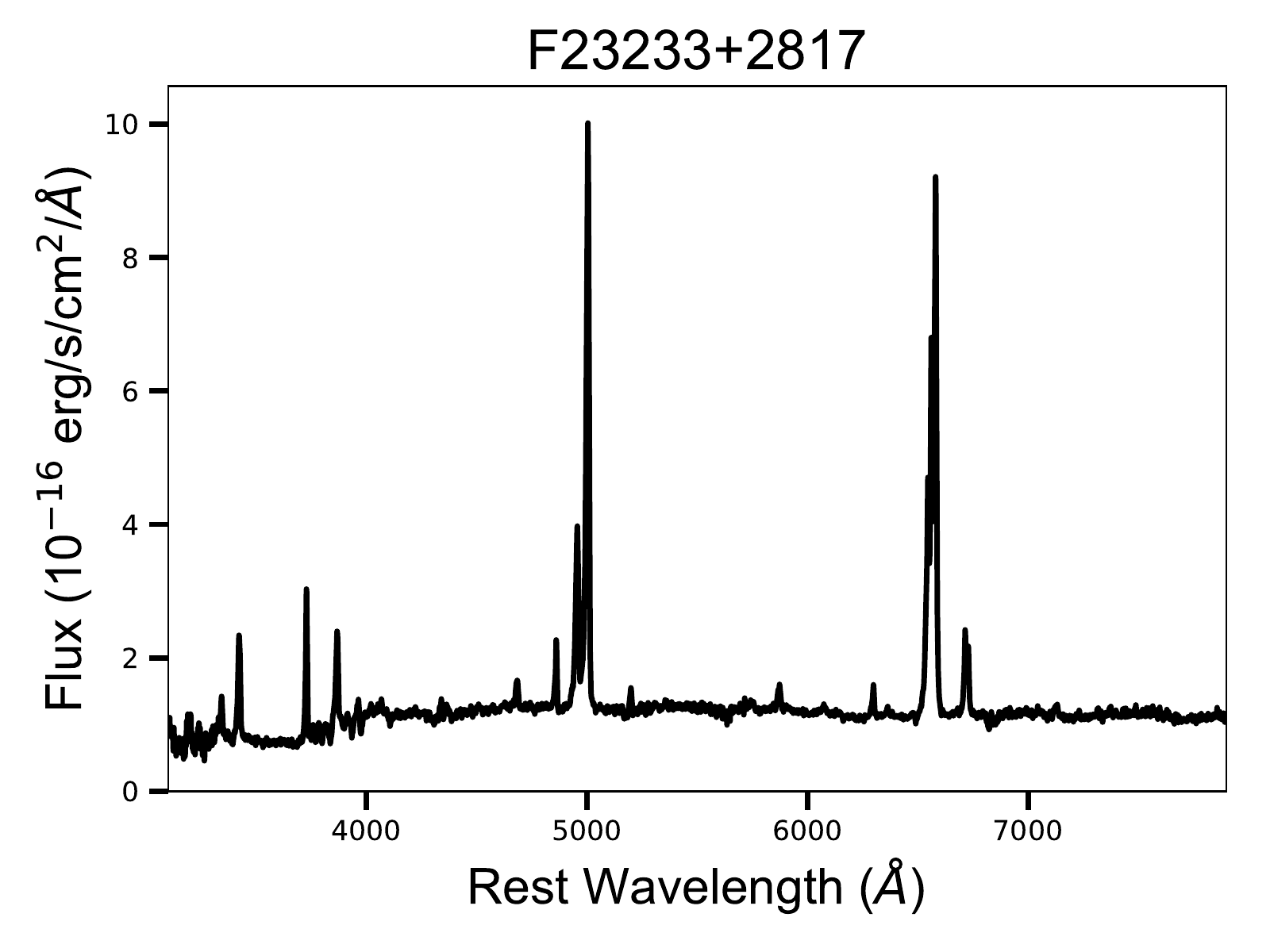}}
	\subfloat{\includegraphics[width = 3in]{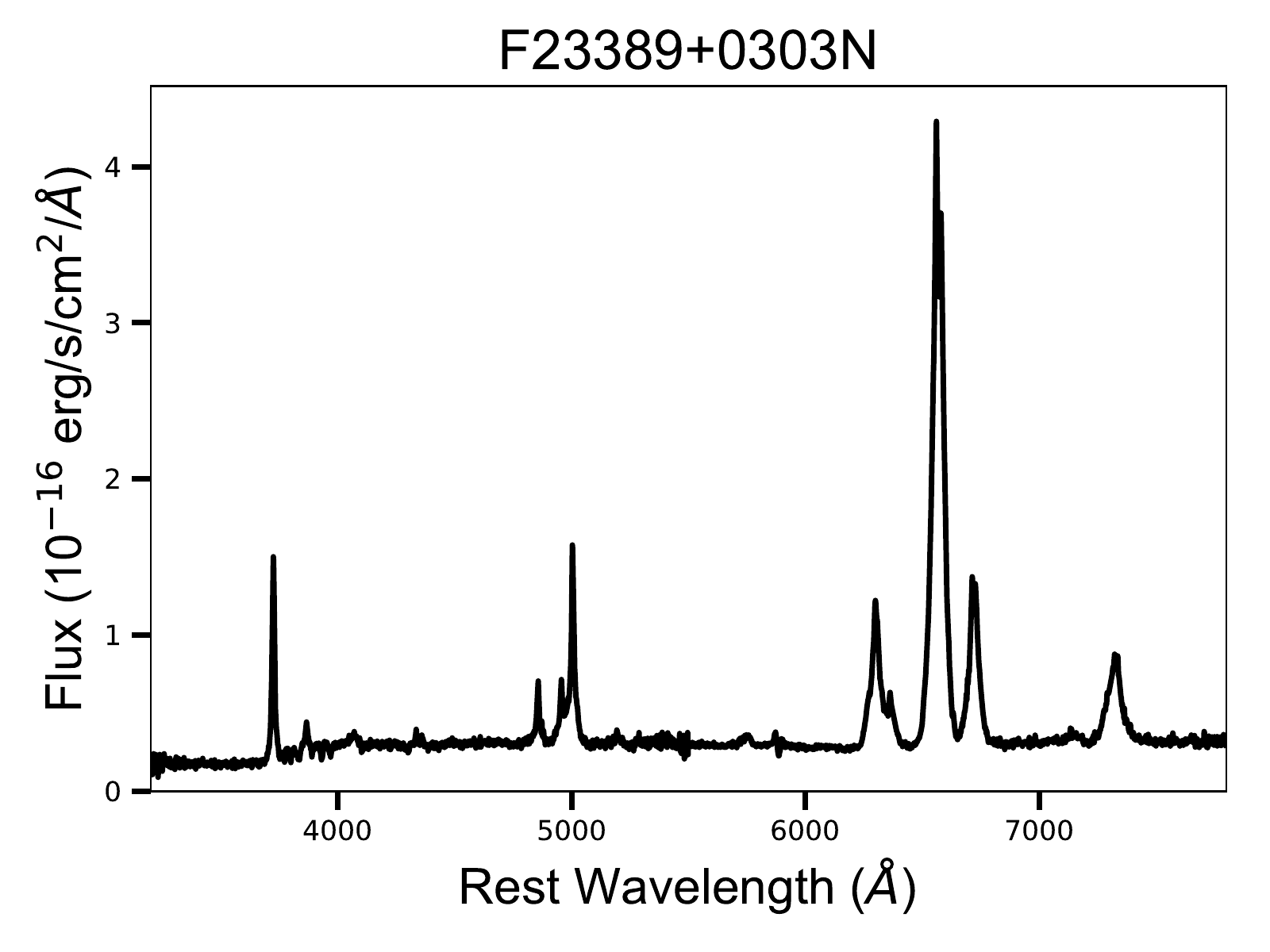}}\\
	\caption{The rest-frame nuclear spectra for the 8 ULIRGs considered in this paper}
	\label{fig:spectra}
\end{figure*}

The ULIRGs in this sample are known to exhibit strong AGN-induced outflows \citep{Rodriguez2013} that are concentrated in their nuclear regions. Therefore we used extraction apertures of  $\sim$5 kpc diameter centred on the nuclei, as in \cite{Rodriguez2009}. Using the pixel-scale of the CCD and the cosmology-derived arcsecond-to-kpc conversion factors, 5$\pm$0.6 kpc apertures were extracted for all objects (see Table \ref{tab:log}, column 8). This resulted in 8 nuclear spectra for analysis, which are shown in Figure \ref{fig:spectra}. In the following sections, we describe the results obtained by fitting the profiles of the emission lines detected in these nuclear spectra. It should be noted that the quality of the data in the vicinity (within $\sim$100 \AA) of the dichroic cut at $\sim$6000\AA\, (observed frame) is relatively poor. Fortunately this does not affect any of our key diagnostic emission lines.   \par

\subsection{HST/STIS Observations}
\label{sec:hst}

In addition to the WHT/ISIS spectroscopy, we also used spectral data taken with the Space Telescope Imaging Spectrograph (STIS) - installed on the \textit{HST} - to provide additional information on the spatial extents and physical conditions of the outflows. HST/STIS observations were available for two of the ULIRGS in this paper: F01004--2237 and F05189--2524 \citep{Farrah2005}. These space-based observations have the advantage of higher spatial resolution, and the narrow slit reduces the level of stellar contamination from the host galaxy, which otherwise reduces the equivalent widths of the weaker trans-auroral lines in the ground-based spectra. \par 

Both objects were observed using the G430L and G750L gratings with the 52X0.2" slit. The main spectroscopic reduction steps were performed by the Space Telescope Science Institute (STScI) STIS pipeline \textit{calstis}. As in \cite{Rose2018}, we then used IRAF and STARLINK packages to improve the bad pixel and cosmic ray removal.

\subsection{Fitting the emission line profiles}
\label{sec:fitting}
Our fitting technique is described in detail in \cite{Rose2018}; however, for completeness we provide an overview here. \par

Prior to modelling the emission line profiles, the spectra were corrected for Galactic extinction using the extinction function of \citealt{Cardelli1989} and E(B-V) values of \cite{Schlafly2011}. They were then shifted to the galaxy rest frame using redshifts measured from the higher order Balmer stellar absorption features (3798\AA{}, 3771\AA{}, and 3750\AA{} where possible). One exception to this is F01004-2237, for which strong non-stellar continuum emission dominates over the stellar absorption features. In this case, the rest-frame was estimated using four narrow off-nuclear emission lines detected in kinematically quiescent regions extending $\sim$10" ($\sim$20 kpc) either side of the nucleus. The redshifts and their associated errors are given in Table \ref{tab:properties}. \par

\begin{figure}
	\includegraphics[width=\columnwidth]{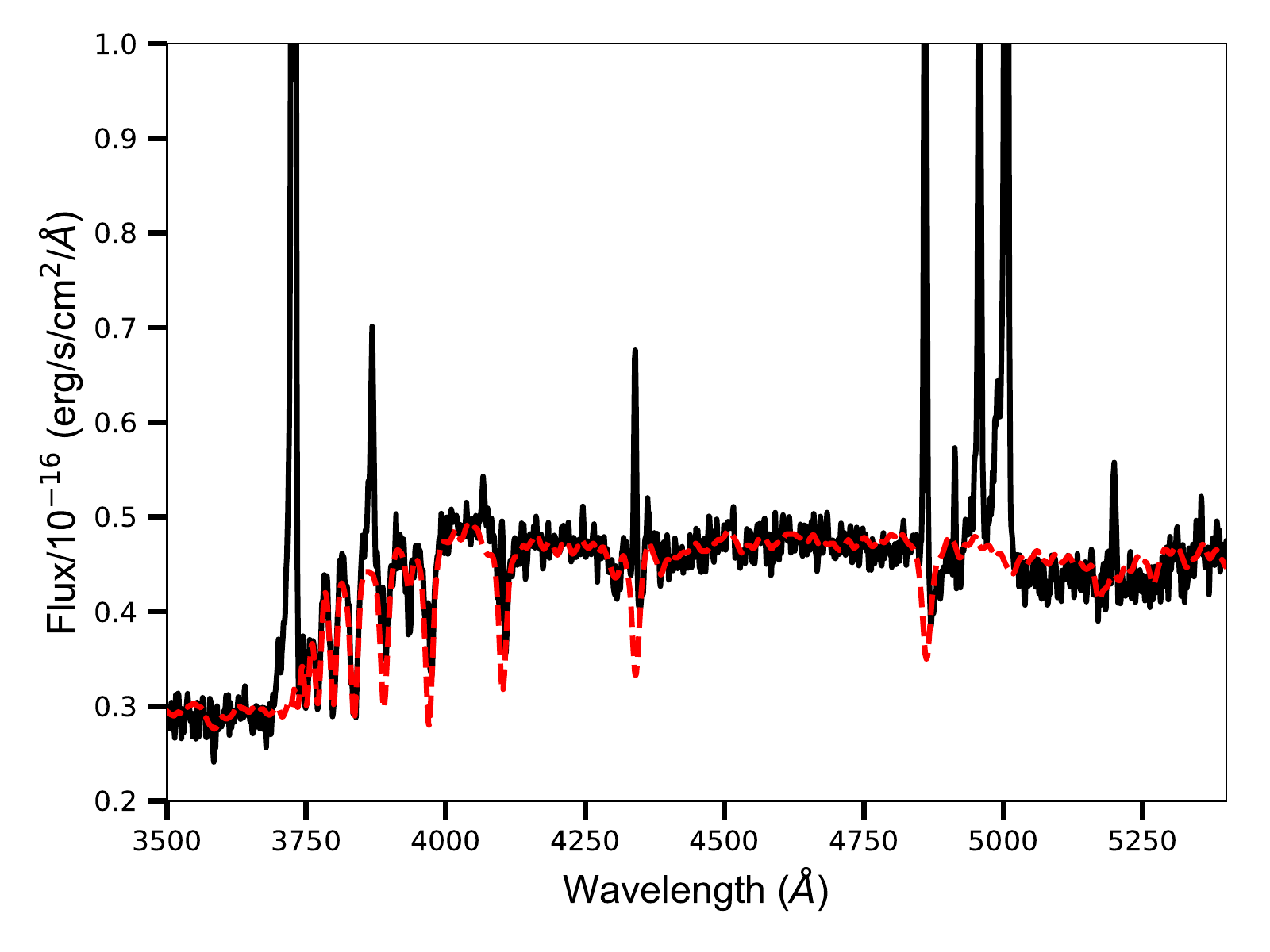}
    \caption{The nuclear spectrum of F17044+6720 (black, solid line), overplotted with the STARLIGHT fit to the stellar continuum (red, dashed line)} 
	\label{fig:fit}
\end{figure}

\begin{figure}
	\includegraphics[width=\columnwidth]{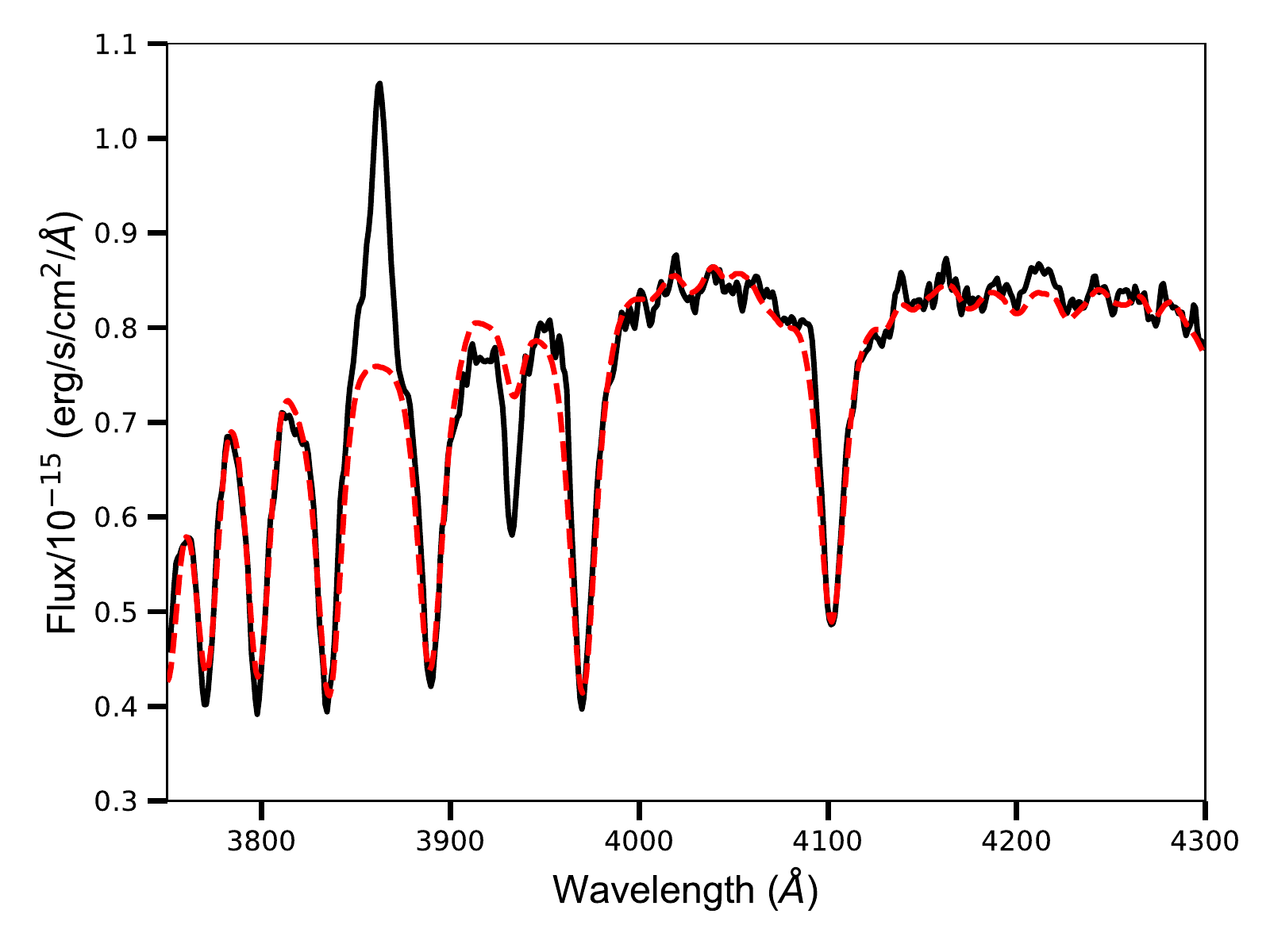}
    \caption{A zoom in on the higher order Balmer absorption features (H$\delta$--H11) in the nuclear spectrum of F05189--2524. Also shown is the NeIII emission line at 3869\AA{} and the CaII K ISM absorption feature at 3933\AA. The black solid line is the data, the red dashed line shows the STARLIGHT fit.} 
	\label{fig:balmer}
\end{figure}

The spectra of six objects were then fitted with the spectral synthesis code STARLIGHT \citep{Fernandes2005}, using the \cite{Bruzual2003} (BC03) solar metallicity base templates in order to facilitate accurate stellar continuum subtraction. The fit to the full optical spectrum of F17044+6720 is shown in Figure \ref{fig:fit}, for reference. A zoom on the part of the spectrum containing the Balmer absorption lines of F05189-2524 is also provided in Figure \ref{fig:balmer}, illustrating the excellent consistency of the STARLIGHT fit with the estimated redshift and the stellar absorption line profiles. Note that no attempt was made to fit F01004-2237. The continuum of this object is highly complex due to the fact that its spectrum is dominated by the emission of a recent tidal disruption event \citep[TDE;][]{Tadhunter2017a}. Similarly, STARLIGHT struggled to fit the continuum in the red arm of F23060+0505 due to an underlying reddened type I AGN component. Therefore, we did not remove the stellar continua in these cases.  \par 

\begin{figure*}
	\vspace{-5mm}
	\subfloat{\includegraphics[width = 3in]{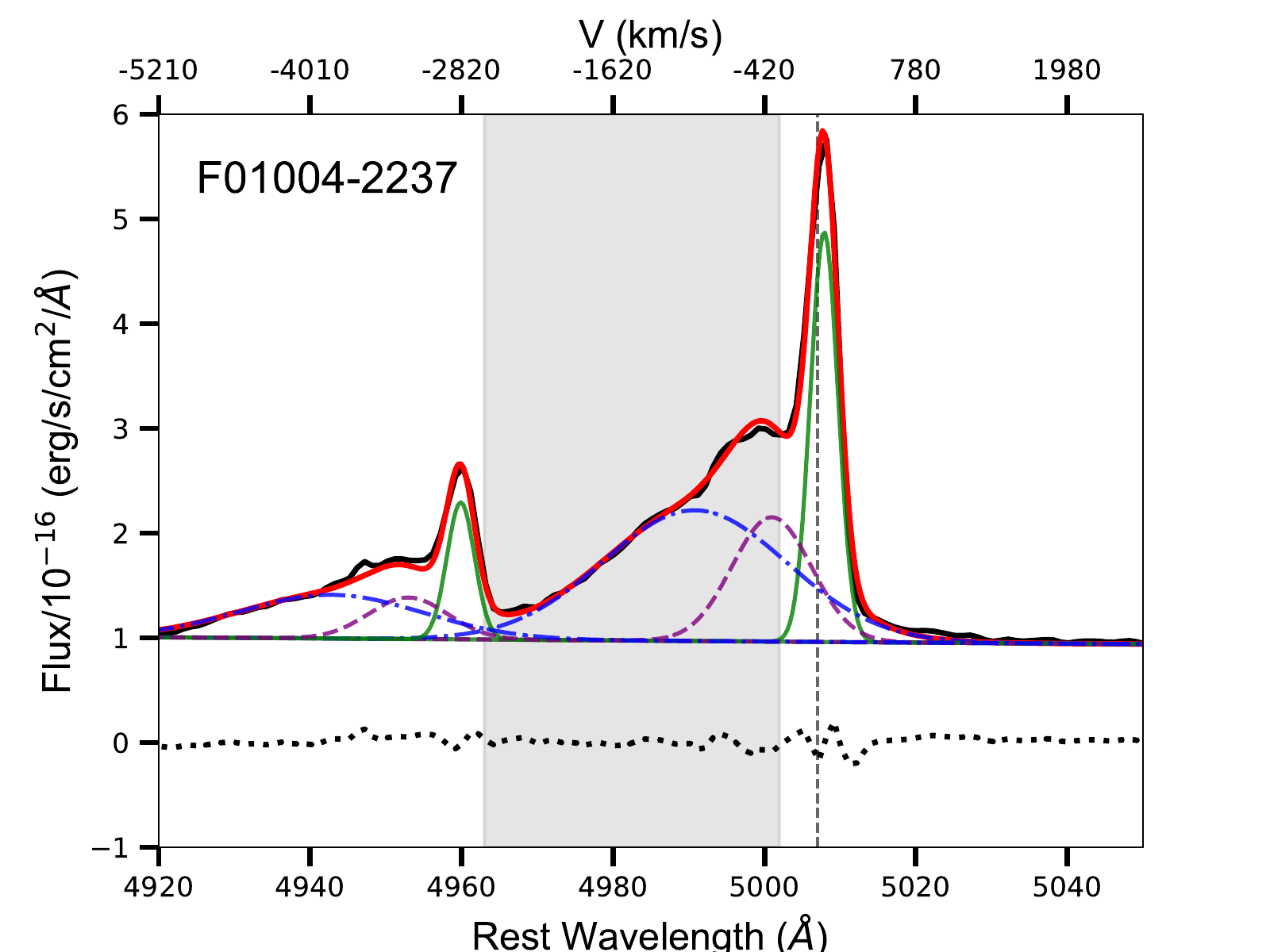}} 
	\subfloat{\includegraphics[width = 3in]{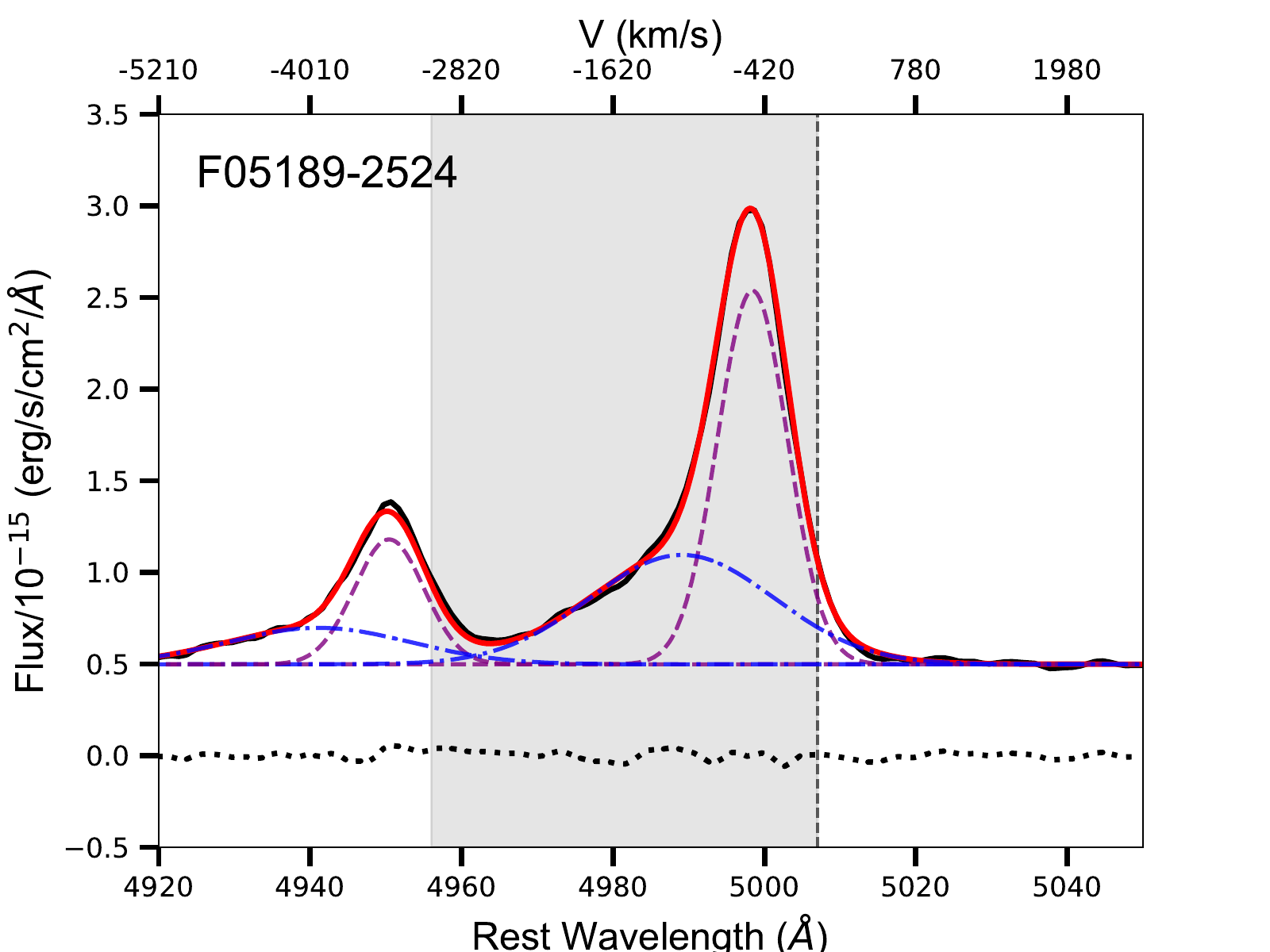}}\\  
	\vspace{-3.6mm}
	\subfloat{\includegraphics[width = 3in]{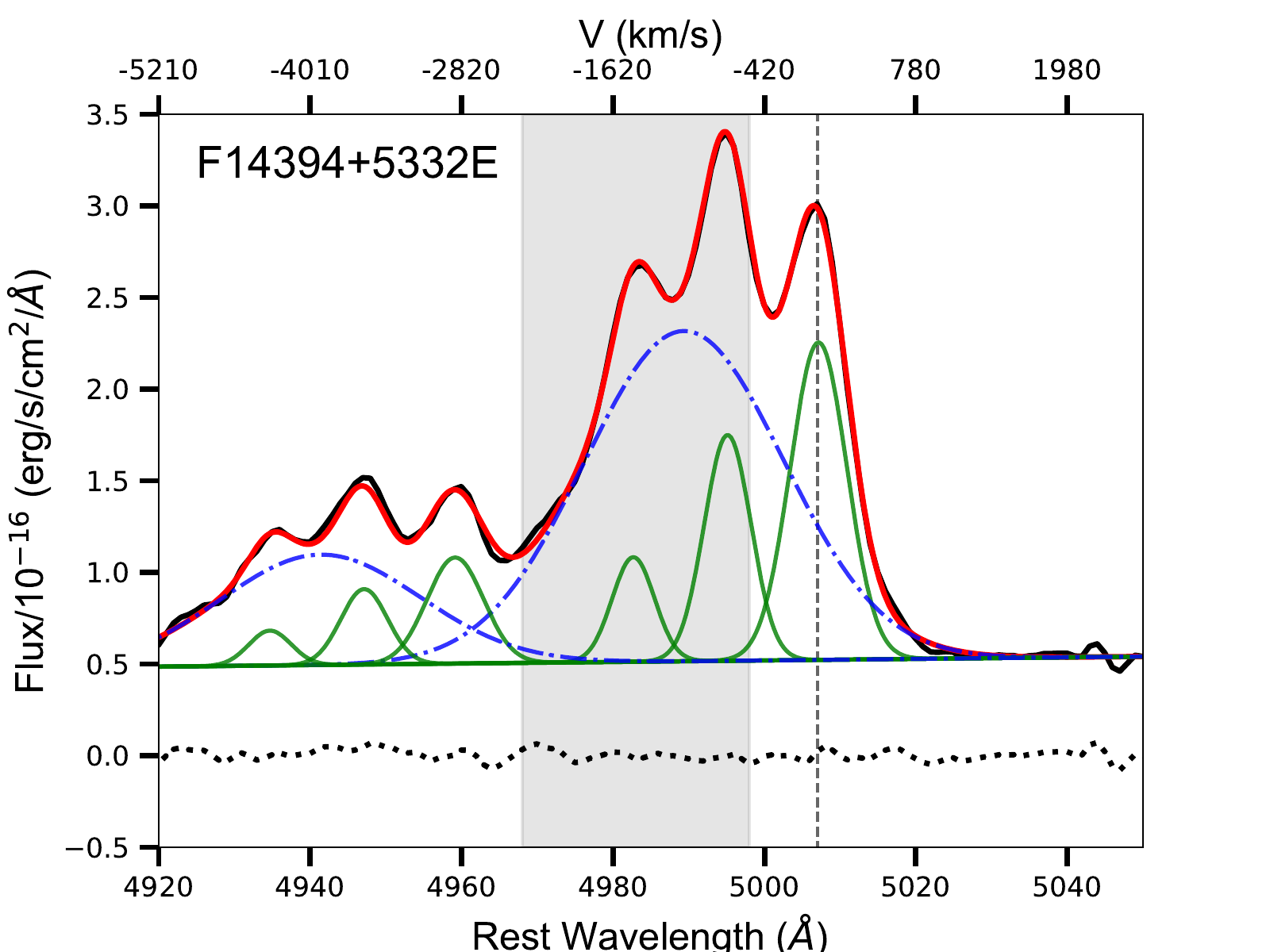}}
	\subfloat{\includegraphics[width = 3in]{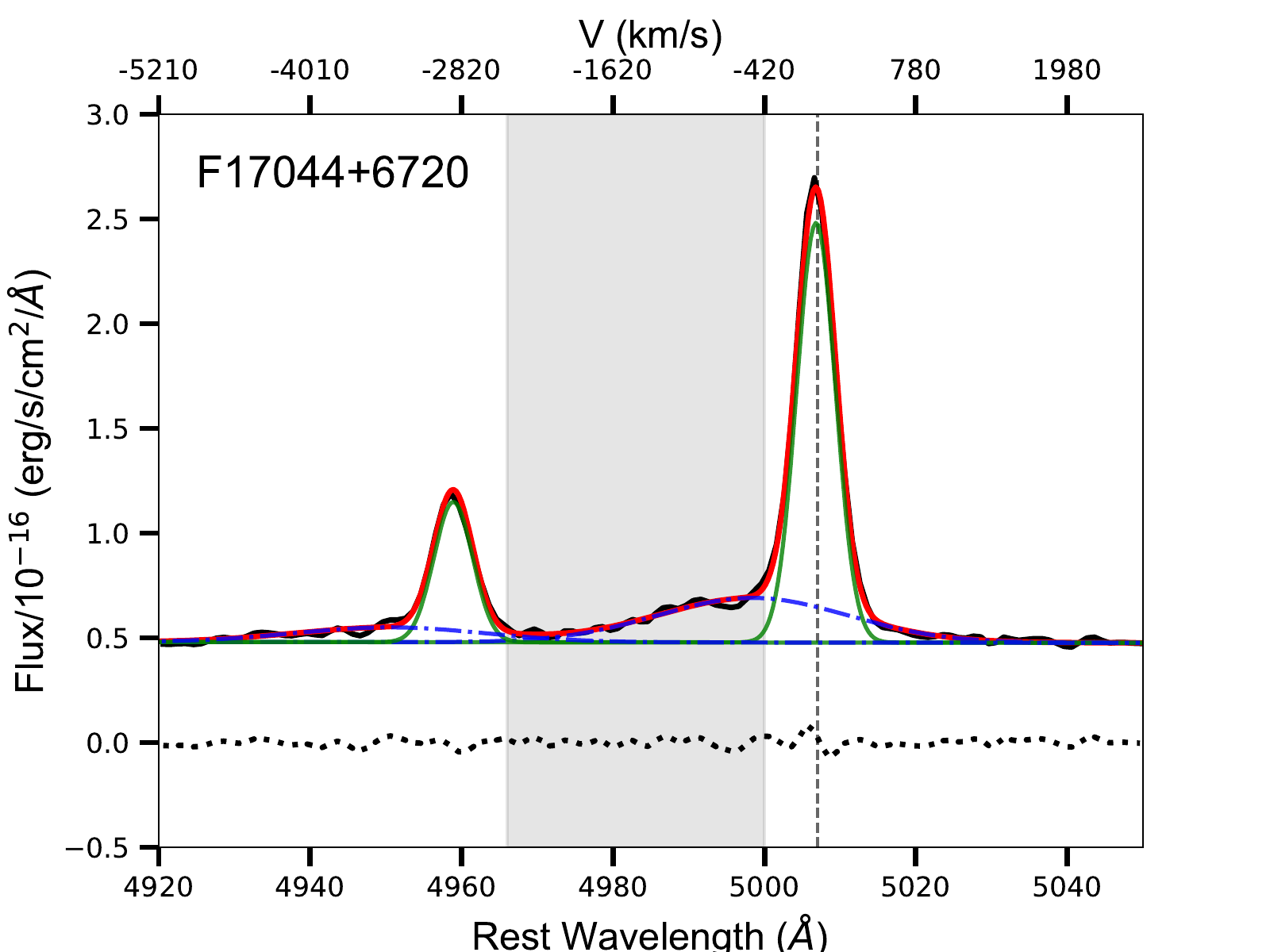}}\\
	\vspace{-3.6mm}
	\subfloat{\includegraphics[width = 3in]{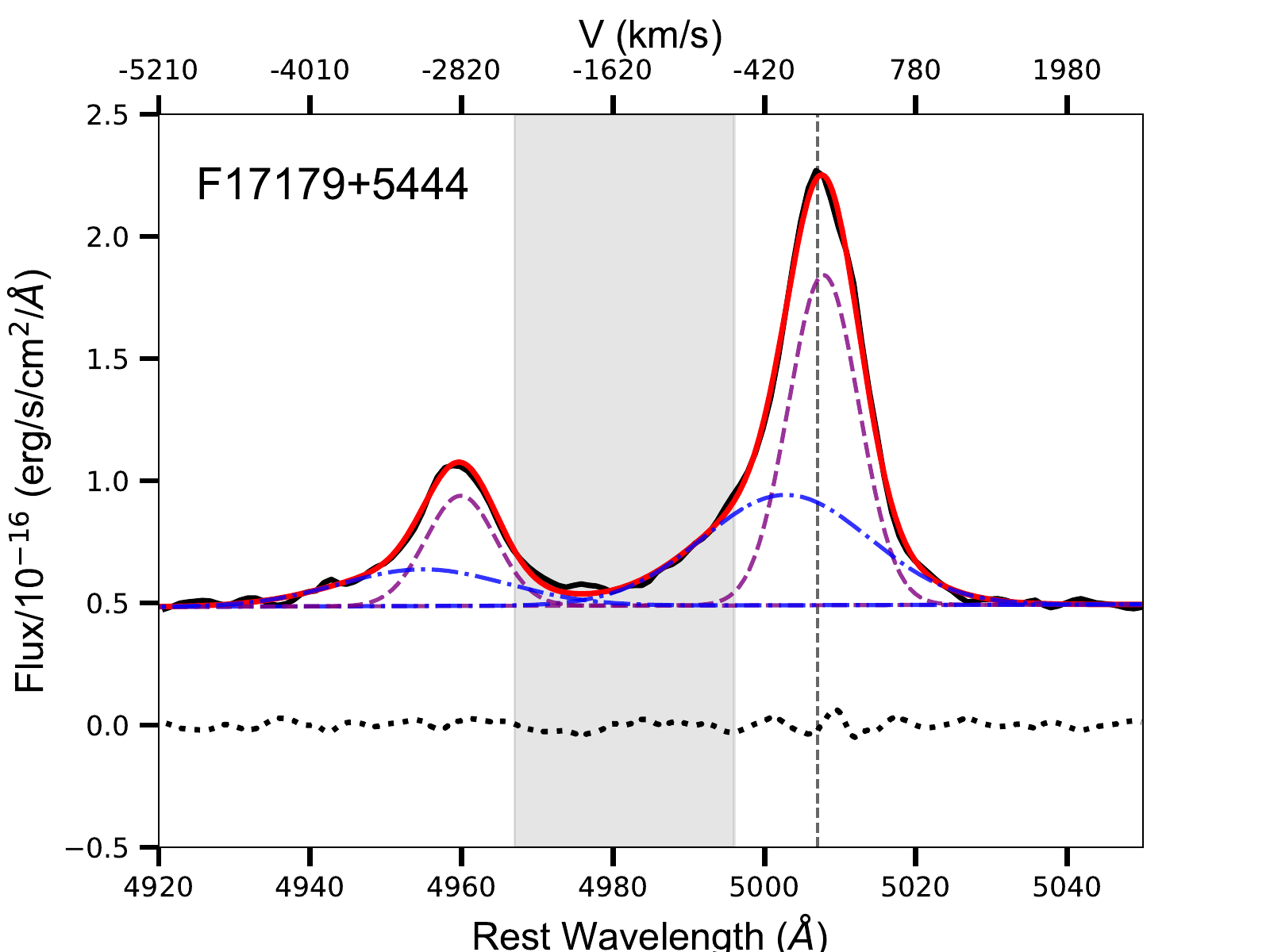}}
	\subfloat{\includegraphics[width = 3in]{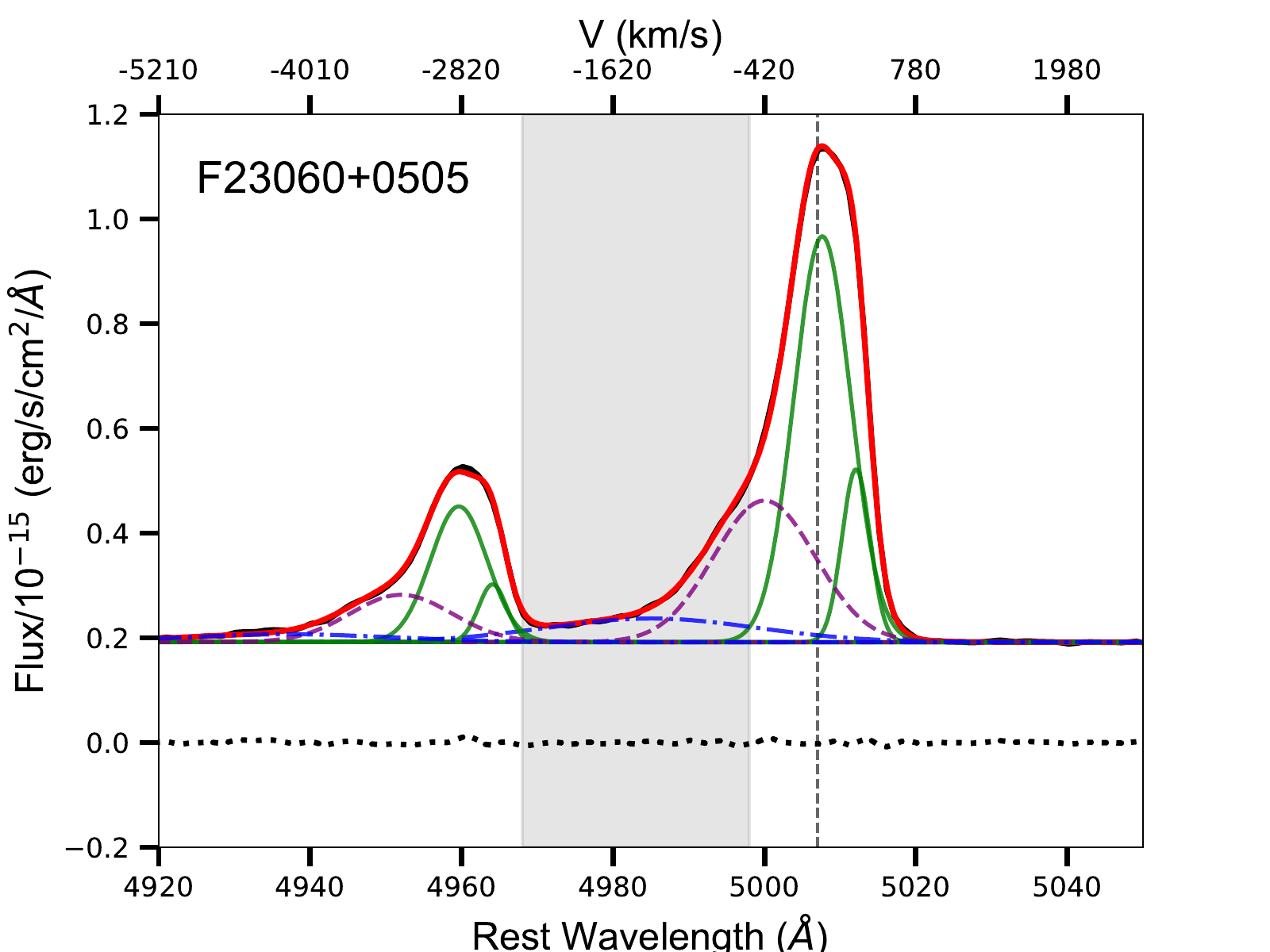}}\\ 
	\vspace{-3.6mm}
	\subfloat{\includegraphics[width = 3in]{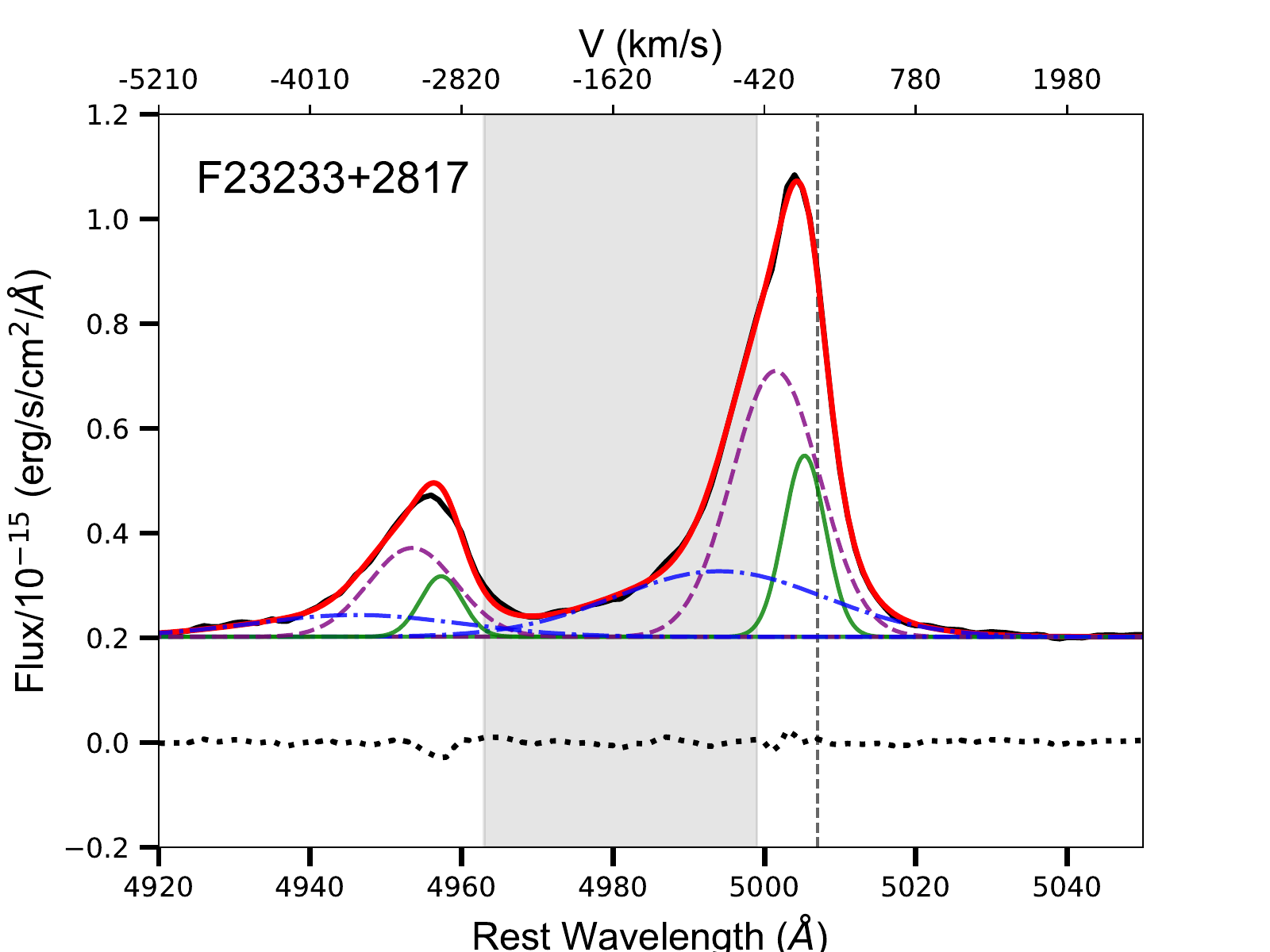}}
	\subfloat{\includegraphics[width = 3in]{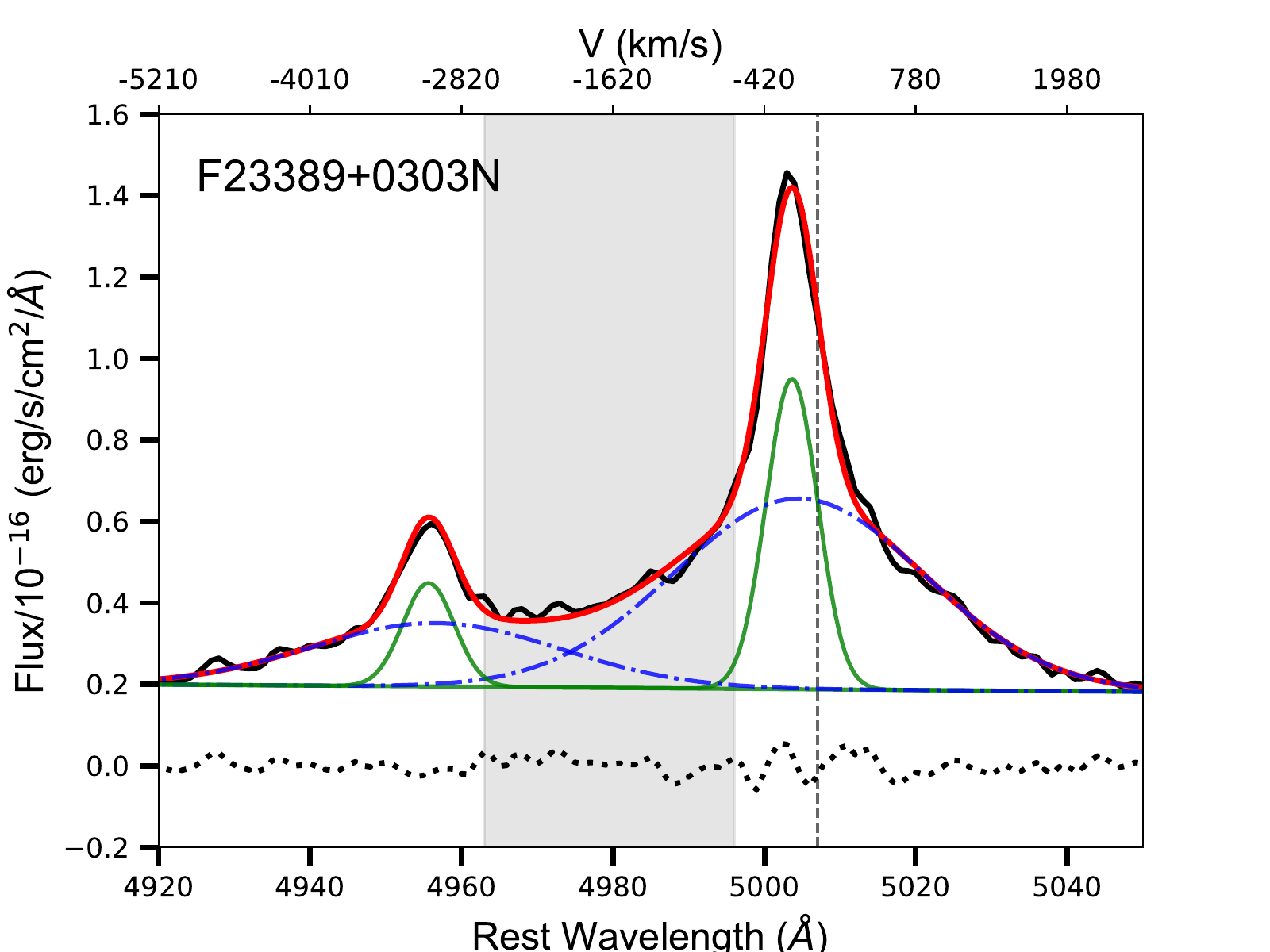}}\\
	\caption{[OIII] profiles and fits for all ULIRGS considered in this paper. The black solid line is the data, and the overall fit is shown by the red solid line. The individual components are coloured as follows: narrow (green, solid); intermediate (purple, dashed); broad (blue, dot-dashed). The residuals of the fit are shown below the profiles (black, dashed). The shaded regions show the velocity ranges extracted when determining the outflow radii. }
	\label{fig:oiii}
\end{figure*}

The [OIII]$\lambda$5007 emission line is the strongest and cleanest high-ionisation line associated with the out-flowing ionised gas. For this reason, we first concentrated our emission line fitting procedure on this line. Our approach was to fit the line profile with the minimum number of Gaussians required to give an acceptable fit. We constrained the fits as much as possible, setting the relative shifts between each component of [OIII]$\lambda\lambda$4959,5007 to be 47.92\AA, and fixing the relative intensities according to those from atomic physics (1:2.99). At least two Gaussian components were required in all cases. The [OIII] fits are shown in Figure \ref{fig:oiii}, along with their residuals. The velocity shifts and widths for each required component are shown in Table \ref{tab:outflowkinematics}. The velocity shifts have been corrected in quadrature for the spectral resolution. For consistency, we label our components based on the line widths (FWHM) following \cite{Rodriguez2013}:

\begin{itemize}
  \item Narrow (N): FWHM < 500 km s$^{-1}$;
  \item Intermediate (I): 500 < FWHM < 1000 km s$^{-1}$;
  \item Broad (B): 1000 < FWHM < 2000 km s$^{-1}$;
  \item Very Broad (VB): FWHM > 2000 km s$^{-1}$.
\end{itemize}

\begin{table*}
	\centering
	\caption{The [OIII] kinematics for the ULIRGs considered in this paper. Columns (3) and (4) give the FWHM and velocity shift of each component. Columns (5) and (6) indicate whether the [OIII] kinematic model worked for the H$\alpha$+[NII], [OII] and [SII] blends. Columns (7) and (8) give the observed [OIII]$\lambda$5007 and H$\beta$ fluxes associated with each kinematic component. 
			}
	\label{tab:outflowkinematics}
		\begin{tabular}{lccccccr} 
			\\
			\hline
			Object & Comp. & FWHM &  $\Delta$V & H$\alpha$ + [NII] & [OII]/[SII] & [OIII]$\lambda$5007 flux & H$\beta$ flux \\
			IRAS  & & kms$^{-1}$ & kms$^{-1}$ & (Y/N) & (Y/N) & (erg s$^{-1}$ cm$^{-2}$) & (erg s$^{-1}$ cm$^{-2}$) \\
			\\
			(1) & (2) & (3) & (4) & (5) & (6) & (7) & (8) \\
			\hline\\ 
			F01004 -- 2237 & N & unresolved & 63$\pm$25 & N  & N & (2.03$\pm$0.04)E-15 & (4.71$\pm$0.02)E-16  \\
			& I & 699 $\pm$ 32 & -361 $\pm$ 35 & & & (1.99$\pm$0.14)E-15 & (6.10$\pm$0.09)E-16  \\ 
			& B & 1586 $\pm$ 38 & -1045 $\pm$ 60 &  & & (3.68$\pm$0.16)E-15 & (1.55$\pm$0.10)E-15  \\ 
			F05189 -- 2524 & I & 582 $\pm$37  & -505 $\pm$ 26  & N & N & (2.48$\pm$0.04)E-14 & (1.93$\pm$0.14)E-15\\
			& B & 1706 $\pm$ 24 & -1072 $\pm$ 44 & & & (1.91$\pm$0.05)E-14 & (1.09$\pm$0.12)E-15 \\
			& I$_{H \beta}$ & 555 $\pm$ 76 & -28 $\pm$ 55 & & & - & (1.53$\pm$0.21)E-15\\
			F14394 + 5332E & N1 & 408 $\pm$11  & 17 $\pm$ 62 & N & N & (1.61$\pm$0.03)E-15 & - \\
			& N2 & 288 $\pm$ 21 & -701 $\pm$ 63 & & & (9.01$\pm$0.44)E-16 & - \\
			& N3 & 242 $\pm$ 34 & -1457 $\pm$ 66 & & & (4.00$\pm$0.32)E-16 & -\\
			& B & 1871 $\pm$ 19 &  -1000 $\pm$ 67 & & & (7.02$\pm$0.10)E-15 & - \\
			& N$_{H \beta}$ & 420 $\pm$ 11 & 17 $\pm$ 63 & & &- & (2.08$\pm$0.06)E-15 \\
			& I$_{H \beta}$ & 986 $\pm$ 42 & -358 $\pm$ 93 & & & - & (1.05$\pm$0.07)E-15\\
	 		& B$_{H \beta}$ & 1927 $\pm$ 20 & -1030 $\pm$ 69 & & & - & (5.42$\pm$0.52)E-16\\
			F17044 + 6720 & N & 218 $\pm$ 12 & -1 $\pm$ 61 & Y & Y & (1.46$\pm$0.01)E-15 & (6.61$\pm$0.10)E-16\\
			& B & 1757 $\pm$ 60 & -503 $\pm$ 85  & & & (7.64$\pm$0.33)E-16 & (1.58$\pm$0.24)E-16\\
			F17179 + 5444 & I & 590 $\pm$ 12 & 58 $\pm$ 62 & Y & Y$^{a}$ & (1.81$\pm$0.04)E-15 & (4.12$\pm$0.18)E-16\\
			& B  & 1530 $\pm$ 33 & -242 $\pm$ 78 & & & (1.43$\pm$0.05)E-15 & (2.20$\pm$0.27)E-16 \\
			F23060 + 0505$^{b}$ & N1 & 147 $\pm$ 37 & 273$\pm$34 & Y$^{c}$  & N & (4.09$\pm$0.45)E-15 & (6.94$\pm$0.87)E-16 \\
			& N2 & 267 $\pm$ 35 & -25 $\pm$ 38 & & & (4.76$\pm$0.49)E-15 & (9.60$\pm$1.02)E-16\\
			& I & 934 $\pm$ 32 & -283 $\pm$ 44 & & & (6.94$\pm$0.28)E-15 & (8.07$\pm$0.59)E-16 \\	
			& B & 1399 $\pm$ 114 & -1220 $\pm$ 146  & & & (1.62$\pm$0.30)E-15 & -   \\
			& BLR$_{H\alpha}$ & 2359 $\pm$ 69 & 393 $\pm$ 102 & & & - & -  \\
			F23233 + 2817 & N & 239 $\pm$ 21 & -92 $\pm$ 27 & Y & Y$^{d}$ & (2.56$\pm$0.14)E-15 & (6.50$\pm$0.58)E-16\\
			& I & 760 $\pm$ 13 & -316 $\pm$ 31 & & & (8.35$\pm$0.22)E-15 & (6.61$\pm$0.92)E-16\\
			& B & 1892 $\pm$ 40 & -785 $\pm$ 54  & & & (4.71$\pm$0.19)E-15 & (4.52$\pm$0.98)E-16   \\
			F23389 + 0303N & N  & 402 $\pm$ 16 & -191 $\pm$ 27 & Y & Y & (7.80$\pm$0.20)E-16 & (3.60$\pm$0.13)E-16\\ 
			& VB & 2346 $\pm$ 38 & -134 $\pm$ 36  &  & & (2.26$\pm$0.25)E-15 & (7.81$\pm$0.25)E-16  \\
		\hline
		\multicolumn{8}{l}{
			\begin{minipage}{0.9\textwidth}~\\
					$^{a}$ The [OIII] model worked for the trans-auroral [OII] blend, but not for the [SII]. In this case, a two component model with different widths and shifts was required for the strong blend. The broad component to the weak trans-auroral [SII] blend was not detected. \\
					$^{b}$ H$\beta$ was fit with just narrow and intermediate components. No broad component was required.\\
					$^{c}$ The [NII] lines were well fit with the [OIII] model. H$\alpha$ was fit with the narrow and intermediate components of the [OIII] model, however the broad component was dominated by a broad-line region component (FWHM > 2000 km s$^{-1}$).\\
					$^{d}$ The [SII] blends did not require a broad component.\\
			\end{minipage}
		}\\
	\end{tabular}
\end{table*} 

We then attempted to apply this [OIII] kinematic model (i.e. relative width and shift w.r.t. to the rest frame for each component) to our other diagnostic emission lines ([OII]$\lambda$3727, [SII]$\lambda\lambda$4068,4076, H$\beta$, [OI]$\lambda$6300, H$\alpha$, [NII]$\lambda\lambda$6548,6583, [SII]$\lambda\lambda$6716,6731 and [OII]$\lambda\lambda$7319,7331). Note that the use of the various diagnostic emission line ratios in the following analyses requires the assumption that the flux of all of these emission lines originates from the same volume of gas. We show this assumption to be valid in \S6. \par 
 
The kinematic components were constrained according to atomic physics as follows \citep[see also][]{Rose2018}:
\begin{itemize}
\item{The [SII](4068/4076) ratios were forced to fall within the range 3.01 < [SII](4068/4076) < 3.28.}
\item{The [OII](7319/7331) ratios were fixed at 1.24.}
\item{ The [NII](6583/6548) ratios were fixed at 2.99.}
\item{Where necessary, the ratios of the broad components of [SII](6717/6731) were fixed to the high-density limit value of 0.44 to ensure a physical fit.}
\end{itemize}\par 

\begin{figure*}
	\vspace{-5mm}
	\subfloat{\includegraphics[width = 3in]{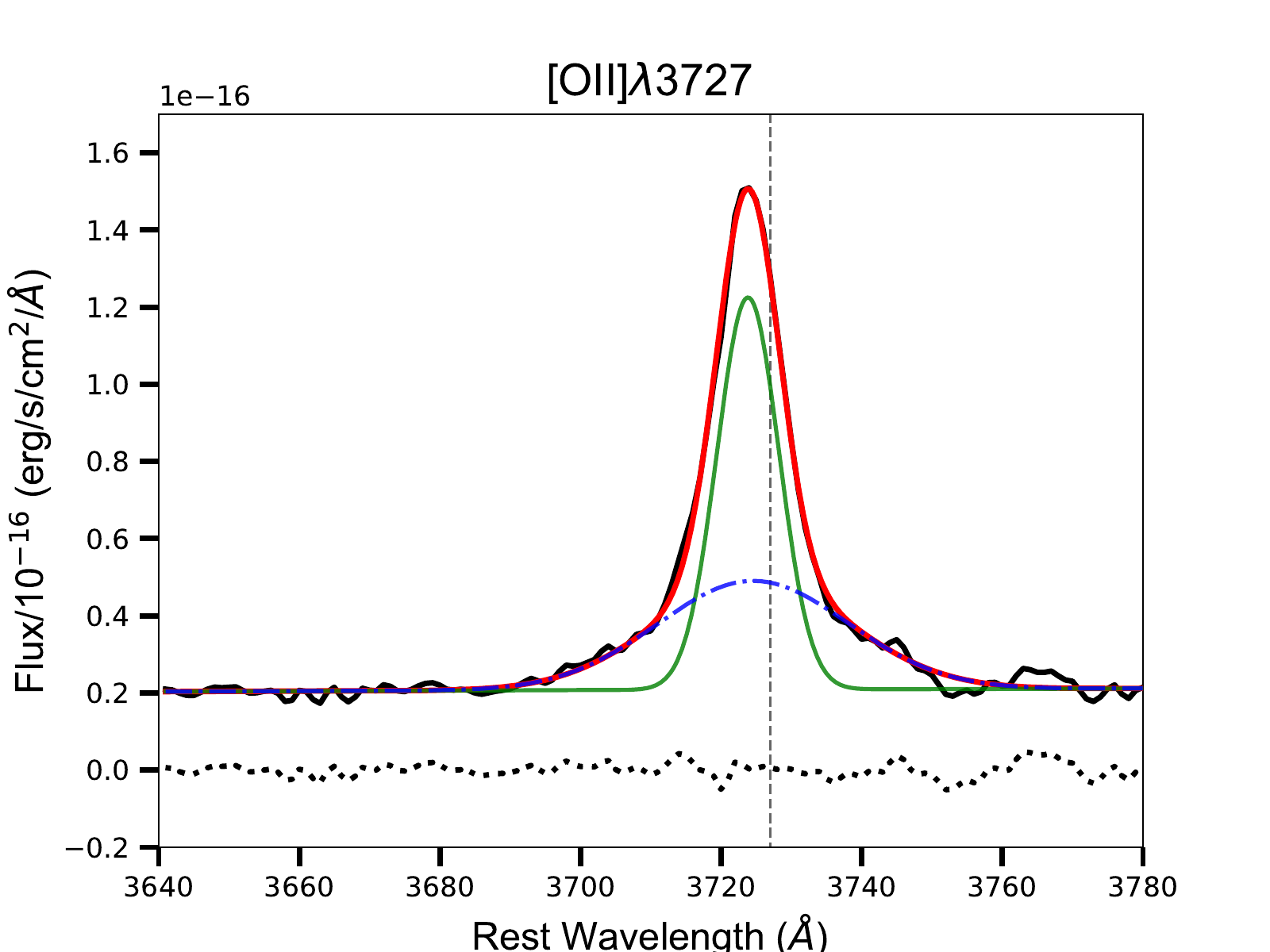}} 
	\subfloat{\includegraphics[width = 3in]{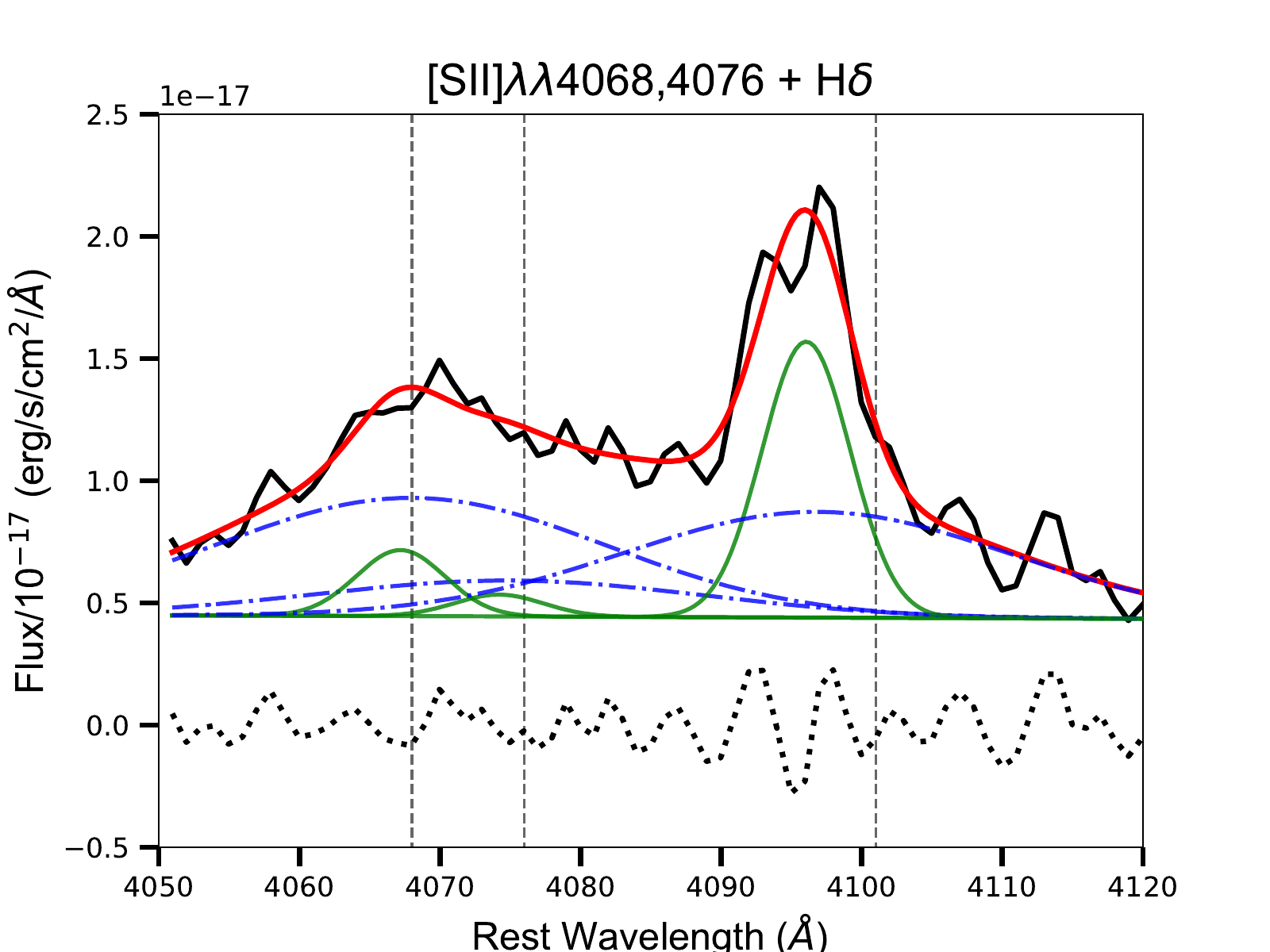}}\\  
	\vspace{-3.6mm}
	\subfloat{\includegraphics[width = 3in]{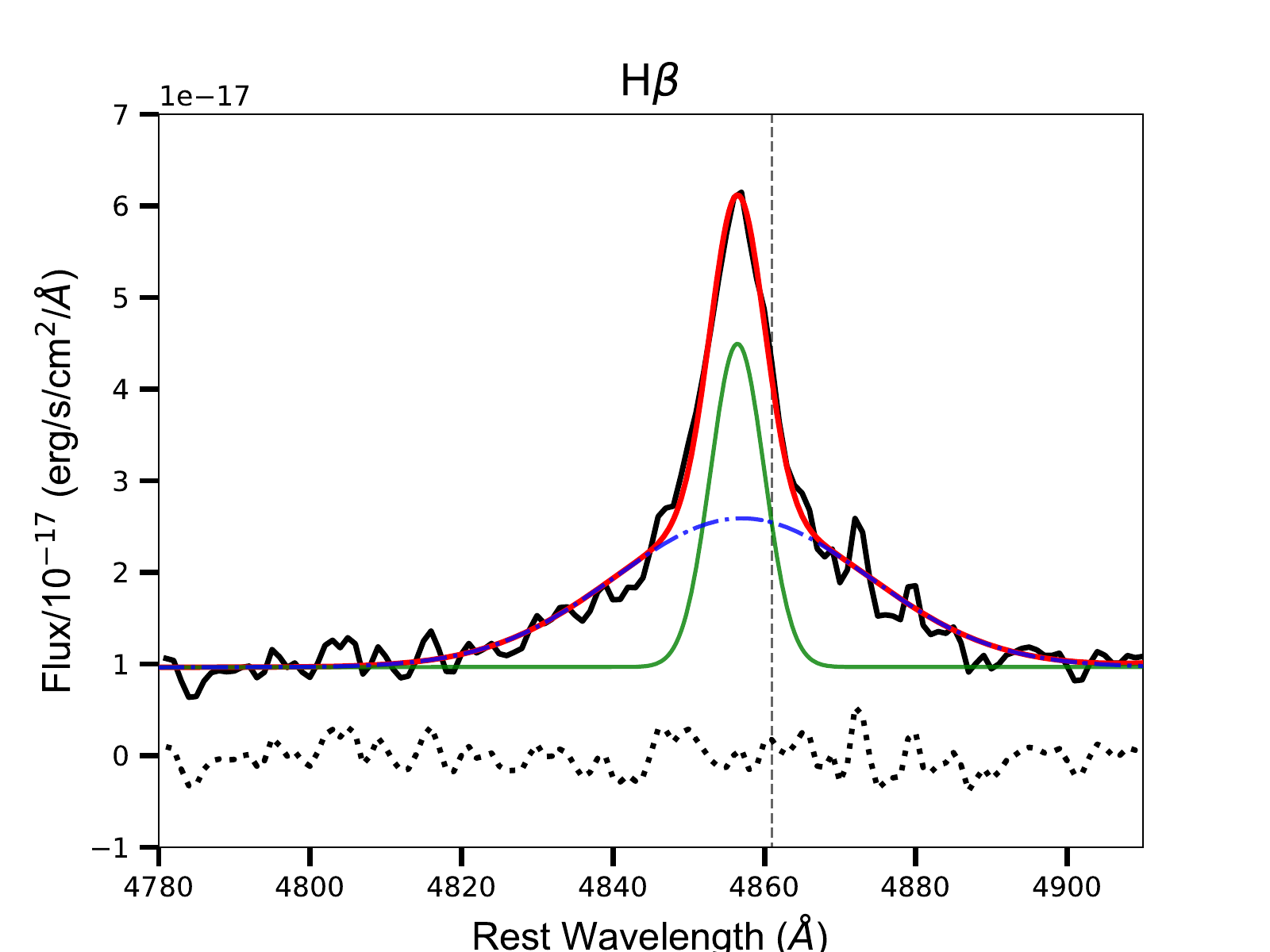}}
	\subfloat{\includegraphics[width = 3in]{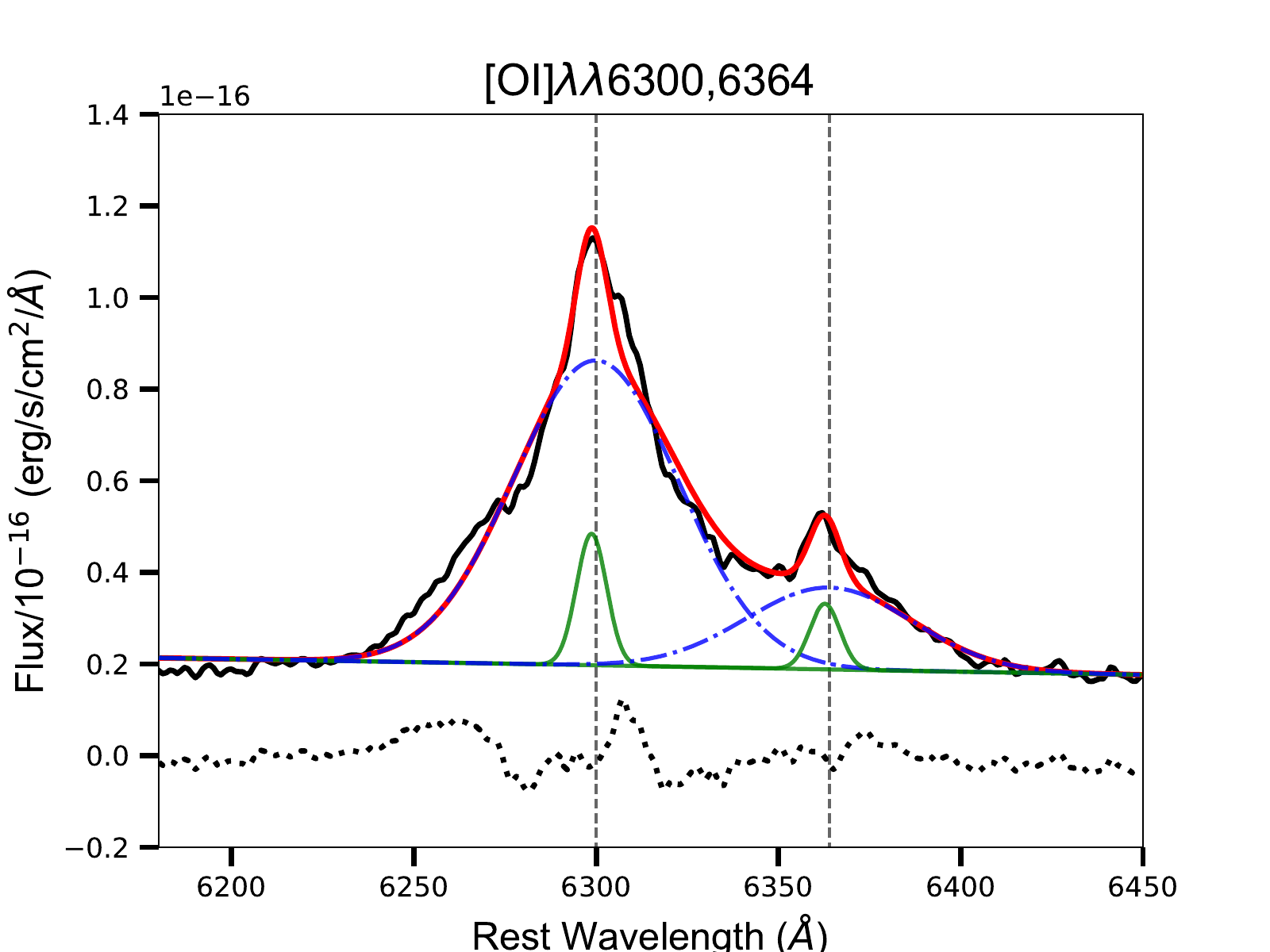}}\\
	\vspace{-3.6mm}
	\subfloat{\includegraphics[width = 3in]{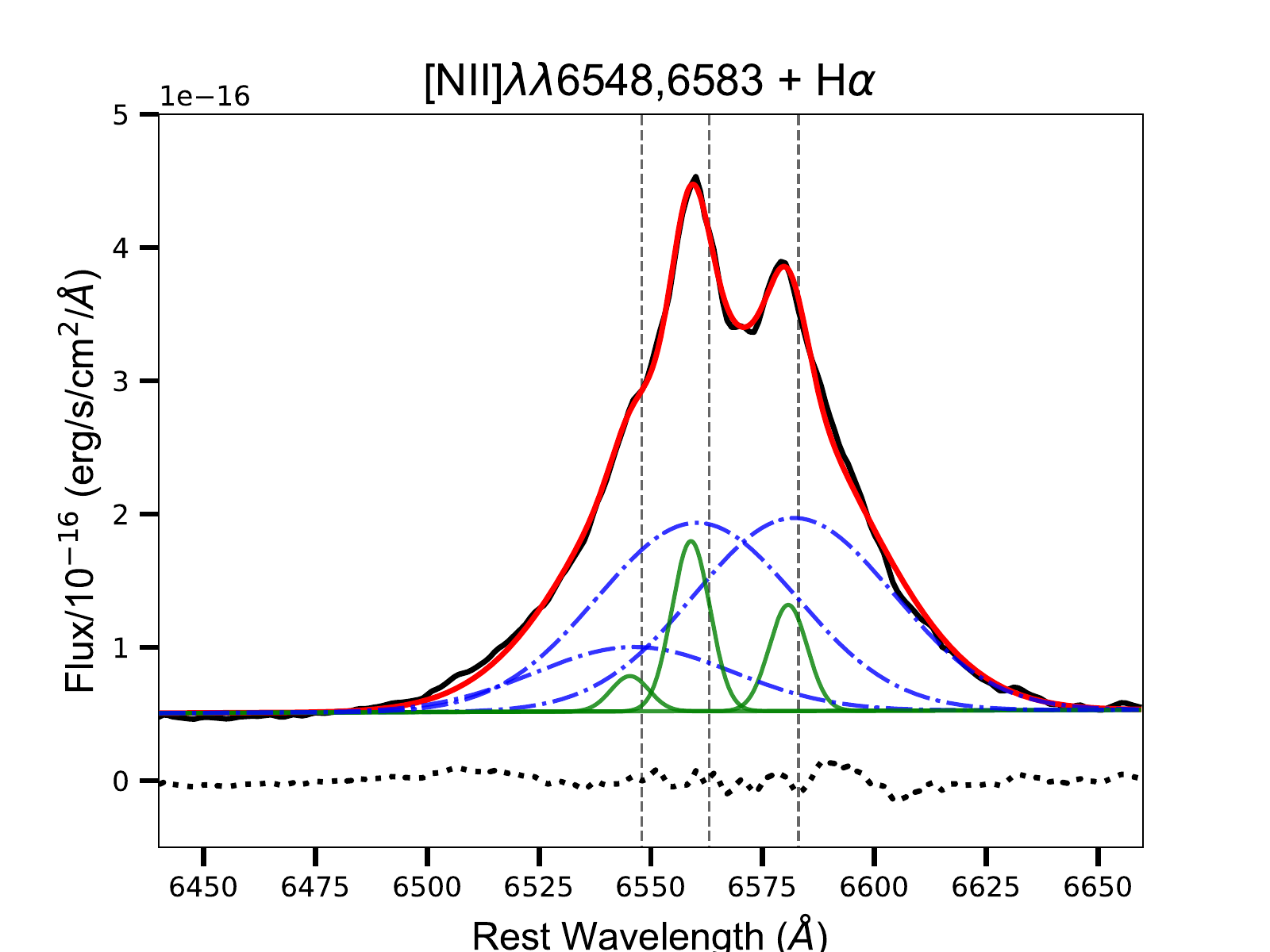}}
	\subfloat{\includegraphics[width = 3in]{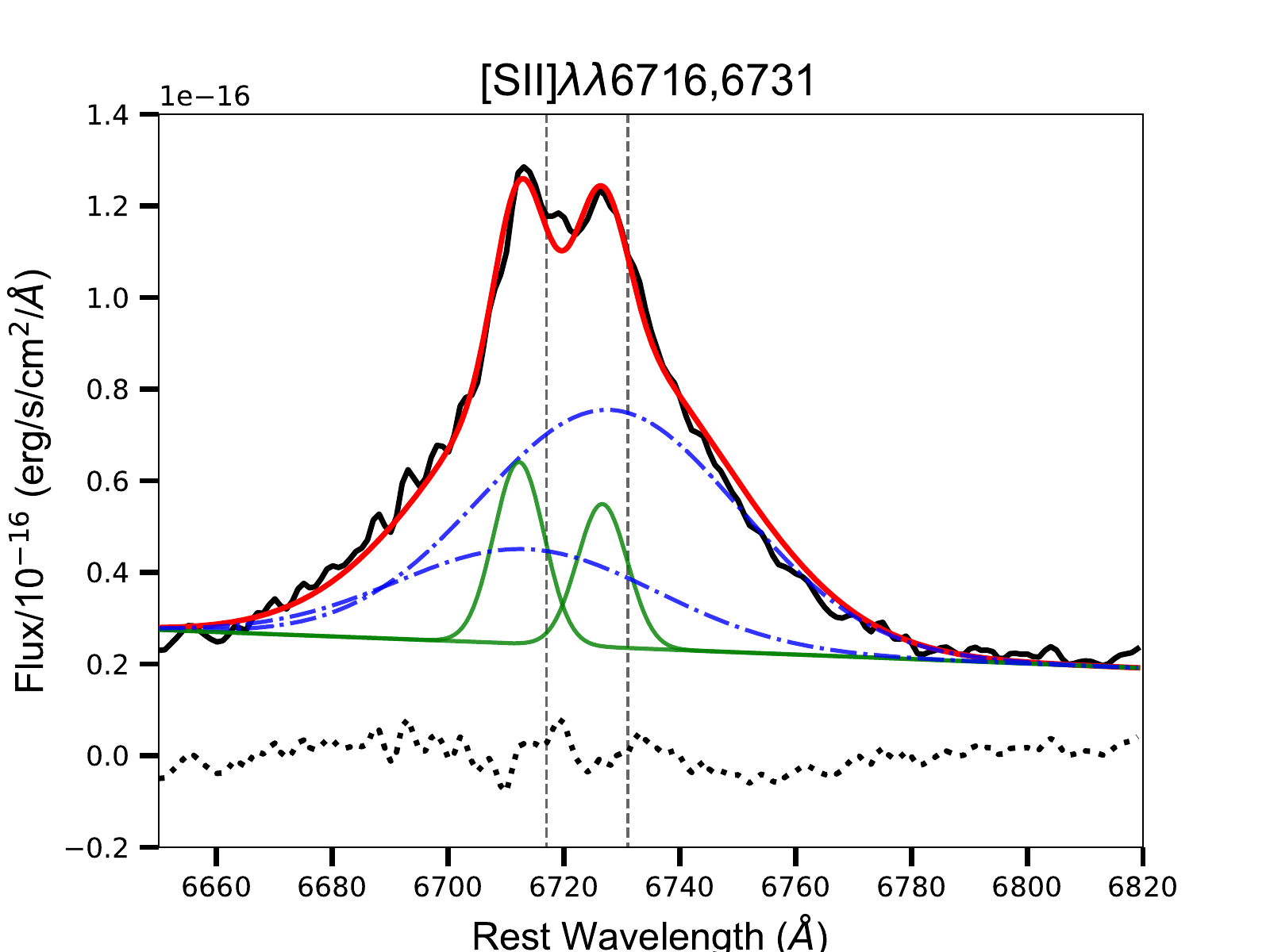}}\\ 
	\vspace{-3.6mm}
	\subfloat{\includegraphics[width = 3in]{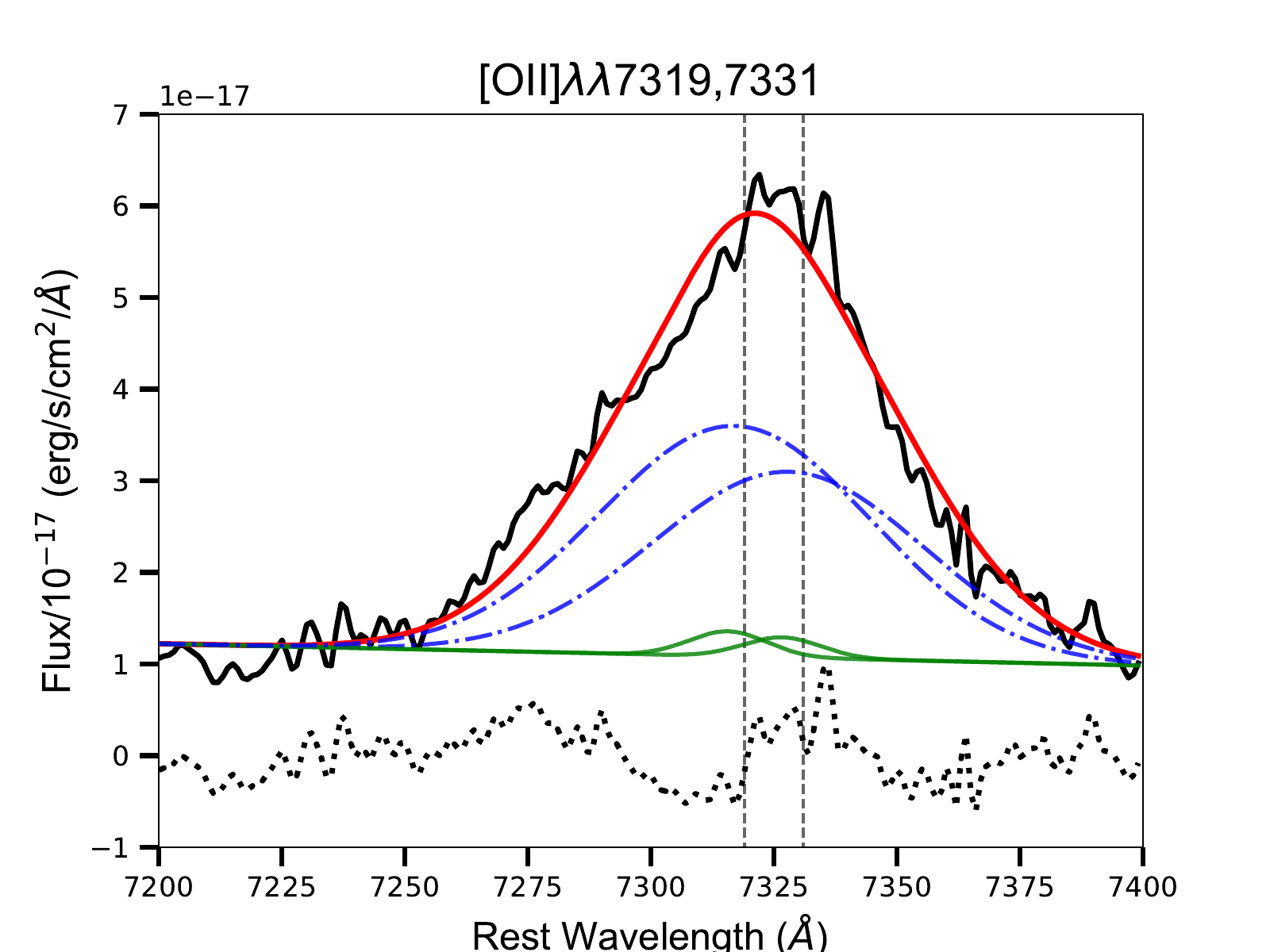}}
	\caption{Emission line profiles and fits for all diagnostic lines for F23389+0303N. The black solid line is the data, and the overall fit is shown by the red solid line. The individual components are coloured as follows: narrow (green, solid); intermediate (purple, dashed); broad (blue, dot-dashed). The residuals of the fit are shown below the profiles (black, dashed).}
	\label{fig:f23389}
\end{figure*}

The [OIII] models were successful for the majority of emission lines in 50\% of our objects. An example of the application of the [OIII] model fits to all the emission lines for F23389+0303N is shown in Figure \ref{fig:f23389}.\par

For F01004--2237, F05189--2524, F14394+5332E and F23060+0505 the [OIII] model did not work for the other lines, and an alternative kinematic model was generated using [OII]$\lambda$3727, or
[SII]$\lambda\lambda$6716,6731 in the case of F14394+5332E.  \par

F01004--2237, as mentioned above, appears to have recently undergone a tidal disruption event, strongly affecting its nuclear spectrum. While the three-component [OIII] model works for H$\beta$, the H$\alpha$ + [NII] blend only requires a two component fit to each line, with the [NII] narrow components very weak in this case. A two-component model (narrow + broad) was required to fit the [OII]$\lambda$3727 and [SII]$\lambda$6725 blends. Note that, although the red trans-auroral [OII]$\lambda\lambda$7319,7331 blend was detected in the WHT/ISIS spectrum, the blue trans-auroral [SII]$\lambda\lambda$4068,4076 blend was not detected due to severe contamination by TDE-related emission lines. It was, however, detected in the pre-TDE HST/STIS spectrum of F01004--2237. Therefore, given that there is good evidence that the slit losses for the HST/STIS and WHT/ISIS spectra were similar \citep{Tadhunter2017a}, the density and reddening estimates presented in \S\ref{sec:density} for this source are based on comparing the total fluxes of the blue [SII] blend measured from the HST/STIS spectrum (single Gaussian fit) with those of the blue [OII], red [OII] (single Gaussian fit) and red [SII] blends measured from the WHT/ISIS spectrum \par  

In the case of F05189--2524, the full [OIII]$\lambda\lambda$4959,5007 profile is blueshifted by more than 500 km s$^{-1}$ relative to the stellar rest frame, similar to the cases of PKS1549--79 \citep{Tadhunter2001} and F15130--1958 \citep{Rose2018}. This behaviour is also seen to a lesser extent in F23233+2817 and F23389+0303N. However, for F05189--2524, an additional rest-frame intermediate component was required for the Balmer, [OI], [OII] and [SII] emission lines in the ground-based WHT spectrum. Although the blueshifted intermediate and broad components were also detected in H$\beta$, the broad component was not detected in the [OI], [OII] and [SII] lines which were fitted with a two-component model consisting of a rest-frame intermediate and a blueshifted intermediate component. \par 

In the STIS spectrum of F05189--2524, only a single Gaussian could be fitted to the trans-auroral [OII] and [SII] lines to estimate the total emission-line flux. However, the relative shift of this component is within 3$\sigma$ of the blueshifted intermediate component detected in the red [SII] and blue [OII] emission lines in the ground-based spectrum. It is therefore reasonable to assume that, due to the narrower slit used for the STIS observations, the measured total flux  samples the blueshifted gas within the outflow.   \par 

The [OIII] profile of F14394+5332E is complex, requiring three narrow components (one rest-frame, two blueshifted) and a broad component \citep[see also][]{Rodriguez2013}. However, the two blueshifted narrow [OIII] lines were not detected in any of the other emission lines in this object. These were instead fitted with a three-component model (one narrow, one intermediate and one broad component), with the exception of the [NII] doublet where no broad component was required. This is likely due to degeneracy with the broad component of H$\alpha$.\par

Finally, the [OIII] profile of F23060+0505 required four components: two narrow, an intermediate, and a broad. No broad component was detected in H$\beta$, nor for the [OI] and [OII] emission lines; and only the narrow components were detected in the [SII] blends, perhaps leading to an underestimation of the outflow density for this object using the trans-auroral line ratios (\S \ref{sec:density}). However,  a broad-line region (BLR) component ($FWHM = 2356\pm69$\,km s$^{-1}$) was required in this case to fit the H$\alpha$+[NII] blend. As discussed by \cite{Rodriguez2013}, the presence of a moderately reddened type I AGN component in this source is confirmed by the relatively red shape of the long-wavelength end of its optical continuum spectrum (see Figure 1), and the detection of a broad P$\alpha$ emission line at near-IR wavelengths \citep{Veilleux1997}.

\section{Outflow Properties}
\label{sec:results}

The ultimate aim of this series of papers is to better quantify the key properties of the AGN-induced outflows in the nuclear regions of local ULIRGs. The necessary calculations require accurate determinations of the outflow kinematics, radii, electron densities and intrinsic reddening. In this section, we present the results of our emission line fitting.\par

\subsection{Emission line kinematics and outflow radii}
\label{sec:radii}
Firstly, we measured the kinematics and radii of the nuclear outflows. Table \ref{tab:outflowkinematics} gives the velocity shifts and widths for the required components of the [OIII]$\lambda$5007 emission line, using the fitting approach described in section \ref{sec:fitting}. Blue-shifted intermediate/broad components were detected in all objects, indicating the presence of ionised nuclear outflows, consistent with the results of \citep{Rodriguez2013}. We concentrated our analyses on the properties of these out-flowing components.   
\par

In order to properly quantify the outflow powers, it is also important to estimate their spatial extents along the slit. Currently the true radial extents of AGN-induced outflows ($R_{out}$) in luminous type II AGN are highly uncertain, with $R_{out}$ measurements ranging from just 0.06 kpc \citep[][submitted]{Tadhunter2018} to >10 kpc \citep{Harrison2012}.\par

To determine the radial extents in our ULIRG sample, we first isolated the out-flowing (broad, blueshifted) components of the [OIII] emission line. Guided by the Gaussian fits to the emission line profiles, we extracted spatial slices between the 95$\rm^{th }$ percentile (v$_{95}$) of the narrow [OIII]$\lambda$4959 component fit and 5$\rm^{th}$ percentile (v$_{05}$) of the narrow [OIII]$\lambda$5007 component fit. In this way, we sampled \textit{only} the out-flowing components, and avoided any significant narrow-line flux not associated with the outflow. In the case of F05189-2524, all components of [OIII] are significantly blue-shifted with respect to the stellar rest-frame. For this target, we extracted all of the [OIII]$\lambda$5007 flux blue-ward of the rest-frame wavelength. The velocity ranges over which the spatial profiles were extracted are given in Table \ref{tab:extent}, column 2, and indicated visually by the shaded regions in Figure \ref{fig:oiii}. \par 

The continuum was then removed by subtracting an average continuum profile created from two $\sim$ 30\AA\ spatial slices, one taken blue-ward and the other red-ward of the [OIII]$\lambda\lambda$4959,5007 blend. The resultant spatial profile was then fitted with a Gaussian. In all cases, a single Gaussian was sufficient. To assess whether we resolve the outflow, the FWHM of this spatial profile was compared to that of the estimated 1D seeing. The outflow was considered spatially resolved if the difference between the measured spatial FWHM and the 1D seeing FWHM was greater than 3 times the estimated uncertainty in the difference (i.e. 3$\sigma$). \par

In the spatially resolved cases, the 1D seeing FWHM was subtracted from the spatial FWHM in quadrature, converted to kpc using the pixel scale and angular scale on the sky consistent with our assumed cosmology, and then halved to find the radius. See \S3.1 in \cite{Rose2018} for a full discussion of this technique. \par

\begin{table*}
	\centering
	\caption{Radial extent of the broad, outflowing [OIII]$\lambda$5007 gas. Column (2): The velocity range over which the spatial slice was extracted. Columns (3) and (4) give the spatial FWHM for the out-flowing and 1D seeing (extracted from a standard star over the same wavelength range) respectively. Column (6) shows the calculated outflow radii using the WHT, and columns (7) and (8) show the radii measured using HST observations for comparison.
			}
	\label{tab:extent}
	\begin{tabular}{lccccccr}
		\hline
		Object name & [OIII]$\lambda$5007 range & FWHM$_{[OIII]}$ & FWHM$_{1D}$ & Resolved WHT? & & R$_{[OIII]}$ & \\
		IRAS & (km $\rm s^{-1}$) & (arcsec) & (arcsec) & (Y/N)  &  & (kpc) & \\
		& & & & & WHT/ISIS & HST/STIS & HST/ACS
		\\
		(1) & (2) & (3) & (4) & (5) & (6) & (7) & (8)\\
		\hline\\
		F01004 -- 2237 & -2591 -- -278 & 1.623 $\pm$ 0.017 & 1.430 $\pm$ 0.123 & N &  < 1.1 & 0.111 $\pm$ 0.004 & -- \\
		F05189 -- 2524$^{a}$ & -3016 -- 0 & 1.774 $\pm$ 0.014 & 1.634 $\pm$ 0.052 & N & < 0.3 & 0.079 $\pm$ 0.002 & --\\
		F14394 + 5332E & -2315 -- -524 & 1.489 $\pm$ 0.010 & 1.248 $\pm$ 0.073 & Y & 0.75 $\pm$ 0.12 & -- & 0.840 $\pm$ 0.008$^{b}$\\
		F17044 + 6720 & -2453 -- -422 & 2.044 $\pm$ 0.047 & 1.610 $\pm$ 0.069 & Y & 1.45 $\pm$ 0.18 & -- & 1.184 $\pm$ 0.006 \\
		F17179 + 5444 & -2327 -- -614   & 1.796 $\pm$ 0.046 & 1.600 $\pm$ 0.060 & N & < 1.0 & -- & 0.112 $\pm$ 0.007\\
		F23060 + 0505 & -2375 -- -524 & 1.469 $\pm$ 0.022 & 1.073 $\pm$ 0.110 & Y & 1.41 $\pm$ 0.19 & -- & -- \\
		F23233 + 2817 & -2609 -- -446 & 1.485 $\pm$ 0.014 & 1.167 $\pm$ 0.154 & N & < 1.1 & -- & --\\
		F23389 + 0303N & -2603 -- -655  & 1.542 $\pm$ 0.018 & 1.441 $\pm$ 0.100 & N & < 1.2 & --& --\\
		\hline
		\multicolumn{8}{l}{
			\begin{minipage}{\textwidth}~\\
			$^{a}$ The entire [OIII]$\lambda \lambda$5007,4959 profile for this component is blueshifted by > 500 km s$^{-1}$ i.e. only the outflowing component has been detected, so the spatial profile is extracted over the entire velocity range of [OIII]$\lambda$5007 blue-ward of the rest-frame.\\
			$^{b}${Note that the HST/ACS measurement assumes that the AGN is located in the dust lane that bisects the eastern nucleus of the system.}
			\end{minipage}
		}\\
	\end{tabular}
\end{table*} 

Using the ground-based spectroscopy, the outflow regions are spatially resolved for three objects, with the other objects unresolved compared to the seeing. In the latter cases, the quoted radii are upper limits, derived by calculating the radius (FWHM$_{[OIII]}$) at which the outflow would have been significantly resolved at the 3$\sigma$ level compared to the 1D seeing (FWHM$_{1D}$):

\begin{equation}
FWHM_{[OIII]} < \sqrt{(FWHM_{1D}+3\sigma)^{2} - (FWHM_{1D})^{2}}.
\end{equation}
\\

Table \ref{tab:extent} gives the FWHM for the spatial [OIII] and 1D seeing profiles for each object (columns 3 and 4), along with the estimated radii (columns 6 - 8). \par 

Two objects (F01004--2237 and F05189--2524) have space-based STIS spectroscopy available for comparison (column 7). Using the same spatial extraction technique, but considering the line spread function instead of atmospheric seeing, we calculate additional estimates for the outflow radii \citep[see][]{Rose2018}. The radii for both objects are relatively small, consistent with the WHT/ISIS upper limits. \par 

Estimates of the outflow radii for a further three objects (F14394+5332E, F17044+6720 and F17179+5444) are also available from HST/ACS narrow-band imaging, presented in \citealt{Tadhunter2018} (submitted). The HST-based flux-weighted mean radius estimates are given in column 8.  There is remarkable consistency between the HST/ACS estimates and the resolved WHT/ISIS estimates for F14374+5332E and F17044+6720. The ACS imaging estimate for F17179+5444 is also consistent with the upper limit from the WHT spectroscopy. \par

Note that, while we can only determine an upper limit on the outflow radius for F23389+0303N leading to lower limits on the mass outflow rates and kinetic powers in this case, we also provide estimates of the outflow properties based on the radius of the radio lobes (r = 0.415 kpc) from \citealt{Nagar2003}. Although this involves the assumption that the outflows are jet-driven in this object, generally the emission-line outflows associated with compact radio sources of similar power to F23389+0303N are found to be co-spatial with the radio sources \citep[e.g.][Rose et al. in prep]{Batcheldor2007, Tadhunter2014} 

Overall, consistent with the results of \cite{Rose2018} and \citealt{Tadhunter2018} (submitted), the outflows in our sample of ULIRGs are compact, with all radius estimates in the range 0.08 < $R_{out}$ < 1.5 kpc.

\subsection{Electron density and reddening}
\label{sec:density}
Next, we calculated the electron densities and intrinsic reddenings of the outflows. As discussed in \cite{Rose2018}, previous attempts to calculate mass outflow rates and kinematic powers in AGN have been severely limited by the lack of accurate estimates of the electron density. For AGN, the most commonly used optical diagnostics are the [SII]$\lambda\lambda6716,6731$ and [OII]$\lambda\lambda3726,3729$ doublet ratios.  However, these are only sensitive to relatively low densities ($\rm10^{2} < n_{e} < 10^{3.5} cm^{-3}$) and are therefore ineffective for the higher density clouds that are also expected to be associated with outflows. Furthermore, the highly blue- or redshifted, broad components can cause degeneracies in the fitting, due to severe blending of the emission line profiles. These degeneracies can also affect the H$\alpha$ + [NII] blend, leading to significant uncertainties in the intrinsic reddening calculated from the Balmer decrement.\par

For this reason, we have obtained spectra optimised for the trans-auroral [SII]$\lambda\lambda$4068,4076 and [OII]$\lambda\lambda$7319,7331 features. We combined these emission line fluxes with the [SII]$\lambda\lambda6716,6731$ and [OII]$\lambda\lambda3726,3729$ as follows:

\begin{equation}
    TR([OII]) = F(3726 + 3729)/F(7319 + 7331); \rm and
	\label{eq:troii}
\end{equation}

\begin{equation}
   TR([SII]) = F(4068 + 4076)/F(6716 + 6731).
	\label{eq:trsii}
\end{equation}
\\
\noindent Combining the fluxes in this way gives a density diagnostic that does not suffer from the problems with degeneracies that affect the [SII](6716/6731) and [OII](3726/3729) diagnostics. This is because we are comparing the total fluxes of the widely separated blends, rather than the fluxes of the individual components within them. This technique, introduced by \cite{Holt2011}, is also sensitive to a higher range of densities ($\rm10^{2} < n_{e} < 10^{6.5} cm^{-3}$). Another advantage is that these diagnostics give a simultaneous estimate of the intrinsic dust reddening, E(B-V), of the emitting clouds. \par

\begin{figure}
	\includegraphics[width=\columnwidth]{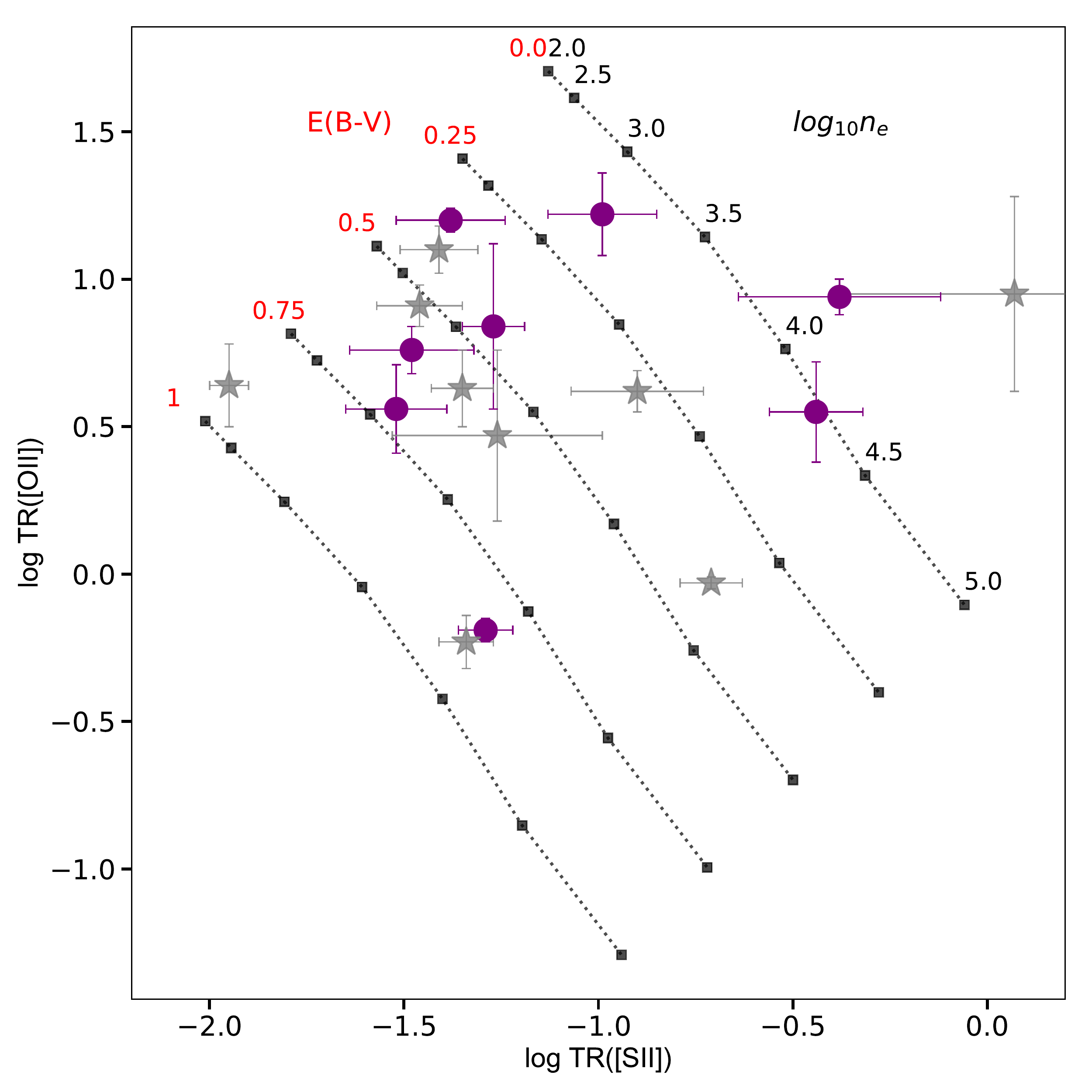}
    \caption{Trans-auroral line ratios for the total emission line fluxes. The purple circles are the results from this paper. Overplotted as grey stars are the results from \citep{Rose2018}. 
    		} 
	\label{fig:trans_total}
\end{figure} 

\begin{figure}
	\includegraphics[width=\columnwidth]{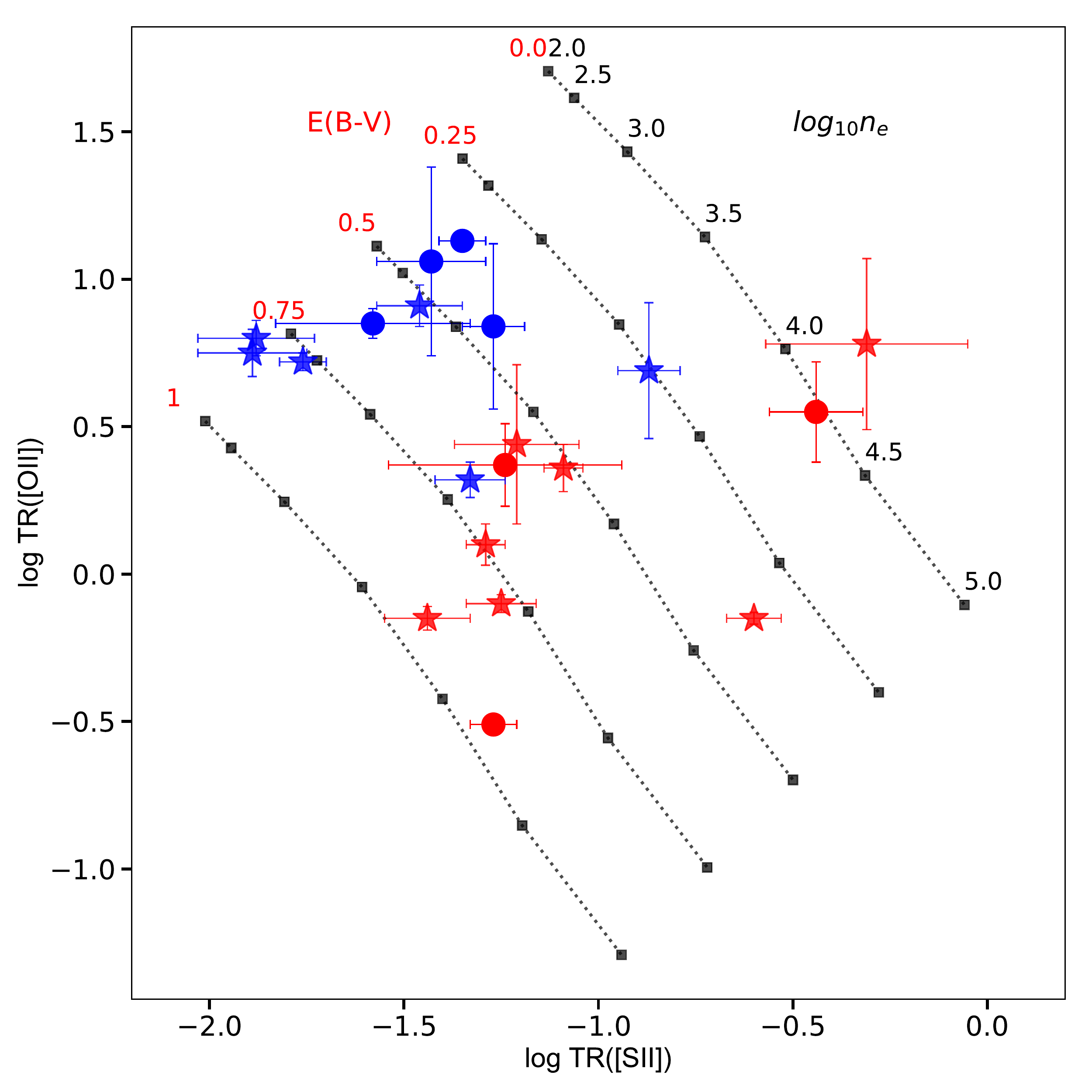}
    \caption{Trans-auroral line ratios for the narrow and broad component fluxes. The solid blue and red circles indicate the narrow and broad components, respectively, for the objects in this paper. Over-plotted as blue and red stars are the narrow and broad ratios from \citep{Rose2018}.
    		} 
	\label{fig:trans_broad}
\end{figure} 

Ideally, we want to be able to resolve the individual trans-auroral kinematic components for all our objects. Unfortunately, the spectral resolution and S/N for WHT/ISIS are lower than for the VLT/XShooter observations used in \cite{Rose2018}. This means that the weaker red [OII] and blue [SII] trans-auroral emission lines for the majority of the objects considered in this paper were not sufficiently strong to allow the separation of narrow and broad/intermediate components. In these cases, a single Gaussian profile was fitted in order to estimate the total emission line fluxes. Note that the densities derived from the total fluxes are generally lower than those derived from the broad component alone (see Figures 8 and 10 in \citealt{Rose2018}). Because the mass outflow rate and kinetic power depends on the inverse of the density, this could lead us to calculate higher values for these outflow properties than \cite{Rose2018}. \par

Figure \ref{fig:trans_total} shows a plot of log TR([OII]) vs. log TR([SII]) for the total emission line fluxes for the 8 objects considered in this paper (purple circles). All fluxes were measured from the WHT/ISIS spectra, with the exception of F01004--2237, for which the HST/STIS spectrum was used to measure the [SII]$\lambda\lambda$4068,4076 flux, and F05189--2524, for which the HST/STIS spectrum was used to measure all emission line fluxes. Also plotted are the measurements from \cite{Rose2018} (grey stars). \par 

Over-plotted is a grid of ratios predicted by AGN photo-ionisation models, fully described by \cite{Rose2018}. The densities based on the total fluxes fall in the range  2.5 < log n$_{e}$(cm$^{-3}$) < 4.5, with the median log n$_{e}$(cm$^{-3}$) = 3.10$\pm$0.29. The intrinsic dust  reddening falls in the range $\rm 0.0 < E(B-V) < 1.0$, with the median E(B-V) = 0.45$\pm$0.14. Table \ref{tab:totals} gives the density and reddening values obtained using the total emission line fluxes for each object. These results are consistent with those obtained by \citet{Rose2018} using XShooter data for the rest of ULIRGs in the QUADROS sample.   \par 

For the objects where narrow and intermediate/broad components of the trans-auroral emission lines could be resolved, the associated density and reddening estimates are also shown separately in Tables \ref{tab:narrows} and \ref{tab:broads}. Note that F23060+0505 is only included in the Table \ref{tab:narrows} because only the narrow components of the trans-auroral emission lines were detected. Similarly, F05189--2524 is only included in Table \ref{tab:broads} because the whole trans-auroral emission line profiles are blue-shifted with respect to the rest frame, with no rest-frame narrow component detected. Therefore we assume that, in this case, all of the detected flux is associated with the outflow. \par 

These narrow and broad flux ratios are plotted in Figure \ref{fig:trans_broad} (blue and red circles, respectively), along with the measurements from \cite{Rose2018} (blue and red stars, respectively). Some overlap between the narrow and broad components can be seen; however, considering the sample as a whole, the densities of the broad, out-flowing components are generally high (3.5 < log n$_{e}$(cm$^{-3}$) < 4.5, median log n$_{e}$(cm$^{-3}$) = 3.8$\pm$0.2). In comparison, the median density for the narrow components is an order of magnitude lower: log n$_{e}$(cm$^{-3}$) = 2.8$\pm$0.3. \par 

Clearly, assuming a single low density (log n$_{e}$(cm$^{-3}$) $\sim$ 2) for AGN-induced outflows, as is common in many outflow studies, is not justified and is likely to lead to some of the higher values of the mass outflow rates and kinetic powers in the literature. \par

Despite the previously mentioned degeneracies in the [NII] + H$\alpha$ fits, as a check we have also determined the intrinsic reddening (E(B-V)) based on the H$\alpha$/H$\beta$ ratios for the individual components. The estimates are shown in column 6 of Tables \ref{tab:totals} to {\ref{tab:broads}. In general, the two techniques are consistent within 3$\sigma$. This is illustrated in Figure \ref{fig:reddening} for the total, narrow and broad fluxes. Interestingly, and consistent with \cite{Rose2018}, we find no clear evidence that the outflowing gas is reddened more than the narrow-line gas associated with the host galaxy. Indeed, the narrow component ratios cover a similar range of reddening as the broad, out-flowing components and are, on average, higher. Based on the reddening estimates derived from the Balmer decrements of the ULIRGs considered in this paper, we find median E(B-V)$\rm_{narrow}$ = 0.52$\pm$0.09; median E(B-V)$\rm_{broad}$ = 0.15$\pm$0.25. \par

As a reference to compare with the trans-auroral estimates, we also measured the density of the gas emitting the narrow-line components using the [SII]$\lambda\lambda$6716/6731 doublet ratio, in which the narrow components are generally strong and well resolved. This technique is not appropriate for the broad components due both to degeneracies in the fit and the fact that in most cases the ratio of the broad components was constrained to the high-density limit during the fitting process. The estimated densities are presented in Table \ref{tab:narrows}, Column 5. For the majority of objects the narrow-component ratios were less than 3$\sigma$ from the low-density limit. For these we provide a 3$\sigma$ upper limit on the density. \par 

In all cases, the density estimates and upper-limits derived from the [SII]$\lambda\lambda$6716/6731 doublet ratios for the narrow components are lower than those derived from the ratios of both the total and broad trans-auroral fluxes, although in the case of F17044+6720 the two estimates are close. These results support the conclusion that the outflows have higher densities than the regions emitting the narrow components.

\begin{figure}
	\includegraphics[width=\columnwidth]{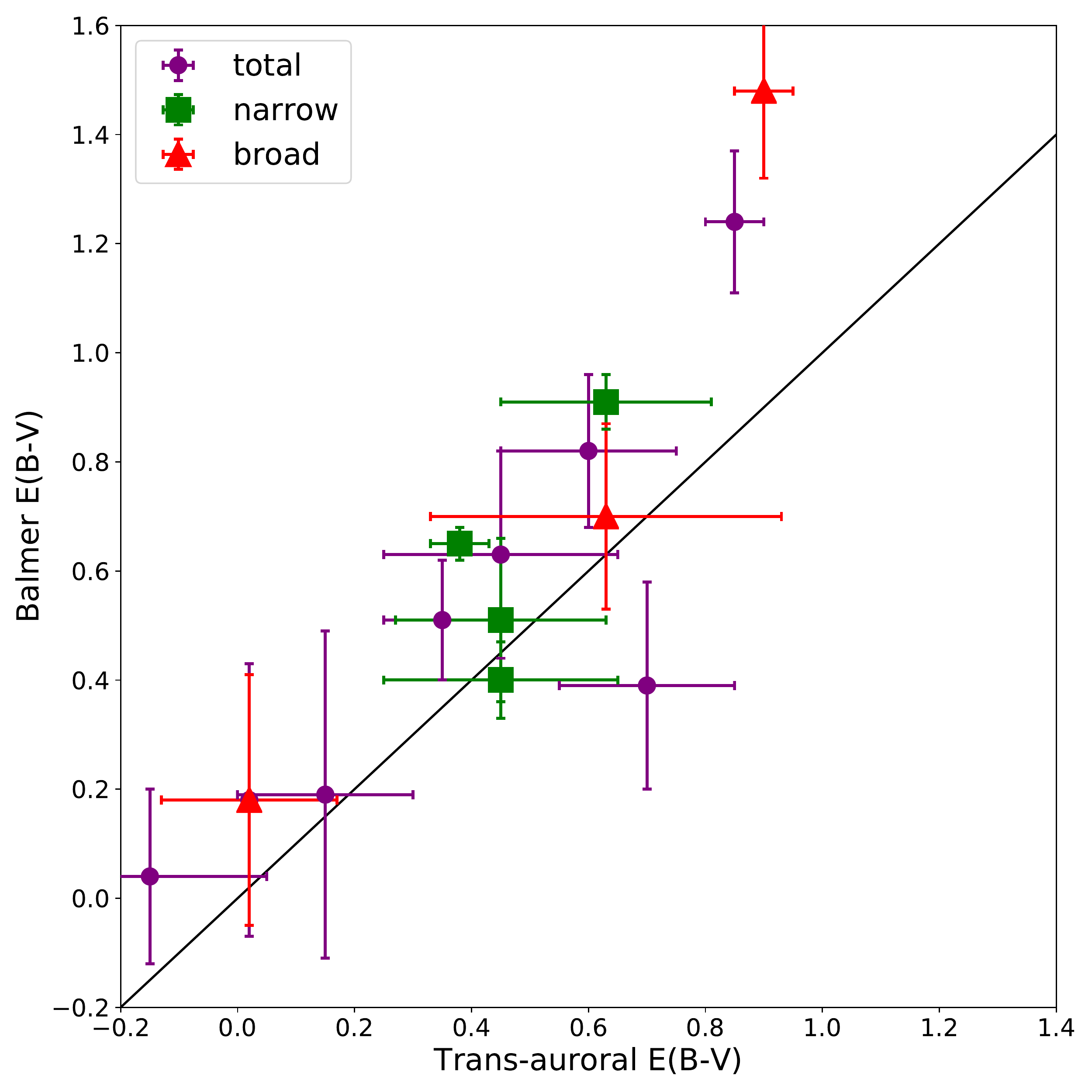}
    \caption{The reddening estimates obtained from the Balmer decrement and trans-auroral line ratios for the emission line fluxes for the different kinematic components. The black solid line represents a one-to-one ratio. The two techniques are consistent to within 3$\sigma$.} 
	\label{fig:reddening}
\end{figure}

\begin{table*}
	\centering
	\caption{The log n$_{e}$ and E(B-V) values determined using the \textit{total} trans-auroral emission-line fluxes. Column (5) shows the E(B-V) estimates obtained using the trans-auroral lines ratios. Column (6) shows the E(B-V) estimates obtained using the H$\alpha$/H$\beta$ Balmer decrement.
			}
	\label{tab:totals}
	\begin{tabular}{lccccr} 
		\hline
		Object name & log[SII] & log[OII] & log(n$_{e}$) & E(B-V)$\rm_{trans}$ &  E(B-V)$\rm_{Bal}$\\
		IRAS & & & (cm$^{-3}$) & &\\
		\\
		(1) & (2) & (3) & (4) & (5)& (6)\\
		\hline\\
		F01004 -- 2237 & -0.38 $\pm$ 0.26 & 0.94 $\pm$ 0.06 & 4.00 $^{+0.25}_{-0.30}$ & -0.15 $\pm$ 0.20 & 0.04 $\pm$ 0.16\\
		\\
		F05189 -- 2524 & -0.44 $ \pm$ 0.12 & 0.55 $\pm$ 0.17 & 4.25 $^{+0.25}_{-0.25}$ & 0.02 $^{+0.15}_{-0.15}$ & 0.29 $\pm$ 0.25 \\
		\\
		F14394 + 5332E & -1.48 $\pm$ 0.16 & 0.76 $\pm$ 0.08 & 2.90 $^{+0.30}_{-0.40}$ & 0.60 $^{+0.15}_{-0.15}$ & 0.82 $\pm$ 0.14 \\
		\\
		F17044 + 6720 & -1.38 $\pm$ 0.14 & 1.20 $\pm$ 0.04 & 2.50 $^{+0.30}_{-0.50}$ & 0.35 $^{+0.10}_{-0.10}$  & 0.51 $\pm$ 0.11\\
		\\
		F17179 + 5444 & -1.52 $\pm$ 0.13 & 0.56 $\pm$ 0.15 & 3.05 $^{+0.30}_{-0.45}$ & 0.70 $^{+0.15}_{-0.15}$ & 0.39 $\pm$ 0.19\\
		\\
		F23060 + 0505 & -1.27 $\pm$ 0.08 & 0.84 $\pm$ 0.28 & 3.10 $^{+0.35}_{-0.50}$ & 0.45 $^{+0.15}_{-0.20}$ & 0.63 $\pm$ 0.19\\
		\\
		F23233 + 2817 & -0.99 $\pm$ 0.14 & 1.22 $\pm$ 0.14 & 3.10 $^{+0.30}_{-0.40}$ & 0.15 $^{+0.15}_{-0.15}$ & 0.19 $\pm$ 0.30\\
		\\
		F23389 + 0303N & -1.29 $\pm$ 0.07 & -0.19 $\pm$ 0.03 & 3.95 $^{+0.10}_{-0.10}$ & 0.85 $^{+0.05}_{-0.05}$ & 1.24 $\pm$ 0.13\\
		\\
		\hline
	\end{tabular}
\end{table*} 

\begin{table*}
	\centering
	\caption{The log n$_{e}$ and E(B-V) values determined using the \textit{narrow} emission-line fluxes. Column (4) shows the density estimates obtained using the narrow component of the trans-auroral emission lines, where detected. Column (5) shows the density estimates based on the narrow [SII]$\lambda\lambda$6716/6731 flux ratio for comparison. Columns (6) and (7) compare the E(B-V) estimates calculated using the trans-auroral [SII] and [OII] emission blends with those calculated using the H$\alpha$/H$\beta$ decrement.
			}
	\label{tab:narrows}
	\begin{tabular}{lcccccr} 
		\hline
		Object name & log[SII] & log[OII] & log(n$_{e}$)$_{\rm trans}$ & log(n$_{e}$)$_{\rm[SII]}$  & E(B-V)$\rm_{trans}$ &  E(B-V)$\rm_{Bal}$\\
		IRAS & & & (cm$^{-3}$) & $\lambda\lambda$6716/6731 (cm$^{-3}$) &\\
		\\
		(1) & (2) & (3) & (4) & (5)& (6) & (7)\\
		\hline\\
		F01004 -- 2237 & - & - & - & < 2.60 & - & 0.75 $\pm$ 0.11\\
		F05189 -- 2524 & -  & - & - & 2.81$^{+0.14}_{-0.14}$ & & 0.47 $\pm$ 0.27 \\
		F14394 + 5332E &-1.58 $\pm$ 0.25 & 0.85 $\pm$ 0.05 & 2.70 $^{+0.35}_{-0.50}$ & < 2.64  & 0.63 $^{+0.17}_{-0.18}$ & 0.91 $\pm$ 0.09 \\
		F17044 + 6720 & - & - & - & 2.49$^{+0.15}_{-0.18}$ & - & 0.65 $\pm$ 0.06\\
		F17179 + 5444 &- & - & - & -  & - & -\\
		F23060 + 0505 & -1.27 $\pm$ 0.08 & 0.84 $\pm$ 0.28 & 3.15 $^{+0.30}_{-0.50}$ & < 2.88 & 0.45 $^{+0.15}_{-0.18}$ & 0.51 $\pm$ 0.18\\
		F23233 + 2817 & - & - & - & < 2.55  & - & 0.52 $\pm$ 0.16\\
		F23389 + 0303N & -1.43 $\pm$ 0.14 & 1.06 $\pm$ 0.32 & 2.60 $^{+0.60}_{-0.60}$  & < 2.91 & 0.45 $^{+0.20}_{-0.20}$ & 0.40 $\pm$ 0.09\\
		\\
		\hline
	\end{tabular}
\end{table*} 

\begin{table*}
	\centering
	\caption{The log n$_{e}$ and E(B-V) values determined using the \textit{broad} trans-auroral emission-line fluxes. Note that the results for F05189-2524 were obtained using HST/STIS data (also in Table \ref{tab:totals}.)
			}
	\label{tab:broads}
	\begin{tabular}{lccccr} 
		\hline
		Object name & log[SII] & log[OII] & log(n$_{e}$) & E(B-V)$\rm_{trans}$ &  E(B-V)$\rm_{Bal}$\\
		IRAS & & & (cm$^{-3}$) & &\\
		\\
		(1) & (2) & (3) & (4) & (5)& (6)\\
		\hline\\
		F01004 -- 2237 & - & - & - & - & -0.20 $\pm$ 0.20\\
		F05189 -- 2524 & -0.44 $ \pm$ 0.12 & 0.55 $\pm$ 0.17 & 4.25 $^{+0.25}_{-0.25}$  & 0.02 $^{+0.15}_{-0.15}$ & 0.18 $\pm$ 0.24 \\
		F14394 + 5332E & -1.24 $\pm$ 0.30 & 0.37 $\pm$ 0.14 & 3.55 $^{+0.45}_{-0.45}$ & 0.63 $^{+0.22}_{-0.30}$ & 0.70 $\pm$ 0.21 \\
		F17044 + 6720 & - & -  & - & - & -0.35 $\pm$ 0.67\\
		F17179 + 5444 &  -  & - & - & - & 0.11 $\pm$ 0.37\\
		F23060 + 0505 &- & - & - & - & 1.10 $\pm$ 0.18\\
		F23233 + 2817 & -& - & - & - & -0.06 $\pm$ 0.45\\
		F23389 + 0303N & -1.27 $\pm$ 0.06 & -0.51 $\pm$ 0.03 & 4.20 $^{+0.10}_{-0.10}$ & 0.90 $^{+0.05}_{-0.05}$ & 1.48 $\pm$ 0.14\\
		\\
		\hline
	\end{tabular}
\end{table*}

\subsection{Bolometric luminosities}
\label{sec:bol}
In order to compare the kinetic powers that we derive for the warm outflows with the total radiative powers of the AGN, their bolometric luminosities ($L_{bol}$) must first be determined. To do this we used the [OIII]$\lambda$5007 emission line luminosity, which has been shown to be a reasonable indicator of the AGN power \citep{Heckman2004, Dicken2014}. We adopted two approaches for determining $ L_{bol}$: 1) the bolometric correction of \cite{Heckman2004}: $L_{bol} = 3500L_{[OIII]}$, where $L_{[OIII]}$ has not been corrected for dust extinction; and 2) the bolometric correction factors of \cite{Lamastra2009} which uses the reddening-corrected $L_{[OIII]}$. The correction factors are 87, 142 and 454 for reddening-corrected luminosities of log$L_{[OIII]}$ = 38-40, 40-42 and 42-44 erg $\rm s^{-1}$, respectively.\par

For our bolometric luminosity determinations we used the total [OIII]$\lambda$5007 fluxes, which involves the assumption that there is little contribution to [OIII]$\lambda$5007 from stellar photo-ionised regions. This is a reasonable assumption for our ULIRGs, based on the BPT diagnostic diagrams presented in \cite{Rodriguez2013}, which suggest that the warm gas is dominated by AGN photo-ionisation. Although F05189--2524 was not included in the latter paper, our own BPT analysis shows the line ratios for this object to also be consistent with AGN photo-ionisation.\par 

We determined our luminosities using the measured [OIII]$\lambda$5007 fluxes. For technique (2), the luminosities were corrected for dust extinction using the total E(B-V) values from Column 5 in Table \ref{tab:totals} and the extinction law of \cite{Calzetti2000}. The calculated luminosities are shown in Table \ref{tab:bol}, where the estimates obtained using the corrections of Heckman and Lamastra are compared in Column 6. The estimates are consistent to within a factor of $\sim$5 for most of the objects; however, for the three objects with the lowest dust extinction (F01004--2237, F05189--2524 and F23233+2817) the Lamastra estimates are over an order of magnitude lower than the Heckman estimates. We argue that the Heckman correction is likely the most appropriate for these cases due to the small amount of reddening. Therefore, we use only the Heckman-derived bolometric luminosities in the following analyses.\par 



\begin{table*}
	\centering
	\caption{[OIII]$\lambda$5007 luminosities and associated bolometric luminosities. Column (2): uncorrected [OIII] luminosity. Column (3): extinction-corrected [OIII] luminosity based on the E(B-V) values derived from the total trans-auroral flux ratios and extinction law of \citealt{Calzetti2000}. Column (4): the [OIII] bolometric luminosity based on the Heckman bolometric correction factor. Column (5): the [OIII] luminosity based on the appropriate Lamastra bolometric correction factor. Column (6): the ratio of the Heckman bolometric luminosity to the Lamastra bolometric luminosity. 
			}
	\label{tab:bol}
	\begin{tabular}{lccccr}
		\hline
		\\
		Object name & L$\rm_{[OIII]-unc}$ & L$\rm_{[OIII]-corr}$ & L$\rm_{bol-Heck}$ & L$\rm_{bol-Lam}$ & Heck/Lam \\
		IRAS & erg s$^{-1}$ & erg s$^{-1}$  & erg s$^{-1}$ & erg s$^{-1}$ & \\
		\\
		(1) & (2) & (3) & (4) & (5) & (6)\\
		\hline\\
		F01004 -- 2237 & (2.5$\pm$0.1)E+41 & (3.0$\pm$1.4)E+41 & (8.8$\pm$0.4)E+44 & (4.2$\pm$2.0)E+43  & 21\\
		F05189 -- 2524 & (1.0$\pm$0.1)E+41 & (1.1$\pm$0.5)E+41 & (3.6$\pm$0.1)E+44 & (1.6$\pm$0.7)E+43 & 23\\
		F14394 + 5332E & (2.6$\pm$0.1)E+41 & (3.1$\pm$1.4)E+42 & (9.0$\pm$0.2)E+44& (1.4$\pm$0.6)E+45& 0.64\\
		F17044 + 6720 & (9.9$\pm$0.2)E+40 & (4.3$\pm$1.4)E+41  & (3.3$\pm$0.1)E+44 & (6.1$\pm$2.0)E+43& 5.7 \\
		F17179 + 5444 & (1.7$\pm$0.1)E+41  & (3.2$\pm$1.5)E+42  & (6.1$\pm$0.1)E+44 & (1.5$\pm$0.7)E+45 & 0.41 \\
		F23060 + 0505 & (1.3$\pm$0.1)E+42 & (8.6$\pm$3.6)E+42 & (4.6$\pm$0.4)E+45 &  (3.9$\pm$1.6)E+45 & 1.2 \\
		F23233 + 2817 & (4.8$\pm$0.2)E+41 & (8.9$\pm$4.0)E+41 & (1.7$\pm$0.1)E+45  & (1.3$\pm$0.6)E+44 & 13 \\
		F23389 + 0303N & (6.7$\pm$0.1)E+41 & (2.3$\pm$0.4)E+43  & (2.4$\pm$0.1)E+45 & (1.1$\pm$0.2)E+46 & 0.22\\
		\hline
	\end{tabular}
\end{table*} 

\section{Mass outflow rates and powers}
\label{sec:powers}

The key parameters for quantifying the importance of AGN-driven outflows are mass outflow rate and kinetic power. Following the arguments of \cite{Rose2018} we  consider the following two cases, based on two different assumptions about the kinematics. 

\subsection{Using flux-weighted mean outflow velocities, ignoring projection effects}

\label{sec:conservative}

The mass outflow rate ($\dot{M}$) is given by:

\begin{equation}
\dot{M} = \frac{L(H\beta)m_{p}v_{out}}{\alpha_{eff}^{H\beta}h\nu_{H\beta}n_{e}R_{out}}
\end{equation}

\noindent where $v_{out}$ is the assumed velocity of the out-flowing gas, measured from the Gaussian fits, $R_{out}$ is the outflow radius, L(H$\beta$) is the intrinsic (i.e. extinction-corrected) H$\beta$ emission-line luminosity for the out-flowing gas (broad + intermediate components), $n_{e}$ is the electron density derived from the trans-auroral emission line ratios, $m_{p}$ is the proton mass, $\alpha_{eff}^{H\beta}$ is the Case B effective recombination coefficient of H$\beta$ for T$_{e}$ = 10$^{4}$K ($\rm\alpha_{eff}^{H\beta} = 3.03\times10^{-14} cm^{3} s^{-1}$), taken from \citealt{Osterbrock2006}) and $h\nu_{H\beta}$ is the energy of an H$\beta$ photon. \par

In this first scenario, we took the mean velocity shift of the out-flowing component of [OIII] with respect to the galaxy rest-frame to represent $v_{out}$. For those objects with two or more out-flowing components, we calculated a flux-weighted mean. For the radius we used the HST estimate where available from Table \ref{tab:extent}. In the three objects where this is not possible, we used the WHT/ISIS estimate, and for F23389+0303N we have used the radial extent of the double-lobed radio source. To calculate L(H$\beta$) we used the appropriate E(B-V) derived from the trans-auroral flux ratios.  \par

The outflow power was then calculated using the following equation:

\begin{equation}
\dot{E} = \frac{\dot{M}}{2}(v_{out}^2 + 3\sigma^{2})
\end{equation}

\noindent where $\sigma$ is the line-of-sight (LoS) velocity dispersion ($\sigma \approx$ FWHM/2.355) calculated using the full-width at half-maximum (FWHM) of [OIII] for the out-flowing gas component. Again, for those objects with more than one out-flowing component, we used a flux-weighted mean FWHM. This technique assumes that all of the emission-line broadening is due to turbulence in the gas and that $v_{out}$ represents the true outflow velocity. Note that, in this method, we \textit{did not} correct for the effects of projection on the measured velocities.\par 

We have also expressed the outflow kinetic power as a fraction of the AGN radiative power ($\dot{F}$) by dividing $\dot{E}$ by $L_{bol}$. The calculated values of $v_{out}$, FWHM, $\dot{M}$, $\dot{E}$ and $\dot{F}$ are presented in Table \ref{tab:cons}. In addition, for the objects where we could only derive upper limits in the outflow radius, the values of the kinetic power, mass outflow rate and AGN fraction given in Table \ref{tab:cons} are lower limits.\par

Note that, unlike \cite{Rose2018}, most of the estimates are based on density values derived from total trans-auroral fluxes (designated ``T" in column 2). This is due to the lower spectral resolution and S/N for WHT/ISIS compared with VLT/XShooter. However, in two cases -- F14394+5332E and F23389+0303N -- we also present estimates based on density values derived from the broad trans-auroral line fluxes alone (designated ``B" in column 2). Moreover, in the case of F05189--2524, the total trans-auroral fluxes are representative of the broad, outflowing components. Overall, we regard the estimates of the outflow properties based on the densities derived from the broad outflowing components in F14394+5332E, F05189--2524 and F23389+0303N as being the most reliable.

In the cases where R$_{[OIII]}$ is resolved, we find mass outflow rates in the range 0.06 < $\dot{M}$ < 6 M$_{\sun}$yr$^{-1}$ and kinetic powers in the range $1.8\times10^{40} < \dot{E} < 4.3\times10^{42}$ erg s$^{-1}$. These ranges are consistent with those found in \cite{Rose2018}, but show little overlap with those calculated by \cite{Harrison2012}, \cite{Liu2013} and \cite{Mcelroy2015} who find larger mass outflow rates and kinetic powers for type-2 AGN using a similar method. Our results are more consistent with those of \cite{Harrison2014}, and overlap very well with the results of \cite{villarmartin2016}. \par

Comparing the calculated outflow kinetic powers with the bolometric luminosities of the AGN, we find values in the range 4$\times$10$^{-3}$ < $\dot{F}$ < 0.5\%. 
\begin{table*}
	\centering
	\caption{Outflow properties for the flux-weighted mean velocity case. Column (2): the trans-auroral fluxes used to calculate the density and reddening values employed in the calculations: T = total, B = broad/intermediate. In the case of F05189--2524, the whole profile is blueshifted with respect to the rest-frame and hence the total flux and broad flux are assumed to be equivalent. Column (3): the flux weighted mean velocity. Column (4): the flux-weighted  mean FWHM. Column (5): the calculated mass outflow rate. Column (6): the calculated kinetic power. Column (7): the kinetic power expressed as a fraction of the bolometric luminosity ($\dot{E}/L_{bol}$).
			}
	\label{tab:cons}
			\begin{tabular}{lcccccr}
				\hline
				\\
				Object name & Comp. & $v_{mean}$ & FWHM  & $\dot{M}$ & $\dot{E}$ & $\dot{F}$ \\
				IRAS & T/B & km s$^{-1}$ & km s$^{-1}$ & M$_{\sun}$ yr$^{-1}$ & erg s$^{-1}$ & \% \\
				\\
				(1) & (2) & (3) & (4) & (5) & (6) & (7)\\
				\hline\\
				F01004 -- 2237 & T & -826$\pm$52 & 1302$\pm$35 &  0.5$^{+1.0}_{-0.3}$ & (2.5$^{+6.2}_{-1.9}$)$\times$10$^{41}$ & (2.8$^{+7.5}_{-2.1}$)$\times$10$^{-2}$\\
				\\
				F05189 -- 2524 & T,B & -688$\pm$31 & 949$\pm$32 & 0.06$^{+0.10}_{-0.04}$ & (1.8$^{+3.4}_{-1.3}$)$\times$10$^{40}$ & (5.1$^{+9.8}_{-3.6}$)$\times$10$^{-3}$\\
				\\
				F14394 + 5332E & T & -990$\pm$89 & 1616$\pm$27 & 5.5$^{+16.9}_{-4.1}$ & (4.2$^{+14.4}_{-3.2}$)$\times$10$^{42}$ & 0.47$^{+1.65}_{-0.36}$\\
				& B & '' & '' & 1.4$^{+6.2}_{-1.3}$ & (1.1$^{+5.3}_{-1.0}$)$\times$10$^{42}$ & 0.12$^{+0.60}_{-0.11}$\\ 
				\\
				F17044 + 6720 & T & -503$\pm$85 & 1754$\pm$60 &  0.3$^{+1.2}_{-0.2}$ & (1.8$^{+8.1}_{-1.3}$)$\times$10$^{41}$ & (5.2$^{+24.0}_{-3.7}$)$\times$10$^{-2}$ \\
				\\
				F17179 + 5444 & T & -241$\pm$78 & 1528$\pm$32 &  3.2$^{+26.0}_{-2.6}$ & (1.3$^{+11.8}_{-1.1}$)$\times$10$^{42}$& 0.22$^{+1.96}_{-0.18}$ \\
				\\
				F23060 + 0505 & T & -461$\pm$63 & 1023$\pm$48 &  0.8$^{+3.2}_{-0.6}$ & (1.9$^{+9.5}_{-1.5}$)$\times$10$^{41}$ & (4.0$^{+22.8}_{-3.4}$)$\times$10$^{-3}$\\
				\\
				F23233 + 2817 & T & -391$\pm$35 & 940$\pm$17 &  > 0.04 & > 6.6$\times$10$^{39}$ & > 4.0$\times$10$^{-4}$\\
				\\
				F23389 + 0303N$^{a}$ & T & -134$\pm$36 & 2345$\pm$37 &  > 0.2 & > 1.4$\times$10$^{41}$ & > 6.0$\times$10$^{-3}$\\
				& B & '' & '' & > 0.1 & > 1.30$\times$10$^{41}$  & > 5.4$\times$10$^{-4}$\\
				\\
				F23389 + 0303N$^{b}$ & T & -134$\pm$36 & 2345$\pm$37 & 0.9$^{+3.8}_{-0.6}$ & (8.9$^{+37.7}_{-5.6}$)$\times$10$^{41}$ & (3.8$^{+16.4}_{-2.4}$)$\times$10$^{-2}$ \\
				& B & '' & '' & 0.7$^{+2.7}_{-0.4}$ & (6.2$^{+26.3}_{-3.9}$)$\times$10$^{41}$ & (2.6$^{+11.4}_{-1.7}$)$\times$10$^{-2}$\\
		\\
		\hline
		\multicolumn{7}{l}{
					\begin{minipage}{0.7\textwidth}~\\
							$^{a}$ These estimates were made using the upper limit on the outflow radius from the WHT/ISIS spectrum. \\
							$^{b}$ These estimates were made using the radius estimate of the radio lobes (R = 415 pc) as a proxy for the outflow radius, taken from \citealt{Nagar2003}. This assumes that the outflows are jet-driven. \\
					\end{minipage}
				}\\
	\end{tabular}
\end{table*} 

\subsection{Using maximal outflow velocities to account for projection effects}
\label{sec:max}

The estimates made in section \ref{sec:conservative} are likely to underestimate the true values due to the fact that use of the mean outflow velocities of the out-flowing components does not take into account LoS projection effects, as discussed in \cite{Rose2018}.\par 

A more physical approach is to assume that the broadening of the emission lines is due entirely to different projections of the velocity vector of the expanding outflow with respect to the LoS, rather than due to turbulence. Therefore, the true velocity of the out-flowing gas is given by the extreme blue wing of the Gaussian profile. This corresponds to the gas travelling directly towards us along the LoS. It follows that, in this scenario, the outflow kinetic power is given by dropping the turbulence term in equation (5), and taking the $v_{05}$ shift of the Gaussian fit to the outflow component relative to the galaxy rest-frame to represent $v_{out}$, rather than the mean shift as before. Note we used $v_{05}$ rather than $v_{00}$ to avoid confusion with the continuum. The estimates derived using this method are presented in Table \ref{tab:exttotals}.

\begin{table*}
	\centering
	\caption{Outflow properties for the maximal velocity case using the $v_{05}$ velocities. Column (2):the trans-auroral fluxes used to calculate the density and reddening values employed in the calculations: T = total, B = broad/intermediate. In the case of F05189--2524, the whole profile is blueshifted with respect to the rest-frame and hence the total flux and broad flux are assumed to be equivalent. Column (3): the $v_{05}$ velocity. Column (4): the calculated mass outflow rate. Column (5): the calculated kinetic power. Column (6): the fraction of AGN bolometric luminosity contained within the outflow ($\dot{E}/L_{bol}$)
			}
		\label{tab:exttotals}
		\begin{tabular}{lccccr}
			\hline
			\\
			Object name & Comp. & $v_{05}$ & $\dot{M}$ & $\dot{E}$ & $\dot{F}$ \\
			IRAS & T/B & km s$^{-1}$ & M$_{\sun}$ yr$^{-1}$ & erg s$^{-1}$ & \% \\
			\\
			(1) & (2) & (3) & (4) & (5) & (6)\\
			\hline\\
			F01004 -- 2237 & T & -1932$\pm$82 & 1.0$^{+2.2}_{-0.7}$ & (1.2$^{+2.9}_{-0.9}$)$\times$10$^{42}$ & 0.14$^{+0.35}_{-0.10}$  \\
			\\
			F05189 -- 2524 & T,B & -1494$\pm$58 &  0.13$^{+0.21}_{-0.09}$ & (9.3$^{+17.0}_{-6.5}$)$\times$10$^{40}$ & (2.6$^{+4.9}_{-1.8}$)$\times$10$^{-2}$ \\
			\\
			F14394 + 5332E & T & -2362$\pm$112 &  13.2$^{+37.4}_{-9.7}$ & (2.3$^{+7.5}_{-1.8}$)$\times$10$^{43}$ & 2.6$^{+8.6}_{-2.0}$
			\\
			& B & `` & 3.4$^{+13.8}_{-3.0}$ & (6.0$^{+27.4}_{-5.4}$)$\times$10$^{42}$ & 0.66$^{+3.12}_{-0.60}$\\
			\\
			F17044 + 6720 & T & -1992$\pm$136 &  1.2$^{+3.8}_{-0.8}$ & (1.5$^{+5.7}_{-1.0}$)$\times$10$^{42}$ & 0.42$^{+1.69}_{-0.30}$    \\
			\\
			F17179 + 5444 & T & -1539$\pm$105 &  20.1$^{+72.9}_{-14.9}$ & (1.5$^{+6.5}_{-1.2}$)$\times$10$^{43}$ & 2.5$^{+10.7}_{-1.9}$ \\
			\\
			F23060 + 0505 & T & -1330$\pm$104 &  2.2$^{+8.1}_{-1.7}$ & (1.2$^{+5.5}_{-1.0}$)$\times$10$^{42}$ & (2.7$^{+13.2}_{-2.2}$)$\times$10$^{-2}$\\
			\\
			F23233 + 2817 & T & -1189$\pm$49 &  > 0.11 & > 4.7$\times$10$^{40}$ & > 9.5$\times$10$^{-2}$ \\
			\\
			F23389 + 0303N$^{a}$ & T & -2126$\pm$67 &  > 3.3 & > 4.4$\times$10$^{42}$ & > 0.18 \\
			& B & ``  &   > 2.3 & > 3.2 $\times$ 10$^{42}$ & > 0.13 \\
			\\
			F23389 + 0303N$^{b}$ & T & -2126$\pm$67 & 14.9$^{+16.0}_{-7.3}$ & (2.1$^{+2.6}_{-1.1}$)$\times$10$^{43}$ & 0.91$^{+1.13}_{-0.48}$ \\
			& B & `` & 10.4$^{+11.1}_{-5.1}$ & (1.5$^{+1.8}_{-0.8}$) $\times$ 10$^{43}$ & 0.63$^{+0.79}_{-0.33}$\\
		\\
		\hline
			\multicolumn{6}{l}{
							\begin{minipage}{0.6\textwidth}~\\
									$^{a}$ These estimates were made using the upper limit on the outflow radius from the WHT/ISIS spectrum. \\
									$^{b}$ These estimates were made using the radius estimate of the radio lobes (R = 415 pc) as a proxy for the outflow radius, taken from \citealt{Nagar2003}. This assumes that the outflows are jet-driven. \\
							\end{minipage}
						}\\
	\end{tabular}
\end{table*}

%
 
Under the maximal velocity assumption, and considering the estimates for the objects in which $R_{[OIII]}$ is resolved, we find mass outflow rates in the range 0.1 < $\dot{M}$ < 20 M$_{\sun}$yr$^{-1}$ and kinetic powers in the range $9.3 \times 10^{40} < \dot{E} < 2.3 \times 10^{43}$ erg s$^{-1}$. These ranges are again consistent with those found in \cite{Rose2018} and the kinetic powers are a factor of $\sim5\times$ higher than those obtained using the first method. \par   

Comparing the outflow kinetic powers with the bolometric luminosities of the AGN, we find values in the range 0.02 < $\dot{F}$ < 3\%. While the fraction of the AGN power contained in the outflow is naturally larger when using higher velocities, even in this maximal velocity case, the kinetic power and mass outflow rates still show little overlap with the results of \cite{Harrison2012}, \cite{Liu2013} and \cite{Mcelroy2015} and fall short of the threshold required ($\dot{F}$ = 5 - 10\%) by the \citealt{DiMatteo2005, Springel2005} models. \par

Initially, these results may suggest that the AGN-induced outflows alone do not contain the necessary energy to significantly impact their host galaxies. However, we note that some caution is required when comparing our $\dot{F}$ results with the predictions of theoretical models for AGN-induced outflows. As discussed in \citealt{Harrison2018}, it may not be appropriate to directly compare the $\dot{E}/L_{bol}$ values required by the models with the $\dot{F}$ values we estimate for the warm outflows in
ULIRGs: first, the $\dot{E}$ used in some of the models \citep[e.g.][]{DiMatteo2005,Springel2005} represents the thermal energy deposited in the near-nuclear gas by the AGN, rather than the kinetic power in the outflow; and second, not all of the kinetic power associated with an inner (e.g. accretion disk) wind generated by an AGN may be transmitted to the cooler, larger-scale outflow that we observe in the optical emission lines, due to radiative cooling, and work against gravity and external pressure as the outflow expands. Indeed, some recent theoretical studies suggest that as
little as 10\% of the nuclear wind power (or $\sim$0.5\% $L_{bol}$) is transmitted to the large-scale outflow \citep{Richings2017}. This is consistent with the direct observational comparisons that have been made between the kinetic powers of the inner high-ionisation wind and outer molecular outflow in the case of the ULIRG F11119+3257 \citep{Tombesi2015,Veilleux2017}. Such a low coupling efficiency is also consistent with the multi-stage outflow model of \citet{Hopkins2010}. Therefore, our estimates of the coupling efficiency for the warm ionised gas (0.02 < $\dot{F}$ < 3 \%) are in line with some of the most recent theoretical predictions.

\subsection{Comparison with neutral and molecular outflows}
\label{sec:otherphases}

Estimates for the mass outflow rates and kinetic powers for the neutral and molecular outflow phases for five of the ULIRGs in the full QUADROS sample are available in the literature. These values are shown in Table \ref{tab:gas}. \par

\cite{Rupke2005c} estimated the properties of the neutral outflows in F05189--2524, F13451--1232 and F23389+0303N based on the NaID absorption line. An additional estimate for the neutral mass outflow rate of F13451--1232, determined using the HI 21cm line, is provided by \cite{morganti2013}. The neutral and ionised mass outflow rates are comparable in the case of F13451--1232, however the estimated  neutral mass outflow rates are significantly higher for the other two objects. The kinetic power in the neutral outflow, on the other hand, is a factor of 2 greater than the ionised outflow in the case of F05189--2524, significantly lower in the case of F13451--1232, and comparable in the case of F23389+0303N. \par

Estimates for the molecular mass outflow rates in F05189--2524, F14378--3651 and F23060+0505 are provided by \cite{Gonzalez2017} and \cite{Cicone2014}. In all cases the molecular mass outflow rates are significantly greater than those of the ionised outflows. \cite{Gonzalez2017} also provides estimates for the kinetic powers of the molecular outflow for F05189--2524 and F14378--3651. In the former case, the molecular kinetic power is significantly larger than that of the ionised outflow, and comparable with the kinetic power of the neutral outflow. In the latter case, the kinetic power measured for the molecular outflow is two orders of magnitude higher than the lower limit derived for the ionised outflow. \par 

The larger mass outflow rates and kinetic powers for the cooler gas components are consistent with the idea that the gas has been accelerated in fast shocks, with  the neutral and molecular gas then accumulating as the warm gas cools behind the shock fronts \citep{Tadhunter2014, Zubovas2014, Morganti2015, Richings2017}.

\begin{table*}
	\centering
	\caption{A comparison between the ionised, neutral and molecular outflow phases, where estimates are available for the QUADROS sample. The sub-scripts are as follows: I = ionised; N = neutral; M = molecular. The measurements for the ionised outflows are taken from this paper and \citealt{Rose2018}. The references for the neutral and molecular outflow measurements are denoted by the super-script letters.  
			}
	\label{tab:gas}
	\begin{tabular}{lcccccr}
		\hline
		\\
		Object name & $\dot{M}_{I}$ & $\dot{M}_{N}$ & $\dot{M}_{M}$ & log($\dot{E}_{I}$) & log($\dot{E}_{N}$) & log($\dot{E}_{M}$) \\
		IRAS & M$_{\sun}yr^{-1}$ & M$_{\sun}yr^{-1}$  & M$_{\sun}yr^{-1}$ & erg s$^{-1}$ & erg s$^{-1}$ & erg s$^{-1}$ \\
		\\
		(1) & (2) & (3) & (4) & (5) & (6) & (7)\\
		\hline\\
		F05189 -- 2524 & 0.13$^{+0.21}_{-0.09}$ & 117$^{a}$ & 270$^{+22}_{-130}$$^{b}$ & 40.97$^{+0.45}_{-0.52}$ & 43.08$^{a}$ & 43.2$^{+0.1}_{-0.2}$ $^{b}$ \\
		\\
		F13451 -- 1232W & 10.5$^{+2.8}_{-2.3}$ & 7.6$^{a}$ & -- & 43.34$\pm$0.11 & 41.67$^{a}$ & -- \\
		&& 16--29$^{c}$ &&&&\\
		\\
		F14378 -- 3651 & > 0.082 & -- & 180$^{+180}_{-33}$ $^{b}$ &  > 40.76 & -- & 43.1$\pm$0.2 $^{b}$  \\
		\\
		F23060 + 0505 & 2.2$^{+8.1}_{-1.7}$ & -- & 1500$^{d}$ &  42.08$^{+0.74}_{-0.78}$ & -- & -- \\
		\\
		F23389 + 0303N & 10.4$^{+11.1}_{-5.1}$ & > 49$^{a}$ & --  & 43.2$\pm$0.3 & > 42.4$^{a}$ & -- \\
		\hline
		\multicolumn{7}{l}{
									\begin{minipage}{0.6\textwidth}~\\
											$^{a}$\citealt{Rupke2005c} \\
											$^{b}$\citealt{Gonzalez2017}\\
											$^{c}$\citealt{morganti2013}\\
											$^{d}$\citealt{Cicone2014}\\
									\end{minipage}
								}\\
	\end{tabular}
\end{table*} 


\section{Links between the warm outflow properties and AGN properties}
\label{sec:links}

Some observational studies have found evidence for  correlations between the properties of the outflows and those of the AGN. For example, \cite{Fiore2017} find evidence for strong correlations between the mass outflow rates and the AGN bolometric luminosities of both molecular and ionised outflows. Similarly, \cite{Cicone2014} show evidence for a correlation between $L_{bol}$ and the kinetic powers of molecular outflows. Given the assumption that our ionised outflows are AGN-driven (see \S \ref{sec:bol}) we have also examined this within our data. \par

In Figure \ref{fig:E_Lbol} we plot the outflow kinetic power ($\dot{E}$) for the maximal $v_{05}$ velocity case against the bolometric luminosity of the AGN for the full QUADROS sample. The red circles represent the results from this paper and the blue stars represent the results of \cite{Rose2018}. We have used the best available estimate of the kinetic power for each object\footnote{For this and the subsequent plots, we have used the estimates of the outflow properties derived from the broad component fluxes of the trans-auroral emission lines, where available. For the objects where the individual kinematic components were unresolved, we have used the estimates derived from the total emission line fluxes.} and have taken the most appropriate bolometric luminosity correction as discussed in \S \ref{sec:bol} in this paper, and \S 3.3 in \cite{Rose2018}. \par 

Over-plotted are three lines corresponding to the fraction of the AGN luminosity contained in the kinetic power of the outflow ($\dot{F}$): 100\% (solid), 5\% (dashed) and 1\% (dotted). Although the  majority of the sample fall well below the 1\% line, a significant number ($\sim$35\%) of the QUADROS ULIRGs fall close to, or above it. For reference, we also over-plot the results of \cite{Fiore2017} for 51 AGN, for which they claim a significant correlation between L$_{bol}$ and $\dot{E}$. Considering only the QUADROS results, for which we have attempted to constrain the outflow parameters in a precise and consistent manner, we find no significant correlation between the kinetic powers of the outflows and the AGN bolometric luminosities - a  p-value of 0.47 means we cannot reject the null hypothesis that the two sets of data are uncorrelated.\footnote{When calculating the Spearman rank-order correlation statistics, we have included the upper/lower limits as if they were measured values.} However, this apparent difference from the results of \cite{Fiore2017} is perhaps not surprising, given the small range of AGN luminosity covered by our sample and the relatively high degree of scatter. \par 

We have also considered potential correlations between the AGN bolometric luminosities and other outflow properties such as the outflow velocities, radii and mass outflow rates, as plotted in Figure \ref{fig:trends}. In all three plots, the red circles, blue stars and grey crosses represent the results from this paper, \cite{Rose2018} and \cite{Fiore2017} respectively. We find no statistically significant correlation between $L_{bol}$ and outflow velocity within the QUADROS sample (p = 0.49). Furthermore, we find little evidence for a significant correlation between $L_{bol}$ and either $\dot{M}$ (p = 0.11) or radius (p = 0.76). Again, this lack of correlation could perhaps be explained by the high scatter and narrow luminosity range. \par

Interestingly, the QUADROS results fall within the scatter of the lower end of the $\dot{E}$ and $\dot{M}$ correlations of \cite{Fiore2017} shown in Figures \ref{fig:E_Lbol} and \ref{fig:trends}c. However, this apparent agreement may be misleading: the lower outflow velocities\footnote{Note that, instead of $v_{05}$, \cite{Fiore2017} used the velocity of the peak of the broad component minus 2$\sigma$ as the outflow velocity.} (see Figure \ref{fig:trends}a) and larger outflow radii (see Figure \ref{fig:trends}b) found by \cite{Fiore2017} may compensate for the lower gas densities (and hence larger total gas masses) assumed in their study. \par    

We note that a lack of correlation between the AGN luminosities and outflow properties is perhaps to be expected. \cite{Zubovas2018} argues that because AGN duty cycles are shorter than the dynamical times of the outflows, any observed outflows greater than around 0.1 kpc in extent are unlikely to have been originally driven by the current phase of AGN activity. Therefore, there is no reason to expect that the currently observed AGN luminosities should correlate with the observed outflow properties, and a high degree of intrinsic scatter is to be expected. Our results appear to be consistent with this conclusion.  



\begin{figure}
	\includegraphics[width=\columnwidth]{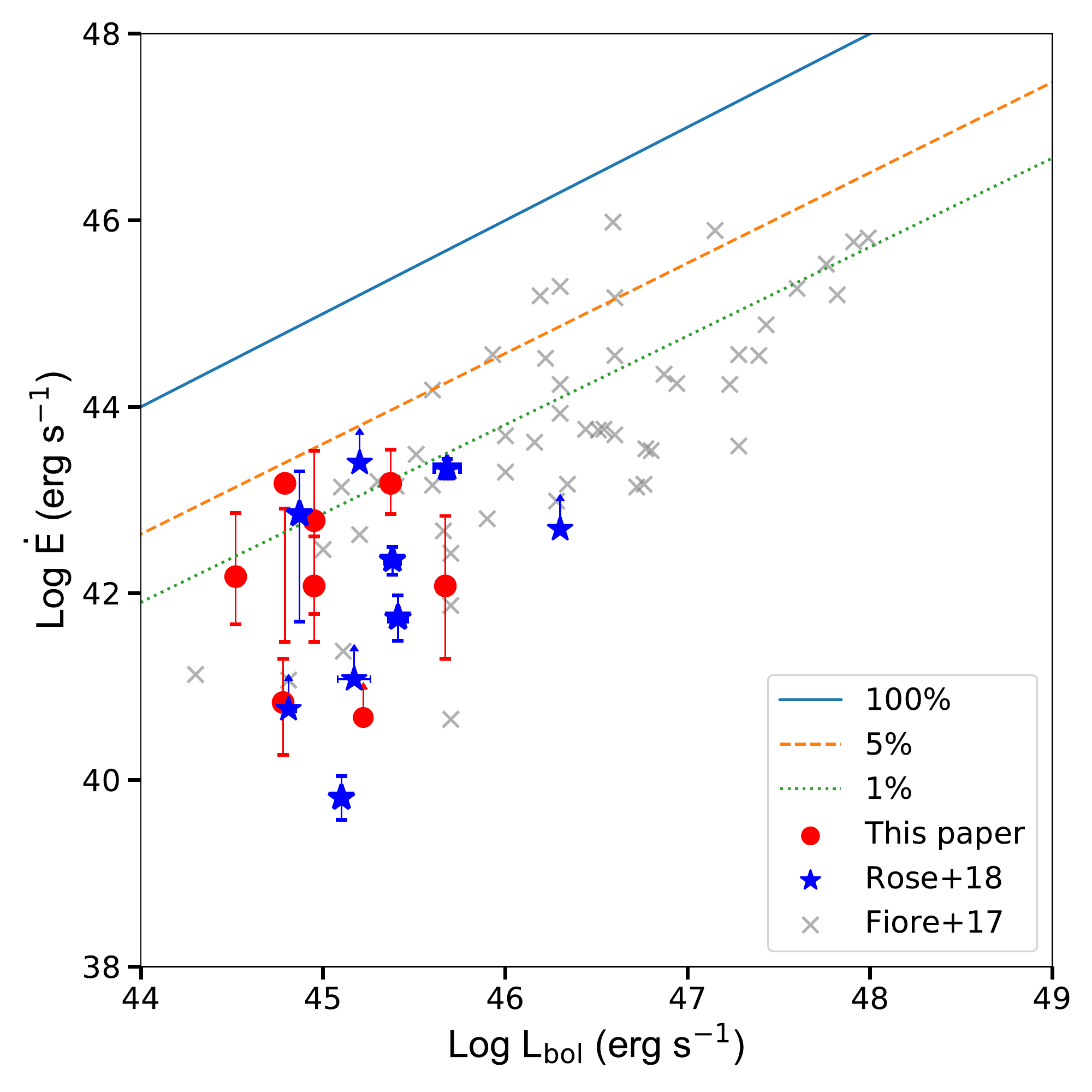}
    \caption{Outflow kinetic power vs. AGN bolometric luminosity for the maximal $v_{05}$ velocity case. The bolometric luminosities were determined using the total emission line flux of [OIII]$\lambda$5007. The kinetic powers of the outflow were determined using the broad, blueshifted components of the H$\beta$ emission line and the best available density estimates (i.e. using the broad trans-auroral fluxes where possible). The red circles and blue stars represent the results from this paper and \citealt{Rose2018} respectively.The over-plotted lines represent $\dot{F}$ = 100\% (solid), 5\% (dashed) and 1\% (dotted). Also over-plotted for comparison are the results of \citealt{Fiore2017} for 51 AGN (grey crosses). Note that caution must be taken when directly comparing the QUADROS and \citealt{Fiore2017} results due to the different methodologies employed for calculation of the outflow parameters. 
    } 
	\label{fig:E_Lbol}
\end{figure}

\begin{figure}
\centering

\subfloat{\includegraphics[width = 2.85in]{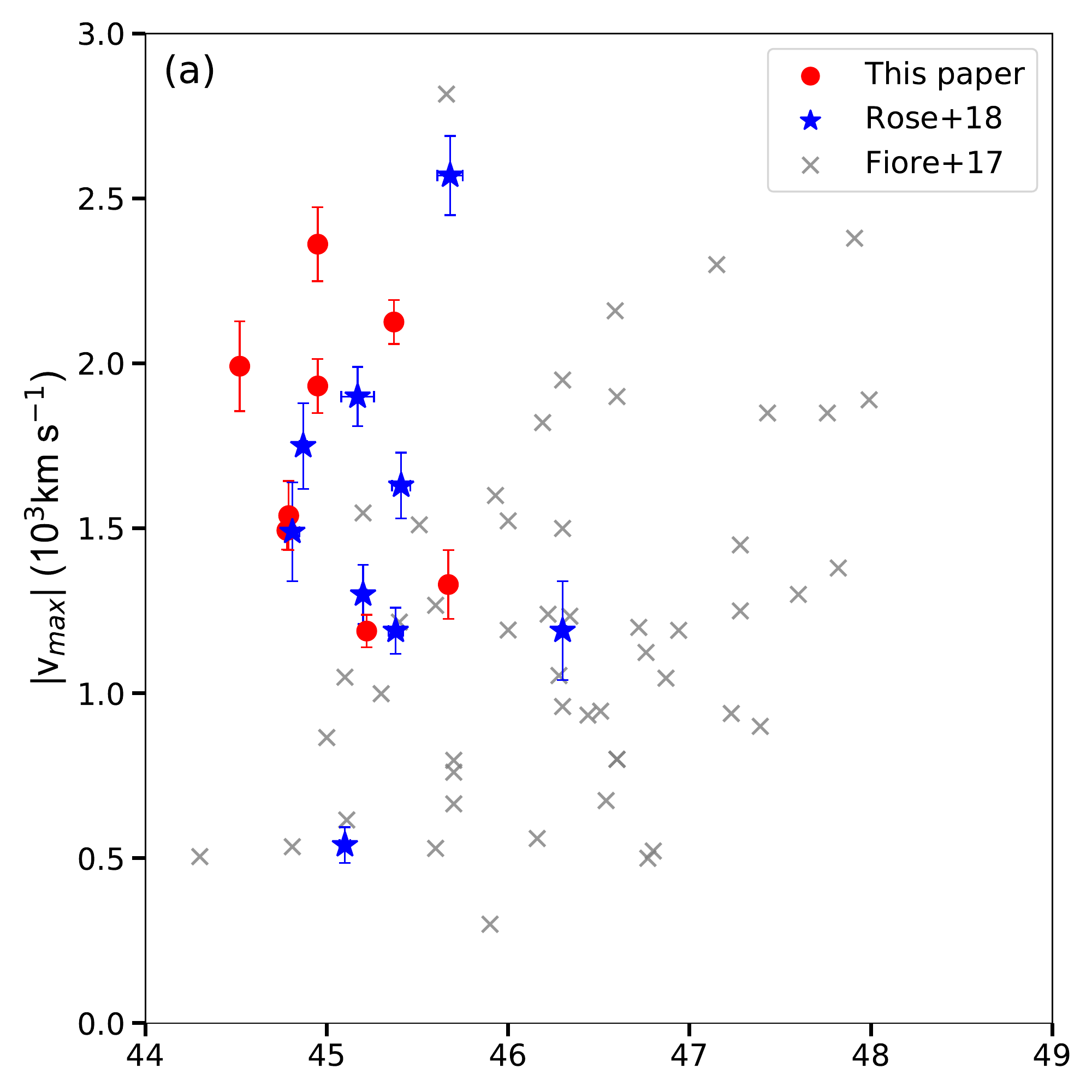}}\\
\vspace{-5.5mm}
\subfloat{\includegraphics[width = 2.85in]{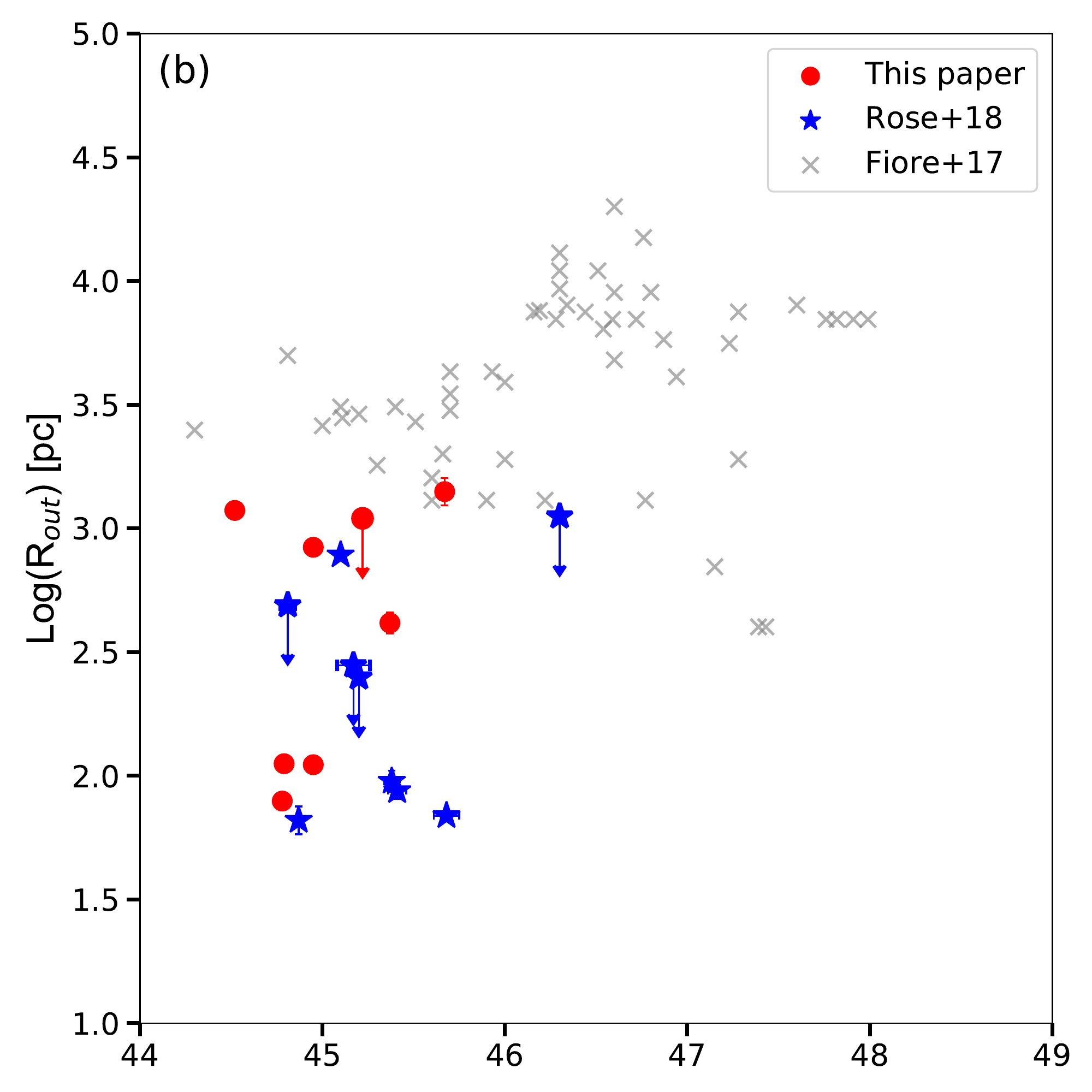}}\\
\vspace{-5.5mm}
\subfloat{\includegraphics[width = 2.85in]{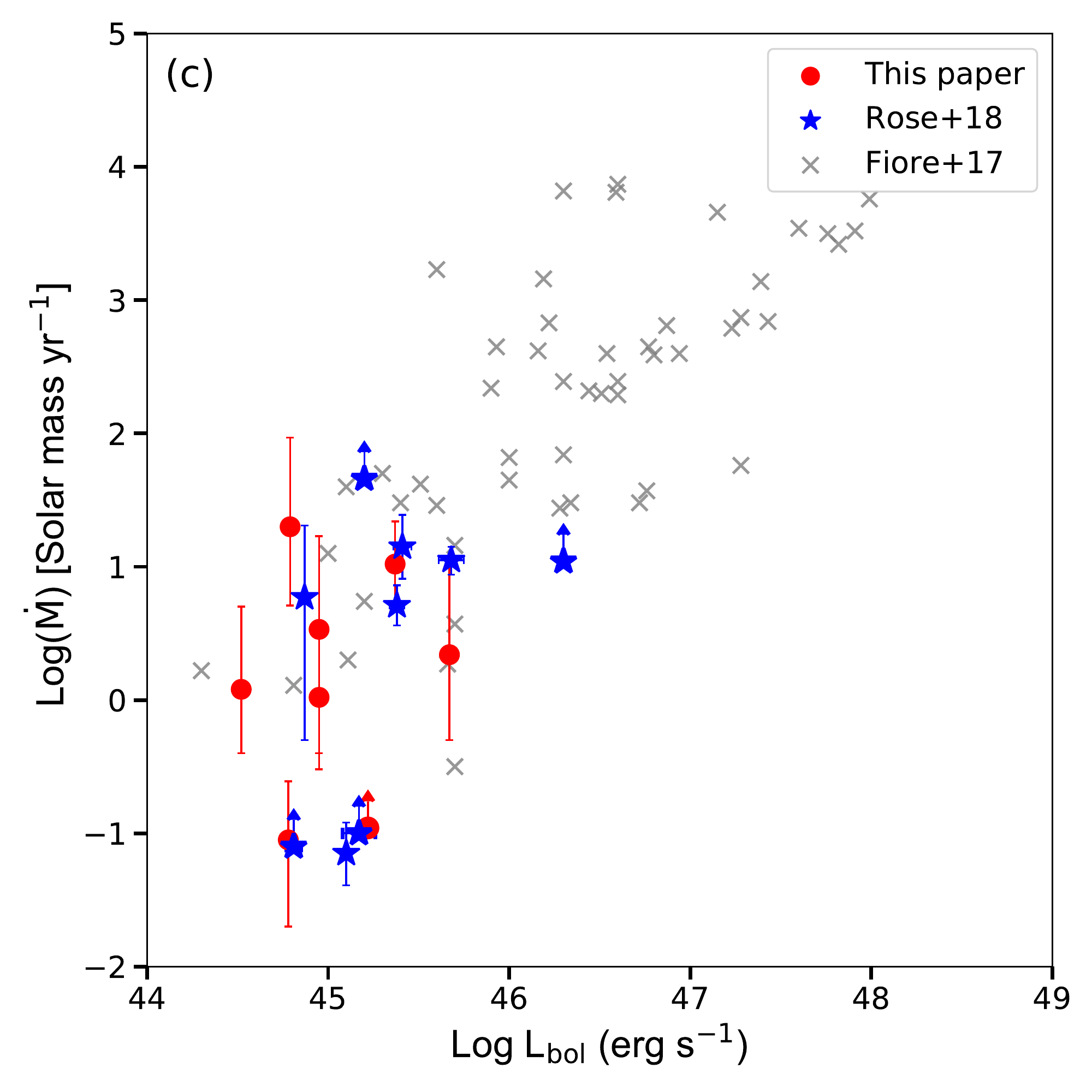}}
\caption{Plots of the outflow properties against log($L_{bol}$) for the full QUADROS sample: (a) $v_{05}$; (b) log($R_{out}$); (c) log($\dot{M}$). The points are as in Figure 9.}
\label{fig:trends} 

\end{figure}





\section{Origin of the trans-auroral emission lines}
\label{sec:origin}

An argument against the use of the trans-auroral lines to measure the densities of the outflows is that the higher critical density species ([SII]$\lambda\lambda$4068,4076 and [OII]$\lambda\lambda$7319,7331) may originate from dense clouds within a lower density outflow, whereas the lower critical density species could instead originate from the low density gas \citep{Sun2017}. If this lower density component contained much of the mass of the outflow, but contributed little to the flux, then this could lead to the trans-auroral diagnostics severely underestimating the mass outflow rates and kinetic powers.\par 
 
However, as we argue in \cite{Rose2018}, any high density clumps would still be expected to radiate strongly in H$\beta$ - used to estimate the gas mass - and the electron densities we measure from the trans-auroral line ratios remain below the critical density, $n_{crit}$, of the [OIII]$\lambda$5007 ($n_{crit} = 7 \times 10^{5}$ cm$^{-3}$), which we use to determine the outflow kinematics and radii. Therefore, depending on the ionisation level, we might also expect the high density clumps to radiate significant [OIII]$\lambda$5007 emission. \par 

To investigate this, we have calculated the theoretical ratios of the high-critical density trans-auroral [OII] and [SII] blends to the [OIII]$\lambda$5007 and H$\beta$ lines from a single, radiation bounded, solar abundance cloud, and compared these ratios to those we observe. For these models we used version C17.00 of Cloudy, last described by \cite{Ferland2013}.

\begin{figure}
	\includegraphics[width=\columnwidth]{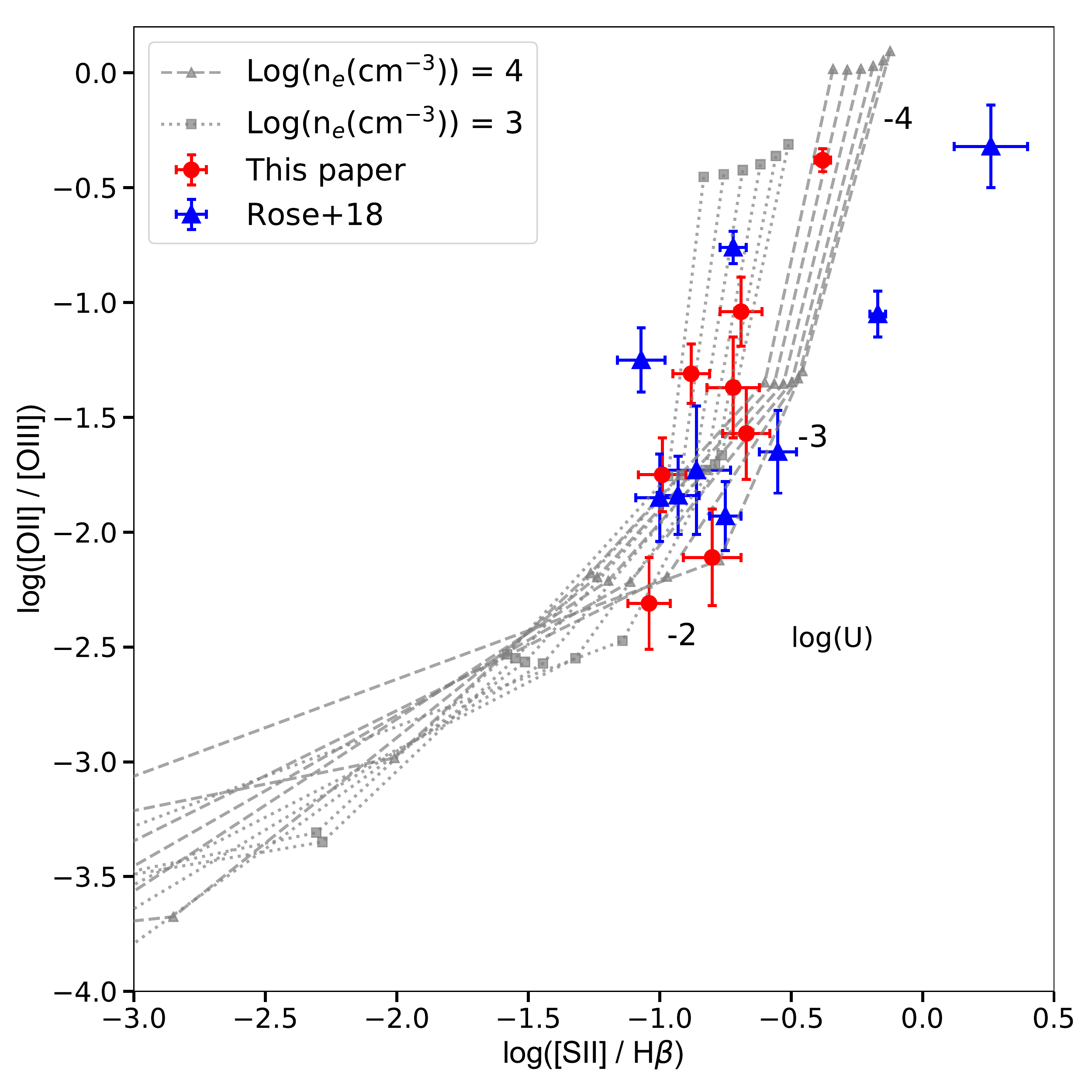}
    \caption{Theoretical photo-ionisation grids modelled with Cloudy for the total emission line flux ratios of log([OII]$\lambda\lambda$7319,7331/[OIII]$\lambda$5007) against log([SII]$\lambda\lambda$4068,4076)/H$\beta$). The dotted lines correspond to a density of 10$^{3}$ cm$^{-3}$. The dashed lines correspond to a density of 10$^{4}$ cm$^{-3}$. The separate lines within these bands correspond to different values of the ionising power-law spectral index in the range -2 < $\alpha$ <-1, and the ionisation parameter, log(U), is indicated by the numbers. Over-plotted are the ratios we measure from this paper (red circles) and \citealt{Rose2018} (blue triangles), respectively. 
    } 
	\label{fig:cloudy}
\end{figure}

Figure \ref{fig:cloudy} plots log([OII]$\lambda\lambda$7319,7331/[OIII]$\lambda$5007) against log([SII]$\lambda\lambda$4068,4076)/H$\beta$). On this plot we present the results from this paper (red circles) along with those of \cite{Rose2018} (blue triangles). Over-plotted as dotted and dashed lines are the Cloudy theoretical grids for densities of 10$^{3}$ cm$^{-3}$ and 10$^{4}$ cm$^{-3}$ respectively. The grids vary in ionisation parameter, log(U), and ionising power-law spectral index, $\alpha$, over the respective ranges: -4 < log(U) < -1 and -2 < $\alpha$ < -1.

Our observed ratios show good consistency with those predicted by the AGN photo-ionisation models. While the displacement between the two sets of models is not particularly large, we find that the positions of the ULIRGs on the respective grids are consistent with the densities we derive from the trans-auroral line ratio analysis. In particular, all ULIRGs considered in this paper (red circles) fall in the range of densities expected from the trans-auroral emission line estimates: 10$^{3}$ < n$_{e}$ < 10$^{4}$ cm$^{-3}$. \par

66\% of the ULIRGs from \cite{Rose2018} (blue triangles) also fall in this range; however in this case the scatter is larger, with three points falling off the grids. The anomalous point to the top right of the grid corresponds to F15462--0450. This object is an un-reddened type I AGN, and the wing of the broad H$\delta$ emission from this AGN overlaps the blue trans-auroral [SII]$\lambda$4073 blend, potentially leading to a higher degree of uncertainty in the flux of this blend and the line ratios derived from it. A further object (F13451+1232W) falls to the right of the log n$_{e}$(cm$^{-3}$) = 4 grid, indicating higher densities. However, this is consistent with the fact that this object shows the highest estimated density of the whole QUADROS sample log n$_{e}$(cm$^{-3}$) = 4.5$\pm$0.2). Finally, one object (F14378--3651) falls to the left of the log n$_{e}$(cm$^{-3}$) = 3 grid, consistent with the relatively low density estimated for this object based on the total trans-auroral emission-line fluxes in \cite{Rose2018}: log n$_{e}$(cm$^{-3}$) $\sim$ 2. \par 

Furthermore, as a consistency check, we have directly estimated the ionisation parameter, U, defined as the ratio of ionizing photon flux to the gas density, for each ULIRG using the following equation:
\begin{equation}
U = \frac{L_{ion}}{c4\pi R_{out}^2n_{e}h\nu_{ion}}
\end{equation}
\\
\noindent where $c$ is the speed of light, $R_{out}$ is the radius of the outflow, $n_{e}$ is the electron density of the gas, $\nu_{ion}$ is the average ionizing photon frequency ($\nu_{ion} = 1.02\times10^{16}$Hz, \citealt{Robinson2000}) and $L_{ion}$ is the proportion AGN luminosity assumed to contribute towards the ionizing continuum. For these estimations, we assumed $L_{ion} = 0.56 L_{bol}$, based on the results of \cite{Netzer2014} for AGN with non-rotating, 10$^{7}$ M$_{\sun}$ black holes. Using equation (6) we find that the ULIRGs cover a range in ionisation parameter of $-1.3 < $log(U)$ < -3.4$. Given our assumptions, this is consistent with the range of log(U) covered by the ULIRG ratios on Figure \ref{fig:cloudy}. Therefore, these results provide strong evidence for the idea that the bulk of the flux of all of our diagnostic emission lines originates from the same dense gas structures. \par 

However, it is not possible to rule out the idea that there exists a lower density, higher filling factor gas component in the outflow that has a higher gas mass but makes a relatively minor contribution to the emission lines fluxes. To see this we consider the case in which a fixed volume $V$ in the outflow contains two gas components with different densities: a higher density component with electron density $n_h$, volume filling factor $f_h$ and electron temperature $T_h$; and a  lower density component with
electron density  $n_l$, volume filling factor $f_l$ and electron temperature $T_l$. For simplicity we assume that the both components comprise of fully ionised pure hydrogen gas. The H$\beta$ luminosities of the two components are then given by:
\begin{equation}
L(H\beta)_l = n_l^2 \alpha_{eff}^{H\beta}(T_l) V f_l h\nu_{H\beta} 
\end{equation}
and
\begin{equation}
L(H\beta)_h = n_h^2 \alpha_{eff}^{H\beta}(T_h) V f_h h\nu_{H\beta} 
\end{equation}
and their total gas masses by
\begin{equation}
M_l = m_p n_l V f_l
\end{equation}
and 
\begin{equation}
M_h = m_p n_h V f_h.
\end{equation}

\noindent Combining these equations, and assuming that $\alpha_{eff}^{H\beta}(T_l) = \alpha_{eff}^{H\beta}(T_h)$,\footnote{Even in the 
extreme case that the low density gas is matter bounded, but the high density gas is radiation bounded, the temperatures 
of the two components, and hence the recombination coefficients (Osterbrock \& Ferland 2006), are unlikely to differ by more
than a factor of $\sim$2.} we find the following expression for the ratio of the volume filling factors:
\begin{equation}
\left( \frac{f_h}{f_l} \right) = \left( \frac{L(H\beta)_l}{L(H\beta)_h} \right)
\left( \frac{M_h}{M_l} \right)^2
\end{equation}
and the ratio of the electron densities is given by:
\begin{equation}
\left( \frac{n_h}{n_l} \right) = \left( \frac{f_l}{f_h} \right) \left( \frac{M_h}{M_l} \right).
\end{equation}

Therefore, it is possible for the low density component to be $10\times$ more massive than the high density component 
($(M_h/M_l) = 0.1$) yet contribute 10\% or less of the H$\beta$ luminosity of the high density component (i.e. $\sfrac{L(H\beta)_l}{L(H\beta)_h} \le 0.1$), provided that the filling factor and density contrasts satisfy $(f_h/f_l) \le 10^{-3}$ and
$(n_h/n_l) \ge 10^2$ respectively. Given that we find typical volume filling factors for the warm outflows in our ULIRG sample in the range 
$10^{-6} < f_h < 10^{-3}$, such a high filling factor contrast is feasible. However, it is not clear why warm gas components with densities, and hence pressures, differing by a factor $\sim$100 should co-exist in the same volume, especially if all the gas has cooled isobarically behind a shock front.



\section{Conclusions}
\label{sec:conclusions}

The results of this study of warm outflows in 8 nearby ULIRGs with WHT/ISIS (QUADROS III) strongly reinforce those obtained for 9 ULIRGs using VLT/Xshooter by \cite{Rose2018} (QUADROS I). After considering the effects of seeing, we find evidence that the ionised outflow regions are compact ($0.08 < R_{out} < 1.5$ kpc, median $R_{out}$ $\sim$ 0.8 kpc). In addition, we find that the outflows can suffer a significant degree of reddening ($0 < E(B-V) < 1$, median E(B-V) $\sim$ 0.5), and have high densities (2.5 < log n$_{e}$(cm$^{-3}$) < 4.5, median log n$_{e}$(cm$^{-3}$) $\sim$ 3.1). \par 

The resultant mass outflow rates (0.1 < $\dot{M}$ < 20 M$_{\sun}$ yr$^{-1}$, median $\dot{M}$ $\sim$ 2  M$_{\sun}$ yr$^{-1}$) and kinetic powers expressed relative to AGN bolometric luminosity (0.02 <  $\dot{F}$ < 3\%, median $\dot{F}$ $\sim$ 0.3\%) are relatively modest. These values are consistent with the theoretical expectations if it is assumed that the inner winds transmit only a modest fraction ($\sim$10\%) of their energy to the large-scale outflows. \par 

We have also used photo-ionisation modelling to show that the bulk of the fluxes of the strong diagnostic emission lines (e.g. 
H$\beta$ and [OIII]$\lambda\lambda$4959,5007) detected in our spectra are plausibly emitted by the same high density clouds that emit the trans-auroral [SII] and [OII] density diagnostic lines. However, we cannot entirely rule out the idea that there exists a lower density, higher filling factor warm outflow component that contributes relatively little to the emission line fluxes, but is substantially more massive. \par

Considering the  QUADROS sample as a whole, we do not find evidence that the properties of the outflows are strongly correlated with the AGN
bolometric luminosities. This lack of correlation may be due in part to the relatively narrow range in $L_{bol}$ covered by our sample,
coupled with the uncertainties on the $L_{bol}$ values themselves. However, given that we have been careful to accurately measure
the key properties of the outflows (radii, densities, reddenings), it is more likely that there is a high degree of intrinsic scatter ($\sim$2 -- 3 orders of magnitude) in $\dot{M}$, $\dot{E}$ and $\dot{F}$ for a given $L_{bol}$. This scatter could be due to different efficiencies in the coupling between the inner winds and the larger-scale ISM, perhaps related to different circum-nuclear environments. Alternatively, it might reflect large-scale variability in the radiative outputs.\par 

Indeed, in the case of F01004-2237 we have direct evidence for just such variability: this object has a high ionisation narrow-line spectrum characteristic of AGN, yet lacks a type I AGN BLR component, despite the fact that the recent detection of a TDE event in its nucleus demonstrates that our line of sight to its central supermassive black hole is relatively unobscured \citep{Tadhunter2017a}. \par 

Ultimately, the results of the QUADROS project emphasise the importance of determining accurate radii, electron densities and reddening values for AGN-driven outflows. The dearth of accurate measurements of these parameters is likely to have been at least partially responsible for the lack of consistency between the results and conclusions of the various studies of AGN outflows in the past. Therefore, to gain a complete multi-phase understanding of the importance of AGN-driven outflows to galaxy evolution, further high-resolution observations designed specifically to measure radii, reddening and densities of outflow regions are required.

\section*{Acknowledgements}
MR \& CT acknowledge support from STFC. RAWS thanks J. Pierce, H. Stevance and K. Tehrani for helpful discussions. We also thank the anonymous referee for useful comments and suggestions that enabled us to improve this work. Based on observations obtained with ISIS on the William Herschel Telescope, operated on the island of La Palma by the Isaac Newton Group in the Spanish Observatorio del Roque de los Muchachos of the Instituto de Astrofisica de Canarias, and observations taken with STIS on the NASA/ESA Hubble Space Telescope. The authors acknowledge the data analysis facilities provided by the Starlink Project, which was run by CCLRC on behalf of PPARC. This research has made use of the NASA/IPAC Extragalactic Database (NED) which is operated by the Jet Propulsion Laboratory, California Institute of Technology, under contract with the National Aeronautics and Space Administration.



\bibliographystyle{mnras}
\bibliography{WHTbiblio} 




\appendix

\section{General description of targets}

In this section we describe the general properties of the targets in our ULIRG sample.
\\
\\
\textbf{F01004--2237.} The single nucleus, compact  morphology of this system suggests it is being observed at, or just after, the peak of the merger \citep{Surace1998}. Historically it has been classified as both an HII galaxy and Seyfert 2 galaxy based on optical diagnostic line ratios -- Although the narrow components appear consistent with stellar photo-ionisation, the broad components are more consistent with ionisation by an AGN \citep{Rodriguez2013}. \par 

Interestingly, our new  optical spectrum of this object shows that it has changed dramatically over the last decade -- completely unexpectedly in the context of this study. Compared to previous spectra taken in 2005, the 2015 spectrum shows the appearance of unusually strong broad helium emission lines \citep{Tadhunter2017a}. A follow-up examination of an archival optical light curve data revealed a significant flare in 2010, providing evidence for the first tidal disruption event (TDE) ever observed in a starburst galaxy.\par 

The [OIII] profile was fit with three kinematic components: one narrow ($\Delta$v = 63$\pm$25 km s$^{-1}$; FWHM unresolved), one intermediate ($\Delta$v = -361$\pm$35 km s$^{-1}$; FWHM = 699$\pm$32 km s$^{-1}$) and one broad ($\Delta$v = -1045$\pm$60 km s$^{-1}$; FWHM = 1586$\pm$38 km s$^{-1}$). Note that, unlike the rest of the objects in this paper, whose host rest-frame redshifts were determined using stellar absorption features, the galaxy rest-frame of F01004--2237 was determined using emission lines from extended quiescent gas either side of the nucleus, because any stellar absorption features that may have been present in the nuclear spectrum were washed out by the TDE continuum. \par

The red [OII]$\lambda$7320 trans-auroral blend is detected in the ground-based WHT/ISIS spectrum, however the blue [SII]$\lambda$4073 blend is washed out by emission lines associated with the TDE. Fortunately, the latter blend is detectable in the HST/STIS spectrum, from which an estimate of the total emission-line flux has been obtained. \par

Based on the STIS spectroscopy, F01004--2237 shows evidence for a compact ($R_{[OIII]}$ = 0.111$\pm$0.004 kpc) ionised nuclear outflow, consistent with the upper limit of 1.1 kpc derived from the ground-based spectrum. Using log n$_{e}$(cm$^{-3}$) = 4.00$^{+0.25}_{-0.30}$ and E(B-V) = 0.04$\pm$0.16, we calculate a mass outflow rate for the maximal velocity case of 1.0$^{+2.2}_{-0.8}$ M$_{\sun}$ yr$^{-1}$. We find the kinetic power of this outflow corresponds to be 0.14$^{+0.35}_{-0.10}$ \% $L_{bol}$.
\\
\\
\textbf{F05189--2524.} The closest ULIRG considered in the paper ($z = 0.0428$), F05189--2524 was not part of the original QUADROS sample, although it fulfils the redshift and spectral selection criteria of that sample; it was observed to fill a gap in the schedule during the observing run. F05189--2524 is a late-stage merger surrounded by tidal debris \citep{Veilleux2002,Veilleux2006}. In HST images it shows a complex nuclear structure, dominated by two bright knots separated by $\sim$0.25 arsec \citep{Surace1998}. It is not yet clear whether these two knots represent a true double nucleus, or a single nucleus bisected by a dust lane; however, the HST/STIS long-slit spectrum --- which cuts through both knots --- shows that the AGN is
associated with the more northerly of the two knots. The extraction aperture for the ground-based optical spectrum, presented in Figure \ref{fig:spectra}, includes both of the bright near-nuclear knots and shows strong Balmer absorption features, as well as emission lines 
with a wide range of ionisation, including strong [NeV] and [FeVII] lines. An in-depth study of the high-ionisation, coronal line outflow detected in the HST/STIS spectrum of this object will be presented in Rose et. al (in prep).\par  

The emission line profile of [OIII] is shown in Figure \ref{fig:oiii}. Notably, there is no rest-frame [OIII] component detected within our WHT/ISIS slit. The entire profile is significantly blueshifted with respect to the host galaxy rest-frame, consistent with \cite{Rupke2005c}, with the best fitting model requiring two Gaussian components: an intermediate component with $\Delta$v = -505$\pm$26 km s$^{-1}$ and FWHM = 582$\pm$37 km s$^{-1}$, and a broad component with $\Delta$v = -1072$\pm$44 km s$^{-1}$ and FWHM = 1706$\pm$24 km s$^{-1}$. In contrast, an additional intermediate component is detected in the Balmer emission lines with $\Delta$v = -28$\pm$55 km s$^{-1}$ and FWHM = 555$\pm$76 km s$^{-1}$, consistent with the host rest-frame.  \par

The high-critical-density trans-auroral [SII] and [OII] blends are barely detected in the WHT/ISIS spectrum, but are detected in the HST/STIS spectrum. The kinematic models (relative shifts and widths of the components) fitted to the emission lines in the WHT/ISIS spectrum also fitted well to the HST/STIS profiles. Therefore we were able to use the trans-auroral ratios from this spectrum to estimate the density and reddening of the warm outflow. \par 

Based on the STIS long-slit spectroscopy, F05189--2524 shows evidence for a compact ($R_{[OIII]}$ = 0.079$\pm$0.002 kpc) ionised nuclear outflow, which is consistent with the 0.3 kpc upper limit derived from the ground-based spectrum. F05189--2524 has the highest estimated density, log n$_{e}$(cm$^{-3}$) = 4.25$\pm$0.25, of all the ULIRGs considered in this paper. Combined with E(B-V) = 0.02$\pm$0.15, this corresponds to a modest mass outflow rate and kinematic power (maximal velocity case: $\dot{M}$ = 0.13$^{+0.21}_{-0.09}$ M$_{\sun}$ yr$^{-1}$; log($\dot{E}$) = 40.3$\pm$0.4 erg s$^{-1}$ = (2.6$^{+4.9}_{-1.8})\times$10$^{-2}$\% $L_{bol}$).

Previous evidence for warm ionised outflows via broad, blueshifted kinematics was reported by \cite{Arribas2014} based on fits to the [NII]+H$\alpha$ blend. This object also shows blueshifted kinematics in OH 119$\mu$m observations \citep{Veilleux2013}, and outflow signatures are detected in the neutral gas based on NaID absorption-line observations \citep{Rupke2005c}. \cite{Rupke2005c} estimate a neutral gas mass outflow rate $\dot{M} = 117$\,M$_{\sun}$ yr$^{-1}$ and kinetic power log($\dot{E}$) = 43.08\, erg s$^{-1}$ --- orders of magnitude higher than the
mass outflow rate and kinetic power we derive for the warm ionised gas in this study. 
\\
\\
\textbf{F14394+5332E.} This object consists of two interacting galaxies separated by $\sim$56 kpc and connected by a bridge of faint, diffuse emission \citep[][submitted]{Kim2002,Tadhunter2018}. The spectroscopic slit for our WHT/ISIS observations was placed on the more westerly of the two nuclei in the eastern galaxy \citep{Kim2002} that contains the optical AGN nucleus, and for which a powerful ionised nuclear outflow has been previously reported in \cite{Rodriguez2013}, based on the [OIII] emission lines \citep[see also][]{Lipari2003}. Evidence for a molecular outflow is also reported in \cite{Veilleux2013}, based on blueshifted OH 119$\mu$m profiles. Our total trans-auroral flux ratio and Balmer decrement measurements both indicate a moderate degree of intrinsic reddening for the warm outflow in this object (E(B-V) $\sim$ 0.6). \par 

The [OIII] emission line profile is spectacular. The best fit requires 3 narrow components (N1: $\Delta$v = 17$\pm$62 km s$^{-1}$; FWHM = 408$\pm$11 km s$^{-1}$; N2: $\Delta$v = -701$\pm$63 km s$^{-1}$; FWHM = 288$\pm$21 km s$^{-1}$; N3: $\Delta$v = -1457$\pm$66 km s$^{-1}$; FWHM = 242$\pm$34 km s$^{-1}$) and one broad component ($\Delta$v = -1000$\pm$67 km s$^{-1}$; FWHM = 1871$\pm$19 km s$^{-1}$). These velocity shifts are given relative to the host galaxy rest-frame, measured from the higher-order Balmer absorption lines.\par 

Given that the two blueshifted narrow components of [OIII] were not detected in any of the other emission lines, an alternative three-component model was created from the red [SII]$\lambda\lambda$6717,6731 blend (N: $\Delta$v = 17$\pm$63 km s$^{-1}$; FWHM = 420$\pm$11 km s$^{-1}$; I: $\Delta$v = -358$\pm$93 km s$^{-1}$; FWHM = 986$\pm$42 km s$^{-1}$; B: $\Delta$v = -1030$\pm$69 km s$^{-1}$; FWHM = 1927$\pm$20 km s$^{-1}$) 

Based on our ground-based analysis, F14394+5332E shows evidence for a resolved ionised outflow with a radius of 0.75 $\pm$ 0.12 kpc. In comparison, the radius estimated from the flux-weighted mean of the HST/ACS imaging is 0.840 $\pm$ 0.008 kpc \citep[][submitted]{Tadhunter2018}. The consistency between these two independent methods is remarkable, given the different techniques involved. The estimated density based on the total trans-auroral flux ratios is log n$_{e}$(cm$^{-3}$) = 2.90$^{+0.30}_{-0.40}$; however, this increases to log n$_{e}$(cm$^{-3}$)  = 3.55$^{+0.45}_{-0.45}$  when considering only the broad, out-flowing component. The narrow component of the trans-auroral lines give a density of log n$_{e}$(cm$^{-3}$) = 2.70$^{+0.35}_{-0.50}$.  \par

We can also place an upper limit on the density of the narrow kinematic component based on a fit to the traditional [SII]$\lambda\lambda$6717,6731  of log n$_{e}$(cm$^{-3}$) < 2.64, which is consistent within the errors with the trans-auroral estimate. Interestingly, the intermediate component of the kinematic fit to the traditional [SII] emission gives a density of log n$_{e}$(cm$^{-3}$) = 3.05$^{+0.21}_{-0.19}$, illustrating the density gradient between the narrow, rest frame component and broad, out-flowing component of the gas. \par

In the maximal velocity case, for E(B-V) = 0.63$^{+0.22}_{-0.30}$, the trans-auroral broad-component density leads to  a mass outflow rate, $\dot{M}$ = 3.4$^{+13.8}_{-3.0}$ M$_{\sun}$ yr$^{-1}$ and kinetic power, $\dot{E}$/L$_{bol}$ = 0.66$^{+3.12}_{-0.60}$\%.  
\\
\\
\textbf{F17044+6720.} This is a highly disturbed system, with two bright nuclear continuum condensations, a prominent series of dust features and a $\sim$15 kpc tidal tail to the north \citep[][submitted]{Tadhunter2018}. The extraction aperture for our WHT/ISIS observations was centred on the 
brightest nuclear continuum condensation, which also contains the AGN nucleus and warm outflow. \par 

Despite the disturbed morphology, the [OIII] emission line profile is one of the least disturbed of the 8 ULIRGs in the paper, with a strong narrow rest-frame component ($\Delta$v = -1.2$\pm$61 km s$^{-1}$; FWHM = 218$\pm$12 km s$^{-1}$) and a weaker, blueshifted broad component associated with the outflow ($\Delta$v = -503$\pm$85 km s$^{-1}$; FWHM = 1757$\pm$60 km s$^{-1}$). The [OIII] kinematic model provided a good fit to all other emission lines. \par 

Based on our ground-based analysis, F17044+6720 shows the most extended ionised outflow out of the 8 ULIRGs considered in this paper, with an estimated radius of 1.45$\pm$0.18 kpc. Similarly to F14394+5332E, we can compare this with the flux-weighted mean estimate from the HST/ACS [OIII] imaging from \citealt{Tadhunter2018} (submitted), which is given as 1.184$\pm$0.006 kpc. The two estimates are consistent to within 2$\sigma$ of the joint error.\par

The density derived from the total trans-auroral emission line ratios is log n$_{e}$(cm$^{-3}$) = 2.50$^{+0.30}_{-0.50}$  --- the lowest density estimated using this technique for any of the objects in this paper; however, the broad component to the trans-auroral lines is extremely weak
in this case, therefore the total trans-auroral fluxes are likely dominated by the narrow component, for which a lower density would be expected. Indeed, the density derived using the traditional [SII]$\lambda\lambda$6717,6731 emission line ratio for the narrow components is consistent with this, giving an estimated density of log n$_{e}$(cm$^{-3}$) = 2.49$^{+0.15}_{-0.18}$. \par 

The mass outflow rate and kinetic power estimates based on the density obtained from the total trans-auroral flux ratios, with E(B-V) = 0.35$\pm$0.10, remain relatively modest despite the low density. For the maximal velocity case we find $\dot{M}$ = 1.2$^{+3.8}_{-0.8}$ M$_{\sun}$ yr$^{-1}$ and $\dot{E}$/L$_{bol}$ = 0.42$^{+1.69}_{-0.30}$ \%.     
\\
\\
\textbf{F17179+5444.} Although the continuum morphology of this single-nucleus object is complex due to a series of dust features that cross the nuclear regions, its [OIII] emission is dominated by
a compact region associated with the brightest nuclear continuum condensation \citep[][submitted]{Tadhunter2018}; the spectroscopic extraction aperture for our WHT/ISIS observations was centred on this region.
While evidence for warm ionised outflows from [OIII] emission] is reported in \cite{Rodriguez2013}, \cite{Rupke2005c} found no evidence for significant outflows using the NaID absorption lines. \par 

F17179+5444 is one of two objects in our sample that can be considered radio-loud, with $L_{1.4GHz} = 10^{25.23}$ W Hz$^{-1}$. To date, however, there have been no reports of jets associated with this radio emission. Our total trans-auroral flux and Balmer diagnostics indicate moderate-to-high dust extinction in this object, with E(B-V) $\sim$0.7. \par 

The [OIII] emission line profile has been fit with two components: one rest-frame component, labelled intermediate using our FWHM criteria ($\Delta$v = 58$\pm$62 km s$^{-1}$; FWHM = 590$\pm$12 km s$^{-1}$), and one blueshifted broad component ($\Delta$v = 242$\pm$78 km s$^{-1}$; FWHM = 1530$\pm$33 km s$^{-1}$). Note, however, that \cite{Rodriguez2013} fitted this profile with three components, splitting the intermediate component into two narrow components and suggesting that this splitting could be due to unresolved rotation. \par

The [OIII] kinematic model worked well for the H$\beta$, [NII]+H$\alpha$, and [OII] emission lines, however it failed to successfully fit the [SII] blends. For these, an alternative two component model was fitted to the red [SII]$\lambda\lambda$6717,6731 blend, where the relative widths and shifts of the broad components were allowed to vary with respect to the [OIII] model to produce an acceptable fit. However, the broad component to the blue [SII]$\lambda\lambda$4068,4076 blend was not detected and only required the narrow component, consistent with the narrow component of the [OIII] model. \par 

Based on our analysis, F17179+5444 shows evidence for a compact ionised nuclear outflow. The outflow is unresolved in our ground-based observations and we therefore derive an upper limit on the radius of 1.0 kpc. This is consistent with the estimate of $r = 0.112\pm0.007$\,kpc for the compact [OIII] nucleus from \citealt{Tadhunter2018} (submitted). \par 

The density we derive from the total trans-auroral lines flux ratios is log n$_{e}$(cm$^{-3}$) = 3.05$^{+0.30}_{-0.45}$. However, because the broad component of the blue [SII] blend was  not detected, this is likely to be dominated by the narrow emission, leading to an under-estimate of the true outflow density. Based on this density, and E(B-V) = 0.70$\pm$0.15, for the maximal velocity case we estimate a mass outflow rate, $\dot{M}$ = 20.1$^{+72.9}_{-14.9}$ M$_{\sun}$ yr$^{-1}$ and kinetic power, $\dot{E}$/$L_{bol}$ = 2.5$^{+10.7}_{-1.9}$\%. The latter are the highest values measured for any object in the QUADROS sample; however, this may be a consequence of the fact that the densities have been under-estimated using
the total trans-auroral flux ratios.
\\
\\
\textbf{F23060+0505.} This object hosts the most powerful AGN out of the 8 ULIRGs in this paper, with its estimated bolometric luminosity of $4.6\times10^{45}$erg s$^{-1}$ placing it in the quasar category. \cite{Kim2002} showed it to be a single-nucleus system with a diffuse tidal feature to the south-west. Our total trans-auroral flux and Balmer decrement diagnostic measurements for the extracted nuclear spectra indicate a moderate degree of intrinsic reddening (E(B-V) $\sim$ 0.5) for the outflowing warm gas. \cite{Cicone2014} report evidence for a massive molecular outflow in this object, based on CO emission. Although they put an upper limit on the mass outflow rate of 1500 M$_{\sun}$ yr$^{-1}$, they note that the observations did not fully qualify as an outflow detection based on their criteria. \par

Despite the classification of this object as a type II object based on its optical spectrum, the steep rise in the continuum at the red end of the optical spectrum, and the detection of broad components ($>$2000 km s$^{-1}$) to both the H$\alpha$ and Pa$\alpha$ emission lines, indicate the presence of a moderately reddened type I AGN component \citep[][this paper]{Veilleux1999, Rodriguez2013}. Indeed, the [NII]+H$\alpha$ blend in our spectrum required a VB component to H$\alpha$, FWHM = 2359$\pm$69 km s$^{-1}$ for an acceptable fit. \par 

A four-component model was required to fit the [OIII] profile: two narrow (N1: $\Delta$v = 273$\pm$34 km s$^{-1}$; FWHM = 147$\pm$37 km s$^{-1}$, N2: $\Delta$v = -25$\pm$38 km s$^{-1}$; FWHM = 267$\pm$35 km s$^{-1}$), one intermediate ($\Delta$v = -283$\pm$44 km s$^{-1}$; FWHM = 934$\pm$32 km s$^{-1}$) and one broad component ($\Delta$v = -1220$\pm$146 km s$^{-1}$; FWHM = 1399$\pm$114 km s$^{-1}$). The [OIII] model did not successfully fit the [OII] and [SII] blends. For these, a three-component model based on [OII]3727 was used. \par

Based on our analysis, F23060+0505 hosts an ionised outflow with a radius of 1.41$\pm$0.19 kpc. No other radius estimates are available for comparison this object; however, given to the consistency between the different methods used for the other objects considered in this paper, we are confident in this value. \par

The density we calculate based on the total trans-auroral emission line ratios is log n$_{e}$(cm$^{-3}$) = 3.10$^{+0.35}_{-0.50}$. However, similarly to F17179+5444, the broad component of the blue [SII]$\lambda\lambda$4068,4076 blend was not detected, meaning this density is likely to be an under-estimate for the outflow region.
Despite this, for the maximal velocity case and E(B-V) = 0.45$^{+0.15}_{-0.20}$, we estimate a relatively low mass outflow rate ($\dot{M}$ = 2.2$^{+8.1}_{-1.7}$ M$_{\sun}$ yr$^{-1}$)m and ratio of kinetic power to bolometric luminosity ($\dot{E}$/$L_{bol}$ = (2.7$^{+13.2}_{-2.2})\times$10$^{-2}$\%). Despite claims of a massive molecular outflow in this object, the outflow in the ionised phase appears modest.
\\
\\
\textbf{F23233+2817.} \cite{Kim2002} and \cite{Veilleux2002} shows this source to have spiral-like morphology surrounding a very compact nucleus, with no visible tidal tails. They note that this is the only object in the original 1 Jy sample with no obvious sign of interaction. Evidence for molecular outflows in this source based on OH 119$\mu$m is presented in \cite{Veilleux2013}. Our trans-auroral and Balmer diagnostics indicate low intrinsic reddening for the source (E(B-V) $\sim$ 0.2) in the extracted nuclear spectrum.\par

The [OIII] profile requires a three component model: one narrow component ($\Delta$v = -92$\pm$27 km s$^{-1}$; FWHM = 239$\pm$21 km s$^{-1}$), one intermediate component ($\Delta$v = -316$\pm$31 km s$^{-1}$; FWHM = 760$\pm$13 km s$^{-1}$) and one broad component ($\Delta$v = -785$\pm$54 km s$^{-1}$; FWHM = 1892$\pm$40 km s$^{-1}$). This model was successful in fitting the other emission lines, however no broad component was detected in the [SII] blends. \par

Based on our analysis, F23233+2817 shows evidence for a relatively compact ionised outflow. The outflow is unresolved compared to the seeing, so we place an upper limit on the radius of 1.1 kpc. The density we obtain from the total trans-auroral emission line ratios is log n$_{e}$(cm$^{-3}$)  = 3.10$^{+0.30}_{-0.40}$. This, however, is likely to be dominated by the low-density narrow component due to the broad component of the [SII] blends being undetected. Nevertheless, based on this density, E(B-V) = 0.15$\pm$0.15 and the maximal velocity assumptions, we obtain lower limits on the mass outflow rate and kinetic powers of $\dot{M}$ > 0.11 M$_{\sun}$ yr$^{-1}$ and $\dot{E}$/$L_{bol}$ > 9.5$\times$10$^{-2}$\%.
\\
\\
\textbf{F23389+0303N.} This is the most extreme source of the 8 ULIRGs in terms of its [OIII] kinematics and trans-auroral ratios. Described as a close binary, separated by 5.2 kpc, by \cite{Kim2002} and \cite{Veilleux2002}, with a short tidal feature to the south, our spectroscopic extraction aperture was centred on the northern nucleus of the system. Evidence for blueshifted kinematics has been reported for neutral Na ID and OH 119$\mu$m in \cite{Rupke2005c} and \cite{Veilleux2013} respectively. \cite{Rupke2005c} place a lower limit on the neutral mass outflow rate of $\dot{M}$ > 49 M$_{\sun}$ yr$^{-1}$, and the kinetic power $\dot{E}$ > 10$^{42.39}$ erg s$^{-1}$. Along with F17179+5444, F23389+0303N is one of two objects in this paper which are radio loud ($L_{1.4GHz} = 10^{25.63}$ W Hz$^{-1}$). A radio map of this object is presented in \cite{Nagar2003}, showing two slightly resolved radio lobes separated by $\sim$ 830 pc.\par

The most striking features of the spectrum of this source, are the strengths and breadths of the [OI]$\lambda$6300 and trans-auroral [OII]$\lambda$7320 emission lines, which are stronger than the broad components to the [OIII] emission lines. It is very unusual to see such relative emission line strengths between these species. The great strength of the high-critical-density trans-auroral [OII] blend immediately suggests that we are sampling high density gas. \par

The [OIII] profile is very broad, requiring two components for an acceptable fit: one narrow ($\Delta$v = -191$\pm$27 km s$^{-1}$; FWHM = 402$\pm$16 km s$^{-1}$) and one very broad ($\Delta$v = -134$\pm$36 km s$^{-1}$; FWHM = 2346$\pm$38 km s$^{-1}$). Both components are blueshifted relative to the stellar rest-frame, however the very broad component, which dominates the flux, is slightly redshifted with respect to the narrow component. This kinematic model was successful at fitting all the other diagnostic emission lines, as illustrated in Figure \ref{fig:f23389}.\par

Our analysis indicates F23389+0303N contains one of the highest density outflows of the QUADROS sample, with the broad trans-auroral ratios leading to log n$_{e}$(cm$^{-3}$) = 4.20$\pm$0.10. It is also a highly reddened object, with E(B-V) $\sim$ 1. The outflow region is unresolved in our ground-based observations, and we place an upper limit on the radius of 1.2 kpc. However if we assume the outflow is jet-driven (see \citealt{Batcheldor2007, Tadhunter2014}, for example), we can use the radius of the radio lobes, r = 0.415 kpc \citep{Nagar2003} as a proxy for the radius of the warm outflow. This estimate, combined with the density derived from the broad trans-auroral component and E(B-V) = 0.90$\pm$0.05, leads to $\dot{M}$ = 10.4$^{+11.1}_{-5.1}$ M$_{\sun}$ yr$^{-1}$ and $\dot{E}$ = 10$^{43.2\pm0.3}$ erg s$^{-1}$ =  0.63$^{+0.79}_{-0.33}$\% $L_{bol}$ for the maximal velocity case.


\bsp	
\label{lastpage}
\end{document}